# Established DFT methods calculation of conjugation disturbed in the presence of torsional hyperconjugation

John N. Sharley, University of Adelaide. arXiv:1512.04171

john.sharley@pobox.com

## Table of Contents

## 1    Abstract

Accurate treatment of amide resonance is important in electronic structure calculation of protein, for Resonance-Assisted Hydrogen Bonding [1-3], RAHB, in the hydrogen bonded chains of backbone amides of protein secondary structures such as beta sheets [4] and alpha helices [5] is determined by amide resonance. Variation in amide resonance is the means by which the hydrogen bonding in these chains is cooperative.

Amide carbonyl orbitals are revealed by Natural Bond Orbital [6-8], NBO, analysis to substantially maintain sigma/pi separation in the presence of torsional hyperconjugative [6] interactions with wavefunction methods but not with established Density Functional Theory [9, 10], DFT, methods. This DFT error is most pronounced with small basis sets such as are used with DFT for proteins to reduce the basis function count. This error disturbs calculation of a range of amide donor-acceptor and steric interactions.

This finding has important implications for the selection of electronic structure methods and basis sets for protein calculations. For example, great caution is needed in interpreting the results of applying established DFT methods to proteins containing any beta sheets. We recommend that every protein DFT calculation be accompanied by NBO assessment of maintenance of amide carbonyl sigma/pi separation and absence of carbonyl bond bending. Further, we propose that these metrics be standard benchmarks of electronic structure methods and basis sets.

## 2  Notation

"->" denotes NBO resonance-type charge transfer and "|" denotes NBO steric exchange repulsion. "(" and ")" enclose specification of an orbital type and follow an atom name for single-center NBOs and a pair of atom names separated by "-" for two-center NBOs.

Examples: N(lp) for the amide nitrogen lone pair NBO, O(lp-p) for the oxygen p-type lone pair NBO, O(lp-s) for the s-rich lone pair NBO, C-O(p)* for the pi carbonyl antibonding orbital NBO and N(lp)->C-O(p)* for the primary amide resonance type charge transfer.

## 3  Overview

NBO analysis provides an optimal account of correlated electron density which is useful for determining differences in electron density. Differences in NBO-derived quantities are not arbitrary with respect to electron density. NBOs are not unitarily equivalent to molecular orbitals [11].

NBO is not committed to maintenance of sigma/pi separation of orbitals of multiple bonds. The NBO account of sigma/pi separation varies according to electron density. NBO analysis may reveal loss of sigma/pi separation in the presence of angular strain or chemical bonding. In the case of double bonds having hyperconjugative interactions, there exists a significant difference between the electron density calculated by established DFT methods versus MP2 [12] and in the extent of sigma/pi separation. This context is not particularly multireference. MP2 largely maintains sigma/pi separation and absence of bond bending, while DFT methods show large variation in the extent of sigma/pi separation and bond bending. This DFT error is sensitive to basis set and in vinylamine is reduced at the correlation consistent basis set [13] level aug-cc-pV5Z though remains significant at aug-cc-pV6Z.

Calculated amide electronic interactions disturbed by this error are not limited to loss of sigma/pi separation and carbonyl bond bending, and include variation of charge transfer from the nitrogen lone pair to carbonyl antibonding orbitals and steric interactions C-O(s)|C-O(p), O(lp-s)|C-O(s) and O(lp-s)|C-O(p).

Vinylamine, ethanamide and polyvaline and polyalanine antiparallel beta sheets are used for these investigations. Ethanamide is used as a minimal example of the amide group having hyperconjugative interactions with the carbonyl group. Results of applying 43 non-double hybrid DFT [14, 15] methods in combination with 9 basis sets are reported. That these DFT issues also apply to protein beta sheets is shown with LC-wPBE [16], though these issues are not limited to that method.

Reduction of sigma/pi separation and increase in bond bending in the amide carbonyl group gives a reduction in calculated amide resonance. This necessarily means that RAHB of hydrogen-bonded chains of protein backbone amide groups is inaccurately calculated. RAHB is cooperative, and errors in calculated amide resonance may also be regarded as cooperative.

These findings have important implications for the selection of electronic structure methods for protein structure. Depending on method/basis set, any calculation of a range of internal amide electronic interactions may have a surprisingly large error. Results of calculation of cooperativity in backbone amide hydrogen bond, HB, chaining within a beta sheet using DFT will be significantly in error even when medium size basis sets are used, and in practice small basis sets have been used due to the secondary structure atom and basis function count. Hyperconjugative interactions with the backbone amide carbonyl group occur in other hydrogen bonded secondary structure types, so this warning applies to proteins generally. Also, we have observed but do not otherwise report on variability in the amino-acid residue-specific extent of loss of sigma/pi separation and increase in bond bending. We recommend that every DFT calculation of protein be accompanied by NBO assessment of the extent of amide carbonyl sigma/pi separation and bond bending.

These findings lead to the proposal that these molecular situations be regarded as standard benchmarks of electronic structure methods and basis sets, and be used for refining parameter values where a method takes parameters.

A second kind of error arising from double bonds having hyperconjugative interactions is also reported here. This second error applies to the amide group, and results in pyramidalization at the amide nitrogen. It occurs depending on basis set when using wavefunction methods. The result of this error may be seen in coordinate geometry, though NBO may be used to quantitate the extent of the shift of nitrogen from sp2 to sp3 hybridization. This error is similar to the benzene non-planarity failures arising from negative out-of-plane bending frequencies reported in [17]. This error also reduces amide resonance since the nitrogen sp3 hybridization is necessarily in competition with amide resonance. The above remarks about accuracy in calculations of RAHB also apply to this error. This second error is corrected by geometry optimization on the Counterpoise Correction [18] potential energy surface with fragment boundary defined at the C-N bond of ethanamide, but not at the C-C bond.

# 4 Introduction

The Natural Bond Orbital analysis procedure has properties that particularly recommend it for the analysis of electron density. Natural orbitals are natural in the sense of being the best possible account of correlated electron density [6], which suggests their use as the standard form for comparison of electron density. The localization of natural orbitals without loss of information as is done by NBO increases suitability for this purpose. NBO is usually directed to analysis of bonding similarity across molecular contexts, but here we use NBO to analyse the bonding differences resulting from different methods applied in the same molecular context.

The present study focuses in particular on the Natural Hybrid Orbitals [6], NHOs, for these describe bond bending and sigma, pi and higher angular momentum character of bonds. There is no built-in bias or geometric preconception embedded in NHO description of bonds, and all aspects of hybridization are converged to an optimal description of electron density [6]. The NBO account of electron density can report reduction in sigma/pi separation and increase in bond bending, depending on the electron density. As mentioned above, differences in NBO-derived quantities are not arbitrary with respect to electron density.

Density Difference Representation [19] is not oriented to chemical bonding and so the chemical significance of density change is not available in that representation.

NBO is the only chemical bonding analysis used in the present study, and justification for this is left to a recent review [20] which is very briefly summarized in the present paragraph. NBO preserves Pauli exclusion and Hermitian energetics and is thus free of the overlapping attributions arising from orbital or wavefunction non-orthogonality. NBO is based only on the *eigen* or intrinsic properties of the interactions of the electronic first-order reduced density matrix (1-RDM), rather than any assumptions concerning geometry or symmetry and is independent of the form of the original wavefunction. This basis in the *eigen* properties of the 1-RDM yields consistency and predictive capacity.

We used NBO to quantify the extent of sigma/pi separation and bond bending in double bonds having hyperconjugative interactions in vinylamine, ethanamide and an antiparallel beta sheet. In vinylamine and ethanamide, torsion is constrained across the single bond central to hyperconjugation during geometry optimization. In beta sheets, no geometry optimization constraints are necessary to maintain the torsion for non-planar hyperconjugation. The surveys reported here use constrained ethanamide as a model for protein backbone amide groups, and the relevance of this model is supported by NBO analysis of fully geometry optimized polyalanine and polyvaline protein secondary structures.

Backbone amide resonance is central to RAHB in protein backbone amide hydrogen bond chaining of certain protein secondary structures including beta sheets. Accurate calculation of RAHB cooperativity is important in calculating protein electronic and geometric structure.

## 5 Methods

Methods used in experiments are as implemented by Gaussian 09 D.01 [21], Orca 3.0.3 [22-24] and TeraChem 1.5K [25-28].

A pre-release version of NBO [29] was used for its XML [30] output option. The XML was queried with XQuery 3.0 [31] or XSLT 3.0 [32] as implemented by Saxon-PE 9.6.0.4 [33], and the results imported into Excel 2013 [34].

Jmol 14.2.2_2014.06.29 [35] was used for visualization of orbitals.

Except where otherwise stated, the default integration grid of the respective quantum chemistry package is used. Except where otherwise stated, the SCF convergence default was used for Gaussian, and VeryTightSCF was specified for Orca. The differing default SCF procedures were used for Gaussian and Orca. Where results differ between packages, the possibility that these differences are due to differences in default parameters is explored. In the case of the second problem mentioned in the overview, the results of Gaussian and Orca differ, and non-default values for integration grids, SCF convergence values and SCF procedures are explored.

## 6 Results and discussion

### 6.1 Vinylamine

As a preliminary note, for charts with many methods named on the horizontal axis, the first of which is Ap1:Figure 25, Gaussian's implementation of standalone functionals are grouped according to the classifications of pure, hybrid, range separated hybrid and double hybrid [36]. VSXC [37] to N12 [38] are pure functionals, B3LYP [39] to MN12-SX [40] are hybrid functionals, wB97 [41] to LC-wPBE are range separated hybrids and B2-PLYP [14] to mPW2-PLYPD [42] are double hybrids. The Gaussian implementation of MP2 and HF appear to the right of mPW2-PLYPD. Orca implementations of methods appear to the right of HF. MP2 appears twice, once as a Gaussian implementation and once as an Orca implementation.

The results appearing in [6] of loss of sigma/pi symmetry in vinylamine at 10 degree C-C-N-H Torsion with B3LYP/6-311++G** are in accord with our results (Figure 1 and Ap1:Figure 25). However, the ~20% loss of p character with B3LYP becomes less than 1% with MP2 (Ap1:Figure 23 and Ap1:Figure 25). Only with the smallest basis sets used, 6-31G** and def2-SVP, does the loss of p character with MP2 approach 1%, and with other basis sets is ~0.5%. Of the basis sets used, 6-311++G** gives the least loss of p character with MP2.

The C-C NBO morphologies for B3LYP/6-311++G** and MP2/61-311++G** can be seen in Figure 6 to Figure 13. These graphics are for 10 degree torsion, which we use for commonality with [6], though greatest loss of sigma/pi symmetry is found at 15 to 30 degrees hyperconjugative torsion depending on basis set except for aug-cc-pVTZ (Figure 1 and Ap1:Figure 23).

NHO deviation from line of centers, bond bending, also differs significantly between DFT methods and wavefunction methods. These deviations are charted as Figure 2, Ap1:Figure 24 and Ap1:Figure 26. For the C-O(pi) NHOs, 90 degrees is no deviation from expectation based on line of centers. For some basis sets with B3LYP the bond bending is as large as 45 degrees from the reference 90, contrasting with the 7 degree maximum bond bending with MP2.

Differences in the polarization of the C-O(sigma) and C-O(pi) NBOs is shown in Ap1:Figure 27 and Ap1:Figure 28. The variable performance of different DFT methods with aug-cc-pVTZ can be seen. Broadly, aug-cc-pVTZ is most associated with greatest difference between DFT methods, and between DFT methods and wavefunction methods. aug-cc-pVDZ offers markedly less DFT divergence, which is frequently observed in this work. As is usual in this work, there is greatly less divergence between properties calculated with correlated wavefunction methods than with DFT methods.

The B3LYP and MP2 results are dichotomous, and nomination of which is in such large error is most desirable. It is not a given at the outset that the error is with B3LYP, for DFT methods are in principle capable of incorporating static correlation and uncorrected MP2 is not. In answering this question we refer to the results of other DFT methods including double hybrid methods [14], corrected MP2 methods [43, 44], DLPNO-CCSD(T) [24] and the multireference MRCI+Q [45], and note that B3LYP loss of symmetry begins to collapse at very large basis sets starting at aug-cc-pV5Z (Figure 3 to Figure 5).

Ap1:Figure 25 to Ap1:Figure 28 show sigma/pi symmetry, bond bending and polarization at a range of DFT methods including hybrid methods, long-range corrected methods [46] and double hybrid methods, and some corrected MP2 methods, the invariably worst-performing method HF [34] and single point calculation at DLPNO-CCSD(T). The double hybrid [14] methods tested have only twice the small error of MP2, but computational cost scales similarly.

The poor performance of HF and single hybrid methods and better performance of double hybrid and MP2 variants yields suggests an underlying mechanism of these errors, that being different treatment of multicenter exchange delocalization. DFT single hybrid and GGA exchange does not have multicenter delocalization, HF's multicenter exchange delocalization is uncorrected, MP2's correlation largely corrects HF's excessive multicenter delocalization [47] and double hybrid DFT's correction by correlation is not quite as good as MP2's.

We anticipate that with optimization at CCSD [48] or above, the loss of p character will be further reduced from the already small MP2 figure. Where DLPNO-CCSD(T) has been used in this work, it has necessarily been used for single point energy calculation, with geometry optimization at MP2 or SCS-MP2 and aug-cc-pVTZ, aug-cc-pVQZ or def2-QZVPP.

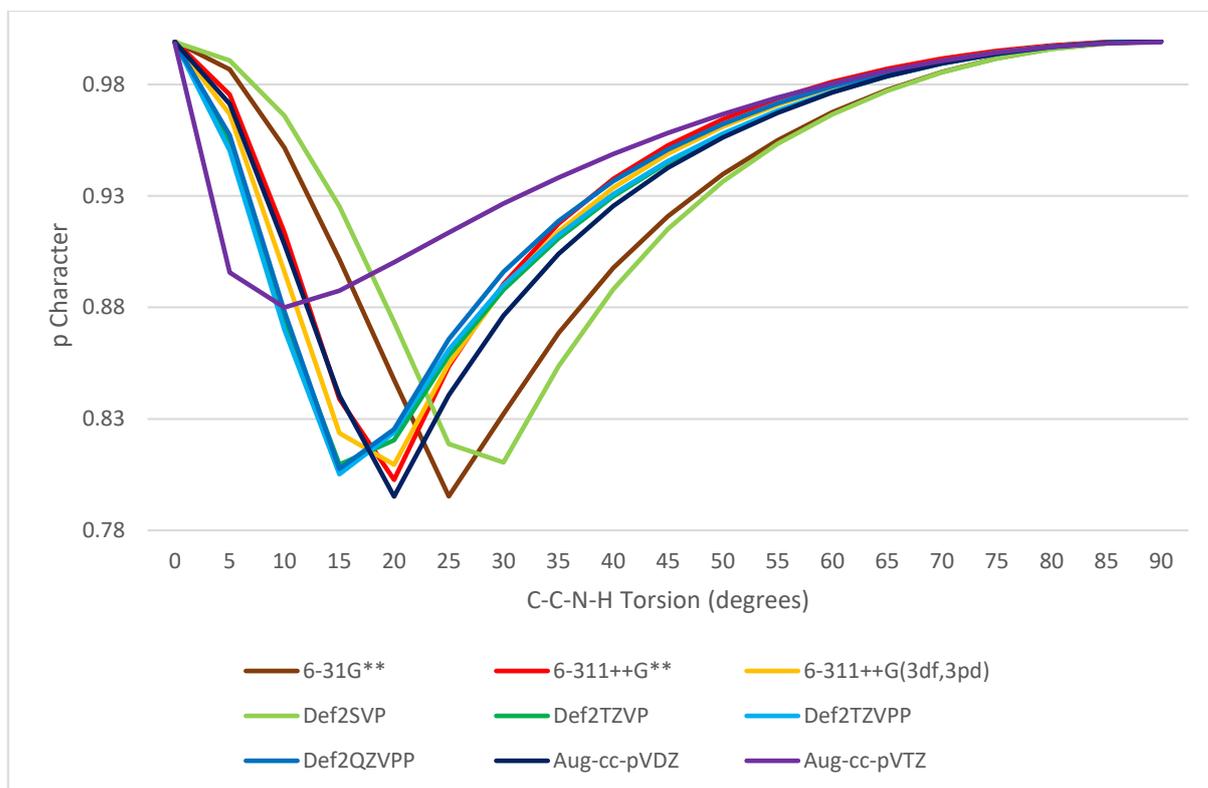

Figure 1. C-C(pi) Central Carbon NHO p Character in Vinylamine at Varying C-C-N-H Torsion and Basis Set with B3LYP

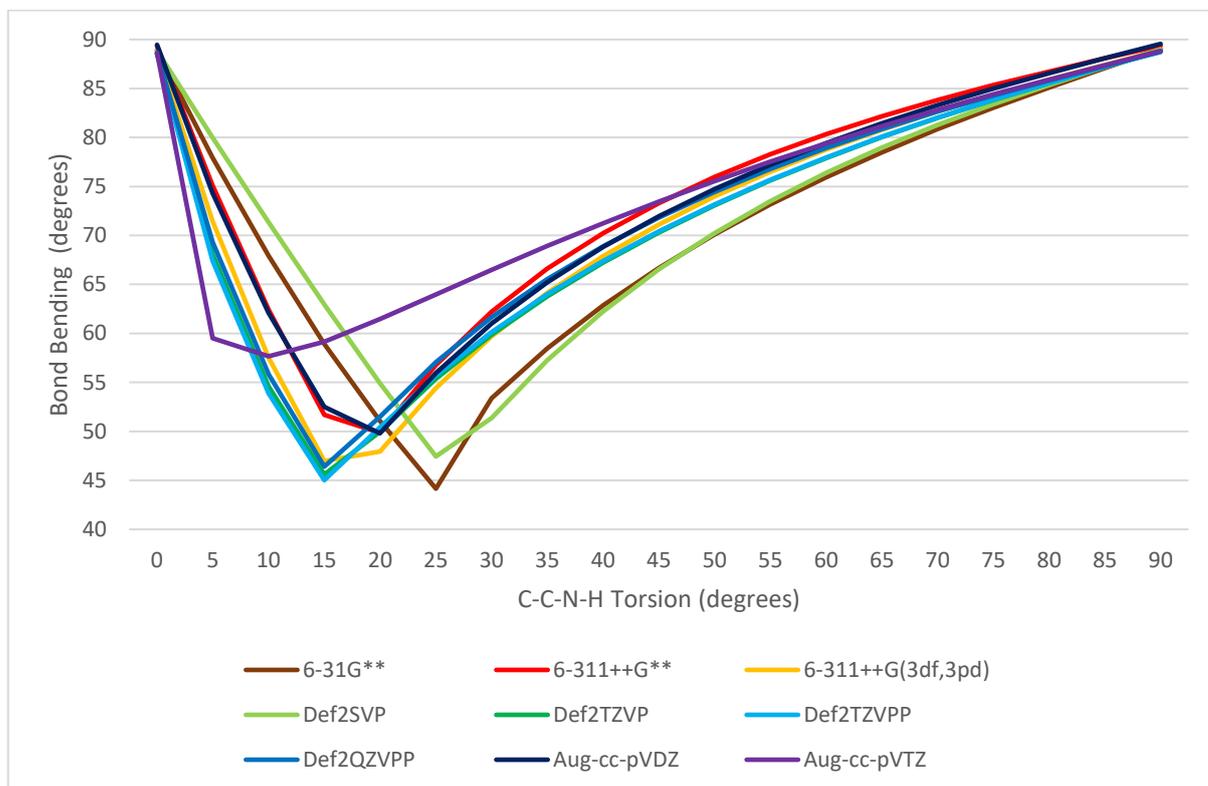

Figure 2. C-C(pi) Central Carbon NHO Bond Bending in Vinylamine at Varying C-C-N-H Torsion and Basis Set with B3LYP

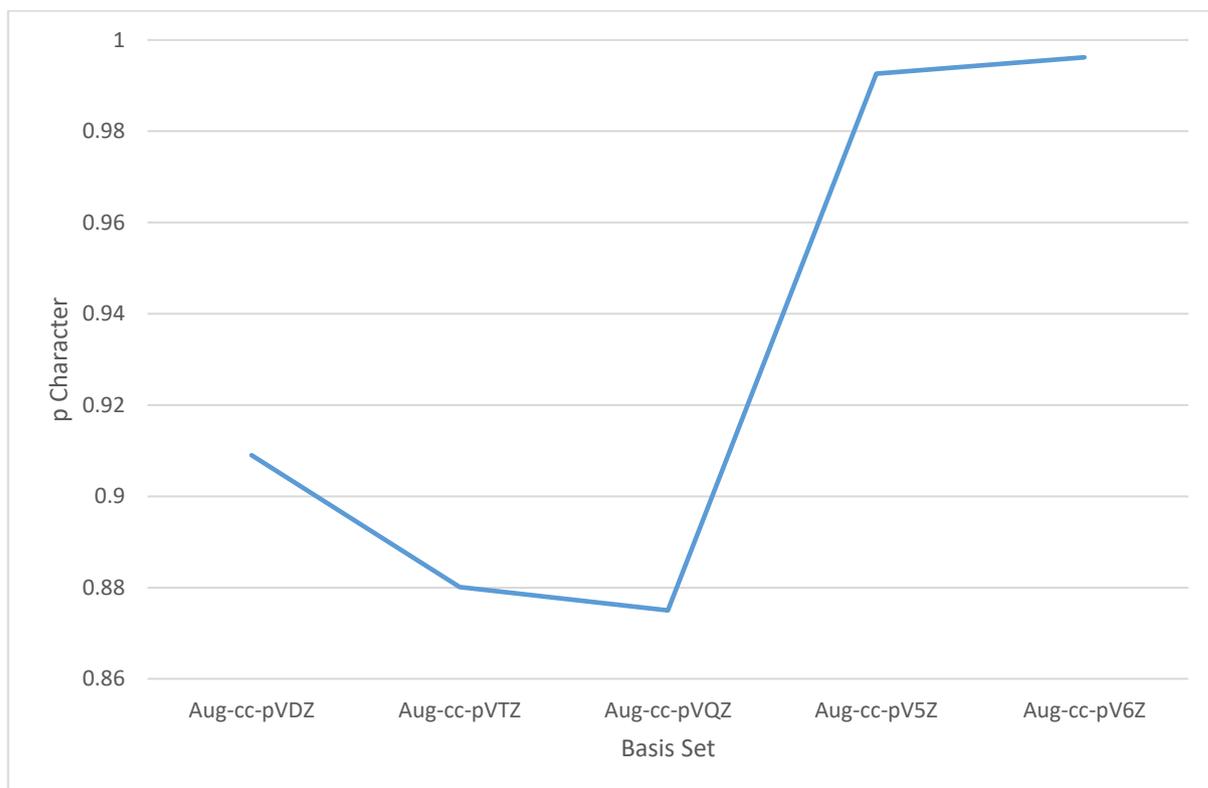

Figure 3. C-C(pi) Central Carbon NHO p Character in Vinylamine at 10 Degree C-C-N-H Torsion with B3LYP and Varying Basis Set

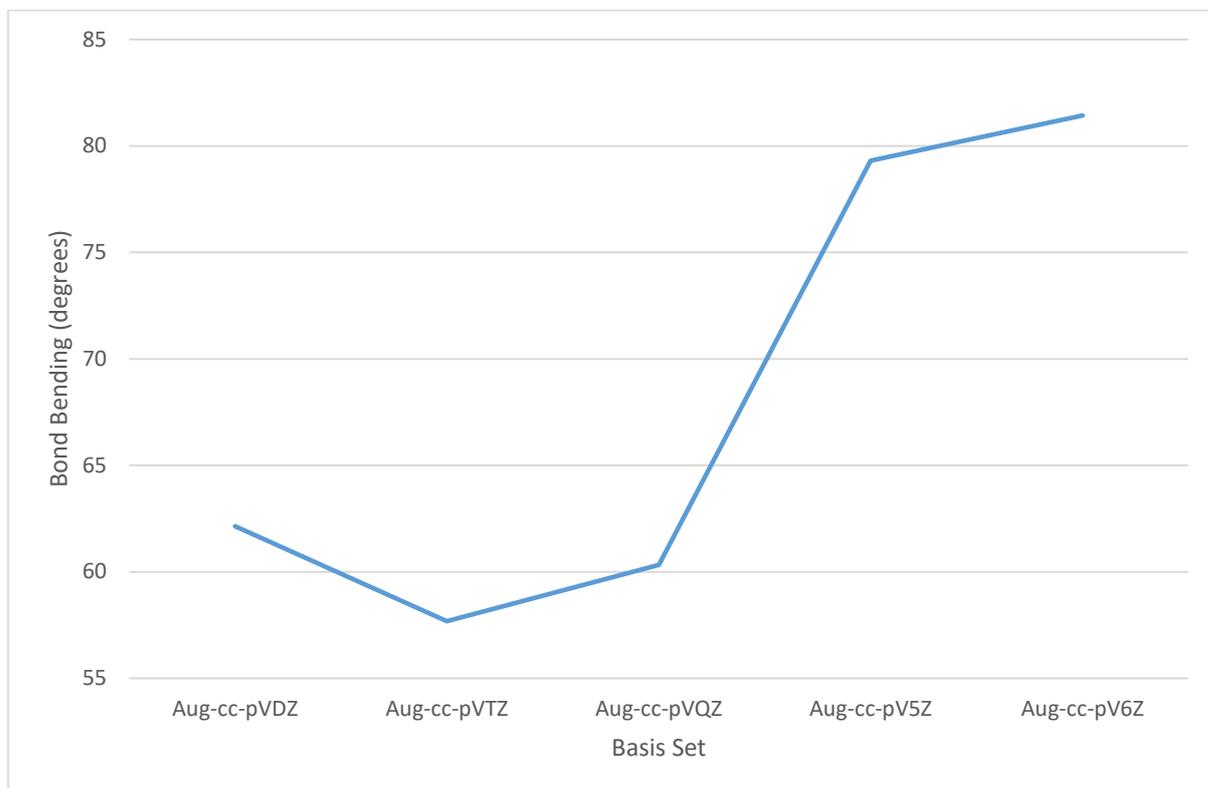

Figure 4. C-C(pi) Central Carbon NHO Bond Bending in Vinylamine at 10 Degree C-C-N-H Torsion with B3LYP and Varying Basis Set

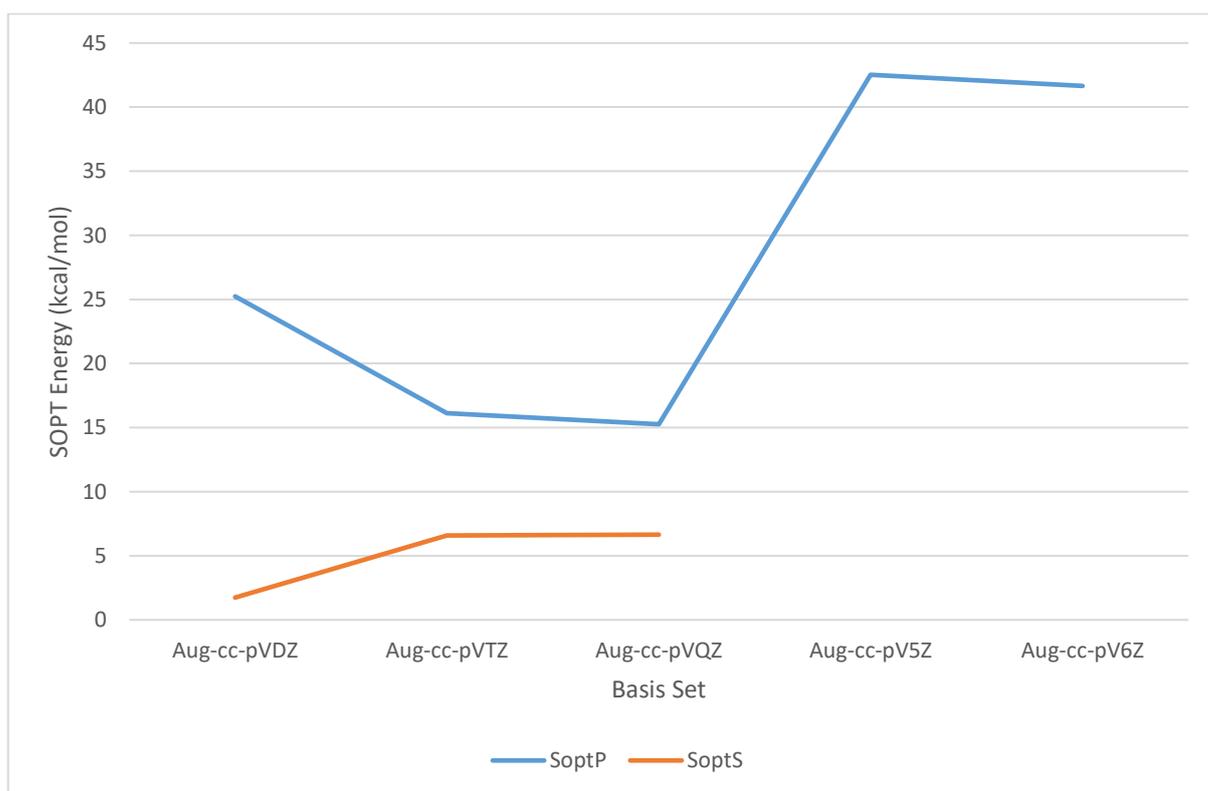

Figure 5. N(lp)->C-C(pi then sigma)* SOPT Energy in Vinylamine at 10 Degree Torsion with B3LYP and Varying Basis Set

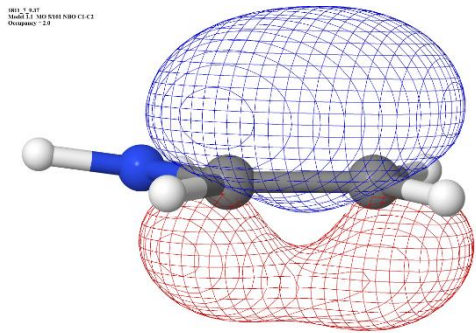

Figure 6. B3LYP/6-311++G** C-C(pi) NBO in Vinylamine at 10 Degree C-C-N-H Torsion

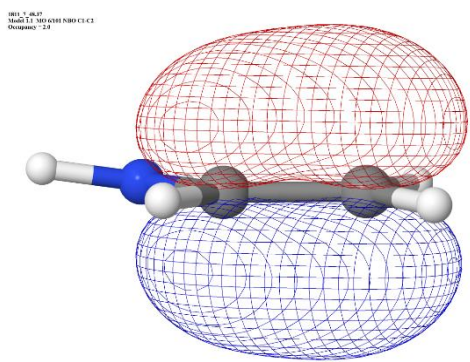

Figure 7. MP2/6-311++G** C-C(pi) NBO in Vinylamine at 10 Degree C-C-N-H Torsion

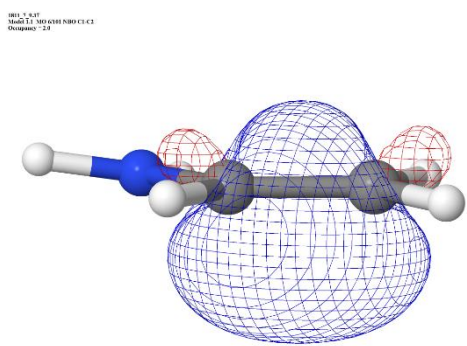

Figure 8. B3LYP/6-311++G** C-C(sigma) NBO in Vinylamine at 10 Degree C-C-N-H Torsion

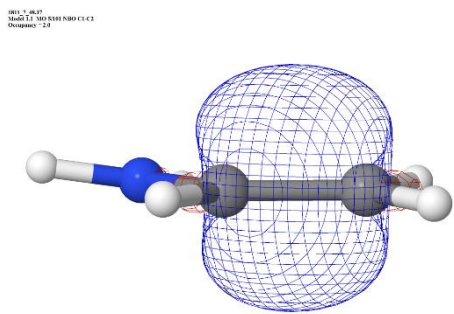

Figure 9. MP2/6-311++G** C-C(sigma) NBO in Vinylamine at 10 Degree C-C-N-H Torsion

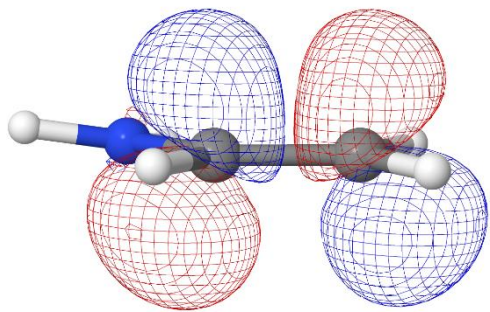

Figure 10. B3LYP/6-311++G** C-C(pi)* NBO in Vinylamine at 10 Degree C-C-N-H Torsion

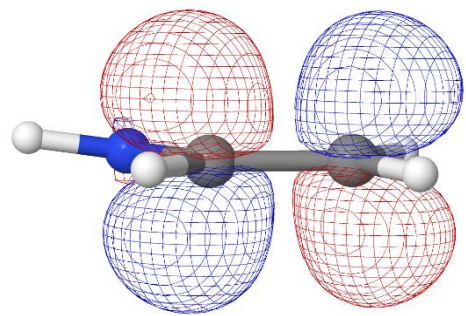

Figure 11. MP2/6-311++G** C-C(pi)* NBO in Vinylamine at 10 Degree C-C-N-H Torsion

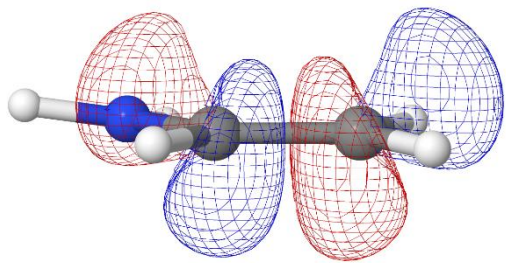

Figure 12. B3LYP/6-311++G** C-C(sigma)* NBO in Vinylamine at 10 Degree Torsion

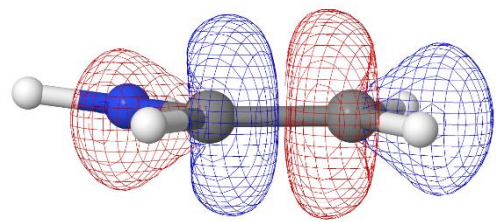

Figure 13. MP2/6-311++G** C-C(sigma)* NBO in Vinylamine at 10 Degree Torsion

## 6.2 Ethanamide

### 6.2.1 Carbonyl orbital disturbances

Ethanamide is the simplest molecule that can demonstrate the hyperconjugative interactions with the amide carbonyl group that may be seen in peptide backbones. In parallel beta strands, the O-C-CA-HA torsion may be taken as 173 degrees and in antiparallel beta strands as -165 degrees [49], being 7 and 15 degrees respectively from being perfectly antiperiplanar. We denote the ethanamide non-amide carbon as CA for commonality with peptides. Given the equivalence of the hydrogens connected to CA in ethanamide, rather than refer to them as HA1, HA2 and HA3, we refer to them generically as H, referring to them by O-C-CA-H dihedral angle.

Figure 14 and Ap1:Figure 29 contrast the difference in loss of sigma/pi symmetry of the carbonyl orbitals when B3LYP or MP2 is used. The y axis scales differ markedly, for the B3LYP loss of p character is ~5% for basis sets 6-31G** and def2-SVP at 175 degrees O-C-CA-H torsion. Of secondary significance, with basis sets def2-TZVP, def2-TZVPP and aug-cc-pVTZ the loss of p character approaches 1.5% at 165 degrees torsion. The data shows that a geometry away from perfectly antiperiplanar is necessary for hyperconjugation to disturb sigma/pi symmetry, and later figures bear out disturbances of other properties also require a geometry away from perfectly antiperiplanar. The 180 to 60 degree sweep of O-C-CA-H dihedrals does not quite complete the cycle of values, for only one H in CA-H is constrained, an arrangement that was selected to mimic peptide backbones in which only one of the substituents of CA is torsionally constrained. The unconstrained H that is moving into antiperiplanar dihedral angle does not remain at precisely 120 degree relative to the constrained H throughout the rotation.

Hyperconjugation between the sigma carbonyl orbitals and CA-H is centered at O-C-CA-H torsion of 180 degrees, and hyperconjugation between the pi carbonyl orbitals is centered at O-C-CA-H torsion of 90 degrees. Disturbances of properties at torsional angles associated with pi carbonyl hyperconjugation can be seen Ap1:Figure 31, Ap1:Figure 32 and Ap1:Figure 35.

Figure 16 shows a reduction in Second-Order Perturbative Analysis, SOPT, of donor-acceptor interactions [6] for the primary amide resonance delocalization of ~61 kcal/mol to ~41 kcal/mol in 5 degrees of torsion from antiperiplanar with B3LYP and basis set 6-31G**, and also large reduction with def2-SVP. Ap1:Figure 22 shows the N(lp) delocalization into C-O(s)* rather than C-O(p)*. However, with 6-311++G**, the variation in amide resonance delocalization throughout the range of torsion is only ~2 kcal/mol. This is a very large variation between basis sets.

Ap1:Figure 19 shows that the steric exchange energy of C-O(s)|C-O(p) increases from 0 to ~25 kcal/mol with 6-31G** in 5 degrees of O-C-CA-H torsion from antiperiplanar. Ap1:Figure 20 shows that in the same torsional range O(lp-s)|C-O(p) increases from 0 to ~7.5 kcal/mol with 6-31G**. Ap1:Figure 21 shows that in this torsional range O(lp-s)|C-O(s) decreases by ~6 kcal/mol with 6-31G**.

N(lp)->C-O(p)* SOPT Energy is shown with varying torsional hyperconjugation for a selection of Minnesota density functional methods [50] in Ap1:Figure 37 and other DFT methods from the Truhlar group in Ap1:Figure 38. MN12-L [51], MN12-SX and M06-HF [52] are less afflicted of error at torsional hyperconjugation than other DFT methods tested here. Each of these three methods is tested with a range of basis sets, and results are shown in Ap1:Figure 58, Ap1:Figure 56 and Ap1:Figure 57 respectively. Of the three methods, MN12-L is the best performer, which is convenient from the point of view of protein calculations since it is a local and hence better scaling method [53]. MN12-SX is a range separated hybrid and M06-HF is a global hybrid [53], so local, range separated versus global nature of methods is not key to alleviation of the error at torsional hyperconjugation. They are all meta-GGA methods, with MN12-L being at only the third rung of Jacob's ladder [54, 55] and having zero percentage exact exchange.

While it is variation in rather than absolute value of N(lp)->C-O(p)* SOPT energy due to torsional hyperconjugation that is the focus of interest here, it must be remarked that the difference in this quantity between the methods at perfectly antiperiplanar geometry is the next concern in assessing DFT methods applied to proteins. At the highest quality basis set used, def2-QZVPP, the difference in N(lp)->C-O(p)* SOPT Energy between MN12-L and M06-HF is greater than 40 kcal/mol. This difference in calculating amide resonance will have large consequences for calculated RAHB. While MN12-L/def2-QZVPP or linear scaling variant thereof might be recommended for application to proteins on the basis of appropriate relative amide resonance response to torsional hyperconjugation, error in calculating the absolute value of amide resonance is a major obstacle. The occupancies of N(lp) and C-O(p)* in ethanamide calculated with wavefunction methods and DFT methods may be compared to gauge amide resonance. N(lp) occupancy, calculated at high quality basis sets, can be seen in Ap1:Figure 50, Ap1:Figure 51 and Ap1:Figure 52 and C-O(p)* occupancy can be seen in Ap1:Figure 53, Ap1:Figure 54 and Ap1:Figure 55. C-O(p)* occupancies are greater with DLPNO-CCSD(T) than with any of the DFT methods shown, with MN12-L having the largest C-O(p)* occupancies of the three favoured Minnesota density functionals. CASSCF(8,7)/MRCI+Q/def2-TZVPP in turn gives greater C-O(p)* occupancy than DLPNO-CCSD(T). While the second configuration weights of CASSCF/MRCI+Q are less than 0.02 (Table 2) indicating the calculation is not particularly multireference, it seems that improvement on even CCSD(T), the gold standard of single reference quantum chemistry, is needed for calculation of amide resonance. Multireference methods that scale to proteins may be necessary for calculation for protein RAHB.

The influence of the D2 [56], D3 [27] and D3BJ [28] empirical dispersion corrections on the torsional hyperconjugative error reported here is shown where available for selected methods and the 6-31G** basis set in Ap1:Figure 42, Ap1:Figure 43 and Ap1:Figure 44 respectively. None of the empirical dispersion schemes tested alleviates the torsional error, though D3BJ alone can be recommended as not making the error any worse.

The influence of the omega parameter of LC-wPBE with 6-31G** on the torsional hyperconjugative error reported here is shown in Ap1:Figure 46. It can be seen that no listed value gives a result usable for proteins, and values less than 0.4 bohr$^{-1}$ give the worst results. Gaussian 09 D.01 has no label for LC-wPBEh [57], so the equivalent LC-PBEhPBE is used. The influence of the HF exchange coefficient on the error is shown for selected values of omega in Ap1:Figure 46, Ap1:Figure 47 and Ap1:Figure 48. At omega=0.6 bohr$^{-1}$, small exchange coefficients are least disfavoured, and at omega=0.4 bohr$^{-1}$ and omega=0.2 bohr$^{-1}$, large exchange coefficients are least disfavoured. It is likely that there exist no values for omega and the exchange coefficient that yield a method useful for proteins.

To explore whether the hyperconjugative error is due more to the exchange functional or to the correlation functional of pure DFT methods, all combinations of exchange functionals and correlation functionals listed in the Gaussian 09 User's Manual [36] tested for this error with 6-31G**. Results are shown for all listed exchange functionals with the PBE correlation functional [58] (Ap1:Figure 60), the VWN correlation functional [59] (Ap1:Figure 61) and the LYP correlation functional [60] (Ap1:Figure 62). With the exception when in combination with the exchange functionals which are quite erratic during torsional change, the choice of correlation functional is relatively unimportant in this test. The G96 exchange functional [61] has the shallowest N(lp)->C-O(p)* SOPT energy wells through O-C-CA-H torsion with all the listed correlation functionals. Results are shown for all listed correlation functionals with the XA exchange functional [62-64] (Ap1:Figure 63), the BRx exchange functional [65] (Figure 64) and the Slater exchange functional [62-64] (Ap1:Figure 65), and in these figures the curves are usually so little different that they partially overlay, again indicating that the choice of correlation functional is secondary in these tests. This is further shown in Ap1:Figure 59 by the use of the three listed exchange functionals that can standalone compared with results when combined with the VWN correlation functional. The torsional hyperconjugation error principally arises of errors in exchange functionals.

The variation in C-O(s)* and C-O(p)* energy levels for selected Minnesota density functionals with 6-31G** during change of O-C-CA-H torsion is shown in Ap1:Figure 39 and Ap1:Figure 40 respectively. The torsional hyperconjugation error is associated with the energy levels of C-O(s)* and C-O(p)* becoming closer. Decrease in the C-O(s)* energy level makes the orbital a more inviting charge transfer acceptor. The torsional hyperconjugative error is likely due to re-hybridization unduly favouring hyperconjugative charge transfer or sigma acceptor orbitals. The C-O(p)* energy level variation for MN12-L for selected basis sets is shown in Ap1:Figure 41. 6-311++G** behaves erratically, though 6-311++G(3df,3pd) performs well as determined by the basis set we suggest be taken as canonical, def2-QZVPP. There are two efficiency merits of MN12-L/def2-QZVPP, being that MN12-L is local and that def2-QZVPP does not have diffuse functions. The accuracy of MN15-L [66] in calculating amide resonance will be investigated in future, though the accuracy demands of calculating resonance in RAHB where errors are cooperative are likely forbidding.

The influence of basis set diffuse functions on the torsional hyperconjugation error is explored for B3LYP (Ap1:Figure 66) and MN12-L (Ap1:Figure 67). The introduction of diffuse functions to 6-31G** to form 6-31++G** is very beneficial. The torsional hyperconjugation error is very large for cc-pVDZ which has no diffuse functions. The calendar variations of correlation consistent basis sets are shown, though the lines are not distinct for apr-cc-pVDZ, may-cc-pVDZ and jun-cc-pVDZ for B3LYP and apr-cc-pVDZ and may-cc-pVDZ for MN12-L. The aug-cc-pVDZ curve has the deepest wells of any of the calendar variations of cc-pVDZ, and the removal of the diffuse functions on hydrogen to form jul-cc-pVDZ somewhat flattens the curve. Further removal of diffuse functions to form the other calendar correlation consistent basis sets makes little difference. A wider selection of Pople basis sets is shown in Ap1:Figure 68, and the introduction of polarization functions is associated with a worsening of the torsional hyperconjugation error. A wider selection of correlation consistent basis sets is shown in Ap1:Figure 69, and it can be seen that very high quality correlation consistent basis sets (aug-cc-pVQZ) give energy wells during O-C-CA-H for primary amide SOPT that are significantly deeper than the 2.7 kcal/mol given by MN12-L/def2-QZVPP. The good performance of MN12-L/def2-QZVPP (though absolute amide resonance irrespective of O-C-CA-H torsion must also be considered) on ethanamide and the poor performance of B3LYP/6-311++G** on vinylamine indicate that diffuse functions are not in general necessary or sufficient to accurately model torsional hyperconjugation in conjugative systems. Any method/basis combination for proteins must be benchmarked on the torsional hyperconjugative problem as described here. Further, the introduction of diffuse functions markedly increases the computational cost of electronic structure methods and increases convergence difficulties, most often prohibiting their use on even the smallest proteins.

### 6.2.2 Pyramidalization at nitrogen

Figure 17 shows how erratic 6-311++G** is with MP2. This accords with the modern practice of using correlation consistent rather than Pople basis sets [67] with wavefunction methods. MP2/6-311++G** is greatly disturbed by correlation associated with each of sigma and pi carbonyl hyperconjugation. Figure 18 shows the loss of O-C-N-H planarity associated with this disturbance. During the rotation, planarity is lost by 15 degrees in both directions from the amide plane. Correlation associated with hyperconjugation makes sp3 pyramidalization of N more favourable than planarity, weakening the amide resonance. At 95 degrees torsion, just 5 degrees from maximum pi hyperconjugation, the dihedral describing the non-planarity changes sign, breaking the symmetry centered at 120 degrees O-C-CA-H torsion. Non-planarity is not limited to Pople basis sets. With def2-TZVP, the non-planarity is quite significant at ~7 degrees, with def2-QZVPP at ~5 and aug-cc-pVTZ at ~3 degrees. At the hyperconjugative torsions at which amide non-planarity occurs, calculated amide resonance is impaired and necessarily so is the RAHB in which the amide might participate and hence cooperativity in hydrogen bonded secondary structures. The resonance of only one amide group in a hydrogen bonded chain need be erroneously calculated for RAHB throughout the chain to be in error. Since these errors all result in reduction of amide resonance, there are no compensating errors. Errors in the calculation of the resonance of multiple groups cooperate in producing error in calculation of the hydrogen bonding of the chain. Due to this cooperativity of errors, it is quite important that amide resonance be accurately calculated. def2-QZVPP and aug-cc-pVTZ are of reasonable size and repute, and this pyramidalization calls into question the use of MP2 for geometry optimization of peptides/proteins. This test is suitable for benchmarking during and after development of MP2 variants. Perhaps CCSD does not fail in this manner, but optimization with CCSD with large basis sets is not yet viable for extensive studies.

The findings of pyramidalization at amide N with MP2 are similar to a report of non-planarity of benzene [17]. Basis sets passing the benzene planarity test by having positive vibrational frequencies at planarity, such as aug-cc-pVTZ, do not fare altogether well in the geometry optimization tests used here although the correlation consistent basis sets are constructed to provide basis set incompleteness error, BSIE, balance. The report attributes the failures to elevated sigma-pi correlation. The report gives that one-electron theories such as DFT are immune to this two-electron BSIE. The errors with DFT described in the present work are then of different origin, but will still be a result of incorrect emphasis on certain correlations. The report points out that Atomic Natural Orbital [68, 69], ANO, basis sets are better again than correlation consistent basis sets, since ANO minimizes basis set superposition errors. We have not used ANO basis sets due to their computational expense. This pyramidalization at MP2/6-311++G** is correctable by geometry optimization on the Counterpoise Correction potential energy surface with fragment boundaries defined at the C-N bond of ethanamide, but not at the C-C bond (Ap2:Table 3).

Ap1:Figure 72 shows that the SCS correction to MP2 slightly attenuates resonance-type delocalization from the nitrogen lone pair NBO, but not in conjunction with Orbital Optimization. The result of DLPNO-CCSD(T) single point calculation over geometry optimized with MP2/aug-cc-pVQZ is in keeping with the geometry optimized SCS-corrected but not orbital-optimized results. This suggests suitability of SCS-MP2 optimization for DLPNO-CCSD(T) single point calculation for such studies. Since the SCS-MP2 and RI-SCS-MP2 results do not differ appreciably, RI-SCS-MP2 is useful for its reduced run-times. The DFT methods that produce results closest to those of SCS-MP2 are VSXC, tHCTH [70], B97D3 [28, 56] and N12. The furthest are by M11-L [71], M06 [72], wB97 and LC-wPBE, and these will underestimate amide resonance. As always, HF is far outlying and greatly underestimates amide resonance. Further light is cast on the irregular M11-L results seen in figures Ap1:Figure 70, Ap1:Figure 71, Ap1:Figure 73 and Ap1:Figure 74 by the O-C-N-H dihedral values seen in Ap1:Figure 75 in which variation with basis set is ~20 degrees. M11-L is subject to the pyramidalization at nitrogen seen with MP2. It can be seen in Figure 37 and Ap1:Figure 38 how low the N(lp)->C-O(p)* SOPT energy of M11-L, M06L, M06, N12 and SOGGA11 become with the noted standardized geometry. Since pyramidalization at amide N and N(lp)->C-O(p)* are necessarily in competition, it is to be expected that low N(lp)->C-O(p)* is associated with pyramidalization at N. It is anticipated that this is the connection between carbonyl orbital disturbances and pyramidalization at N, and how DFT and wavefunction methods can both manifest pyramidalization at N. The ~13 degree pyramidalization seen with Gaussian at MP2/6-311++G** is not seen with Orca even when the NoFrozenCore option is used for greater commonality with Gaussian defaults. The Gaussian option Stable=Opt reveals that the wavefunctions for both M11-L and MP2 are stable at the optimized geometry. Gaussian Stable=Opt also deems the wavefunction associated with the Orca-produced geometry to be stable. The vibrational frequencies check for geometric stability is not applicable since constraints are used. When Gaussian optimization proceeds from the Orca-optimized geometry, the highly pyramidalized geometry results. When Orca optimization proceeds from the Gaussian-optimized geometry, the highly pyramidalized geometry remains. The Orca stationary point is evidently not seen by Gaussian, though the Gaussian stationary point is seen by Orca. Developers and users of MP2 variant methods need to be aware of this variation in optimized geometry, for from it follows significant consequences for amide resonance and flexibility of protein backbones. A pyramidalization of ~13 degrees might be converted to such a pyramidalization in the other direction by backbone strain, giving a ~26 degree flexibility in each protein backbone omega (CA-C-N-CA') torsion. CASSCF [73] (8,7)/MRCI+Q[45] with def2-TZVPP and auxiliary def2-TZVPP/C over the MP2/6-311++G** geometry gives 0.8372 weight on a single configuration in ground state with next configuration weight in ground state being less than 0.02, so this pyramidalized geometry is only weakly associated with multireference character. Ap2:Table 1 gives the results of this multireference method for the non-pyramidalized geometries arising from beta strand psi torsions with MP2/aug-cc-pVQZ, showing good agreement with single reference wavefunction methods, and that the problem is not particularly multireference.

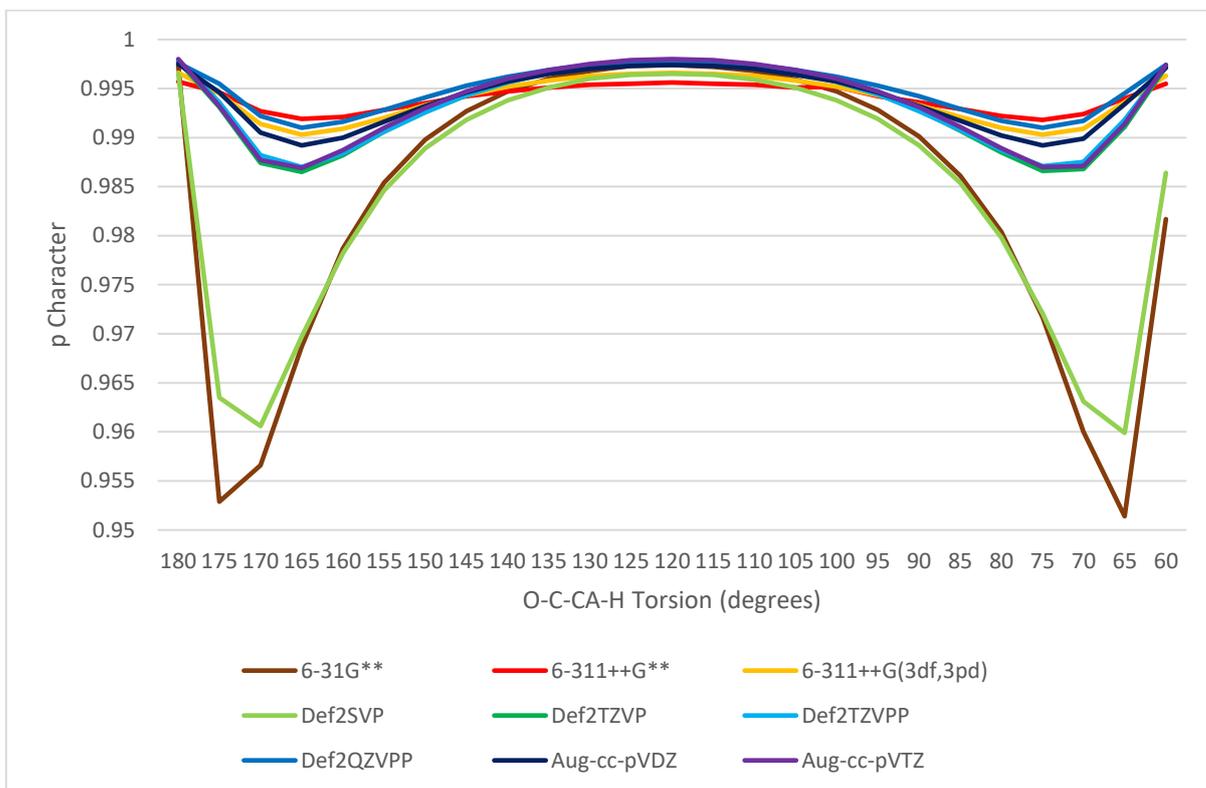

Figure 14. C-O(pi) Carbon NHO p Character in Ethanamide at Varying O-C-CA-H Torsion and Basis Set at B3LYP

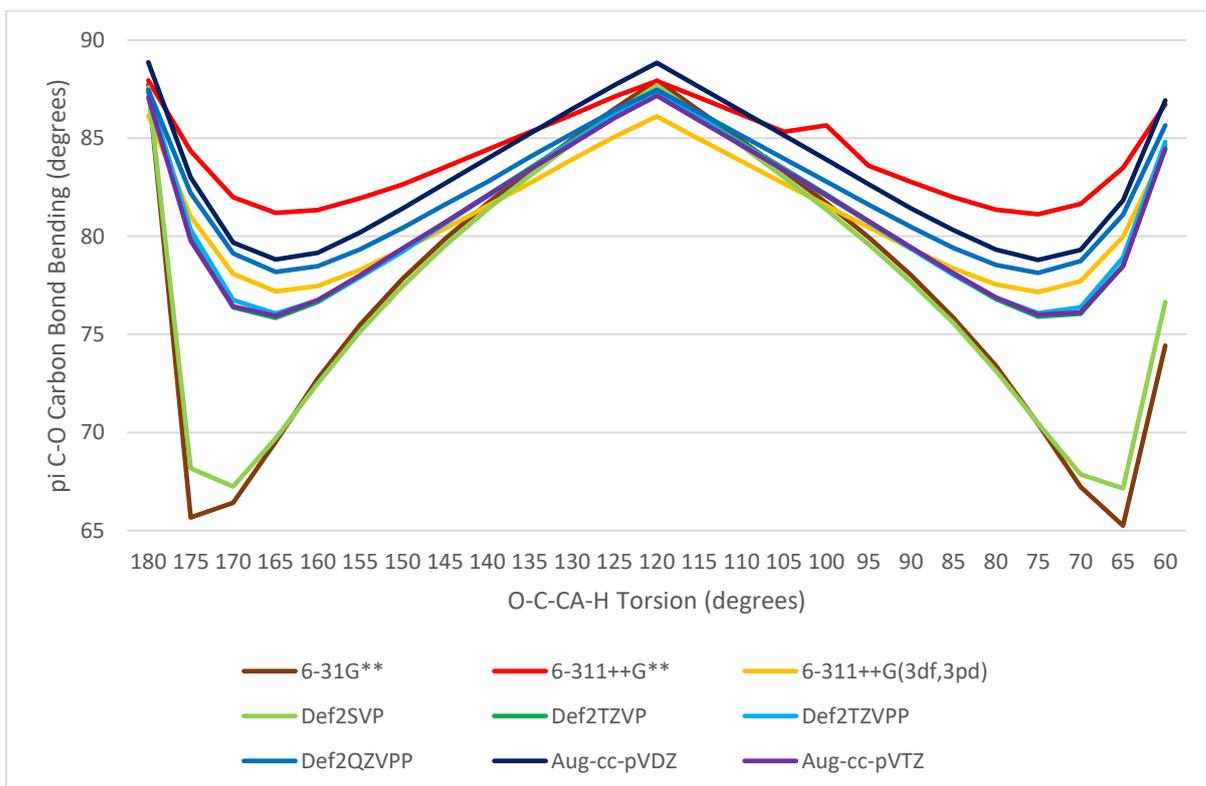

Figure 15. C-O(pi) Carbon NHO Bond Bending in Ethanamide at Varying O-C-CA-H Torsion and Basis Set at B3LYP

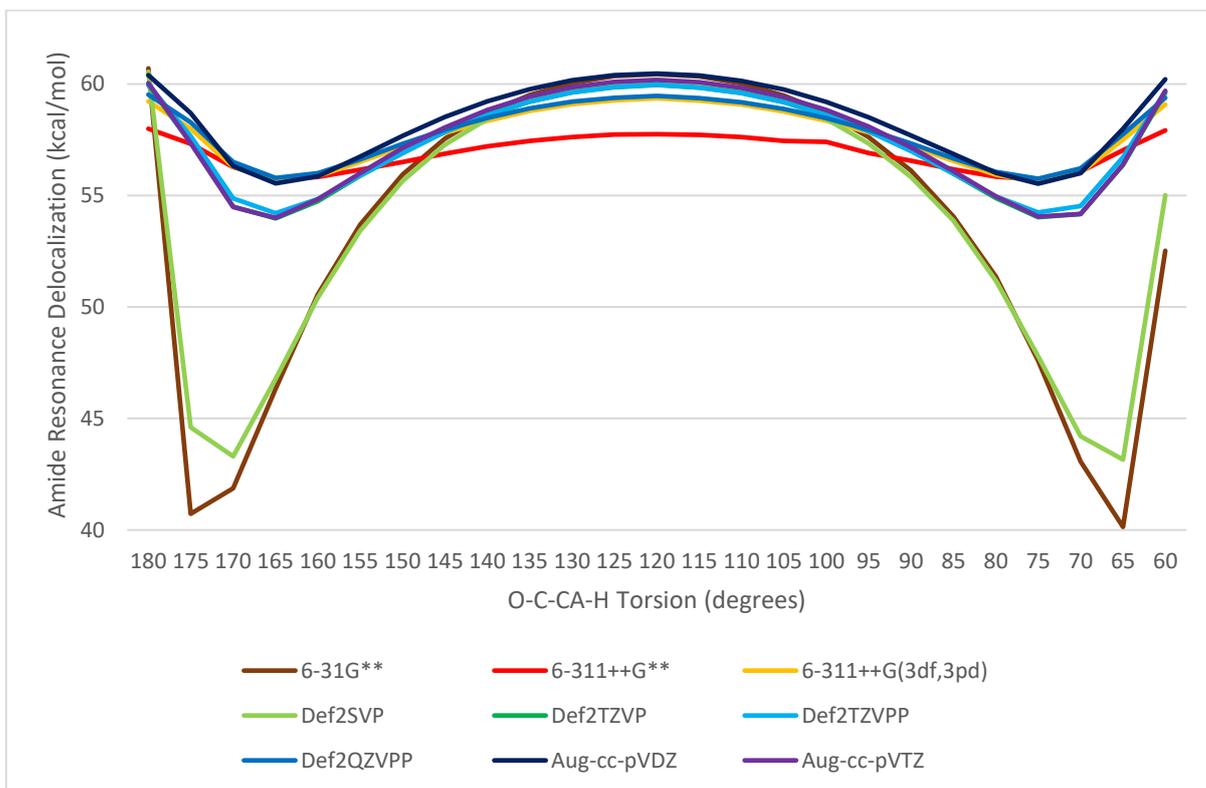

Figure 16. Second-Order Perturbation Theory Primary Amide Resonance Delocalization (N(lp)->C-O(pi)*) in Ethanamide at Varying O-C-CA-H Torsion and Basis Set at B3LYP

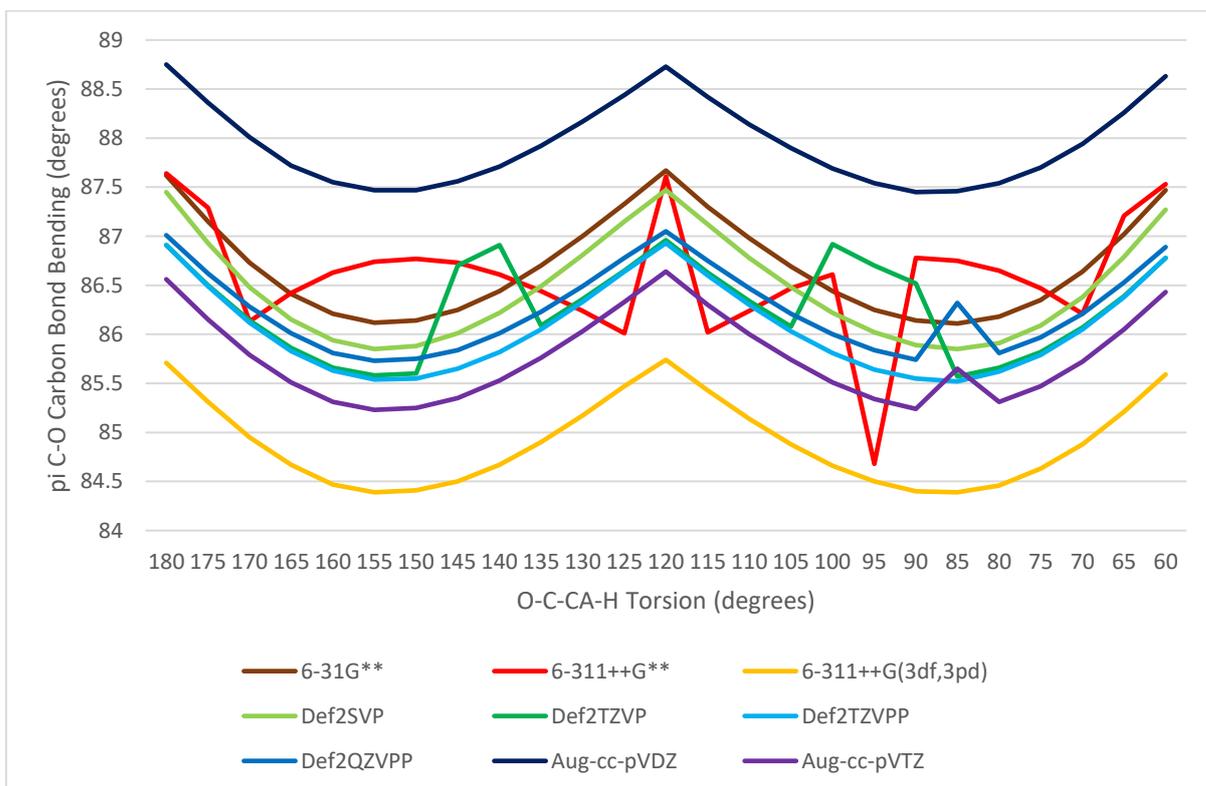

Figure 17. C-O(pi) Carbon NHO Bond Bending in Ethanamide at Varying O-C-CA-H Torsion and Basis Set at MP2

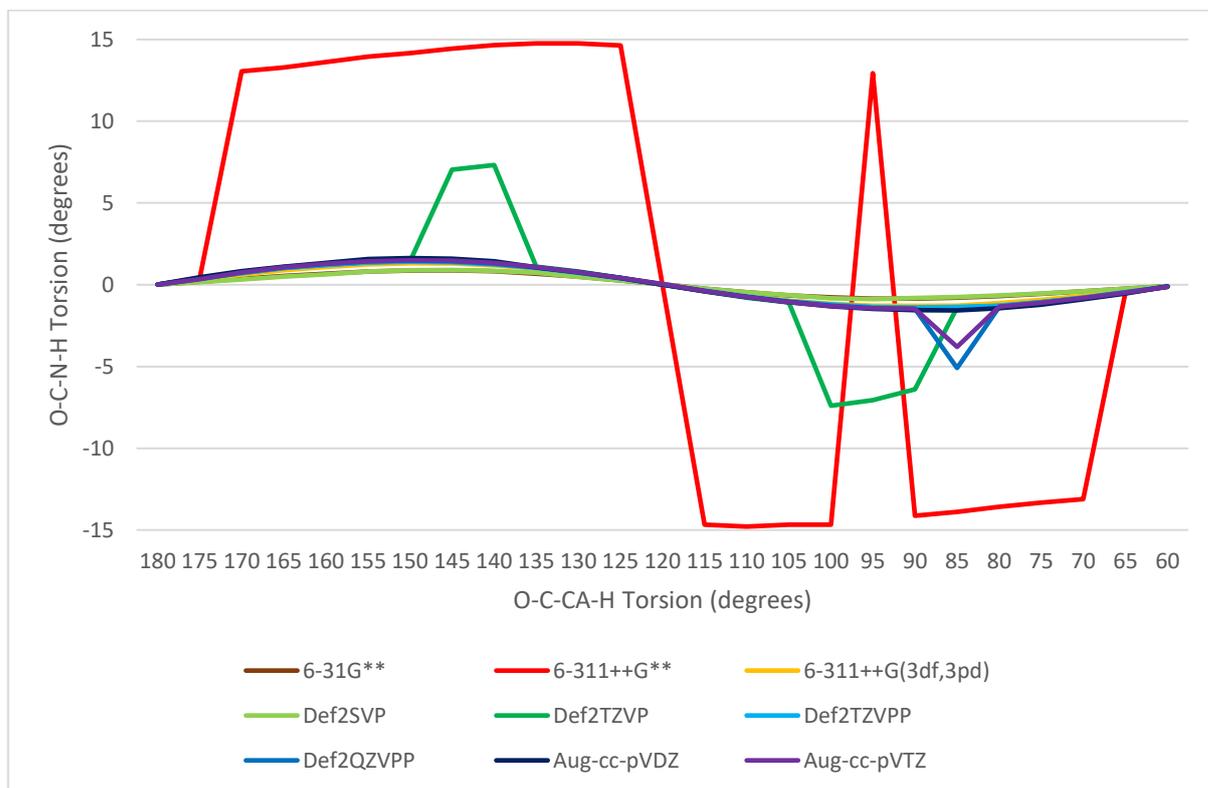

Figure 18. O-C-N-H Torsion in Ethanamide with Varying O-C-CA-H Torsion and Basis Set at MP2 FineGrid

### 6.3 Antiparallel beta sheet

Ap2:Table 4 and Ap2:Table 5 refer to gas phase antiparallel beta sheets which have been geometry optimized with TeraChem 1.5K with LC-wPBE (keyword wpbe in TeraChem), omega=0.4 bohr$^{-1}$ and 6-31G**. While the ethanamide model requires optimization constraints to demonstrate torsional hyperconjugation, these beta sheets require no constraints, that is, they are fully optimized. Ap2:Table 4 refers to a polyvaline structure and Ap2:Table 5 to a polyalanine structure.

By extending ethanamide to species resembling alanine and valine, it can be shown that these backbone errors are very sensitive to sidechain even for uncharged and non-polar residues, so that the utility of methods afflicted of these errors is dubious in the case of heterogeneous amino acid residues. Also, it might be erroneously concluded that this modulation of backbone amide resonance by sidechain is a means by which sidechain determines protein fold.

The variation in the N(lp)->C-O(pi)* SOPT ("SoptP" column) has two sources, one being RAHB which tends to cause amide resonance in hydrogen bonded amide chain to peak in the middle of the chain and be lowest at the ends, and the effects being discussed here, bond bending and loss of local symmetry. The tables mentioned above make the point that this bond bending and loss of local symmetry exists in beta sheets rather than only in the ethanamide model.

These tables are sorted on bond bending ("Dev" column). While bond bending might be expected to be dependent of d orbital involvement to give non-cylindrically symmetric orbitals so the hybrids can point at each other, these tables, particularly Ap2:Table 4, do not strongly bear out this notion.

SOPT kcal/mol values under 1.0 do not appear in these tables. Noteworthy are the variations in N(lp)->C-O(p)* SOPT ("SoptP") and N(lp)->C-O(s)* SOPT ("SoptS"). That the latter is non-zero is evidence of carbonyl orbital disturbance.

There is a strong but imperfect association between the occupancy of C-O(s)* and bond bending and loss of carbonyl sigma/pi symmetry. Perhaps non-double hybrid DFT methods do not deal well with the situation in which there is occupancy of C-O(s)* in the context of large occupancy of C-O(p)*.

In Ap2:Table 5, the least bond bending is associated with O-C-CA-HA dihedral of ~150 degrees, which corresponds to the optimal angle (90 degrees) for hyperconjugation involving C-O(pi)* orbitals. This suggests that this hyperconjugation is not disruptive as hyperconjugation involving C-O(s)*.

Moderate SOPT kcal/mol values are proportional to resonance-type charge transfer [6]. The difference between hyperconjugative charge transfer from the methyl group into C-O(p)* and C-O(s)* has some association with bond bending and loss of sigma/pi symmetry. The variation in individual and collective hyperconjugative SOPT values is less than that of amide resonance, suggestive of a failure of method.

## 7 Conclusion

In contrast to single and multireference wavefunction methods, in the molecules studied, established DFT methods do not substantially maintain sigma/pi separation or bond straightness in double bonds that have hyperconjugative interactions at other than perfectly antiperiplanar geometry. Unlike ethene with C-C torsion, these failures are not due to the problem being strongly multireference. The double hybrid DFT methods are considerably better than other DFT methods, but do not quite attain the accuracy of MP2 methods though they incur similar computational costs. This DFT error is encountered in the study of proteins which include parallel and antiparallel beta sheets and attenuates amide resonance and hence necessarily RAHB in the hydrogen bonded chains of backbone amides. Since RAHB is cooperative, if all errors in calculations of amide resonance give values that are lower than accurate, then the errors themselves may be seen as cooperative. Since every backbone amide in a beta sheet is subject to roughly the same torsional hyperconjugation, hydrogen bonding in the beta sheet is cooperatively reduced. While the smaller basis sets such as D95** [74], 6-31G** and def2-SVP [75] are associated with the largest reductions in amide resonance, these basis sets have been popular because they are the largest that could be used at the atom counts of beta sheets. Great caution is needed in interpreting the results of applying established DFT methods to proteins.

Of the non-double hybrid DFT methods and basis set combinations tested, MN12-L/def2-QZVPP is least disturbed by torsional hyperconjugation. MN15-L and MN15 [76] have not yet been tested. MN12-L being a local functional and def2-QZVPP having no diffuse functions allows for efficiency of calculation, but no non-double hybrid DFT method tested yields occupancies comparable to those calculated by wavefunction methods and correlation consistent basis sets for the orbitals central to amide resonance, N(lp) and C-O(p)*. Though the multireference character of the problem is slight, CASSCF/MRCI+Q gives occupancies associated with greater resonance than does DLPNO-CCSD(T), leading to the concern that multireference methods may be necessary for accurate calculation of RAHB in proteins. Perhaps much improved scaling approximations to full configuration interaction [77, 78] will be useful. DLPNO-CCSD(T) in turn calculates occupancies associated with greater resonance than do non-double hybrid DFT methods, so non-double hybrid DFT methods then fail to accurately calculate absolute amide resonance regardless of O-C-CA-H torsion.

While this work is focused on amide resonance and RAHB, any non-double hybrid DFT calculated amide property is at risk of being significantly inaccurate due to this error. A range of quantities are disturbed by this error along with carbonyl sigma/pi separation and bond straightness, including charge transfers N(lp)->C-O(p)* and N(lp)->C-O(s)* and steric interactions C-O(p)|C-O(s), O(lp-s)|C-O(p) and O(lp-s)|C-O(s).

It can be shown with molecular species extending ethanamide to approximate alanine and valine that the loss of local symmetry with DFT is very sensitive to residue type. With a heterogeneous amino-acid beta sheet, DFT geometry optimization will be erratic rather than reliably wrong. Also, this sensitivity invites the erroneous conclusion that protein folding is influenced by this sidechain modulation of backbone amide resonance.

MP2 has a basis set and optimization implementation dependent propensity for large variation in pyramidalization at the amide nitrogen in ethanamide at C-CA torsion. This pyramidalization necessarily competes with amide resonance. When the erroneous tendency to pyramidalize becomes dominant, calculation will under-estimate RAHB for hydrogen bonded chains in which the amide is potentially involved. Also, the flexibility of a protein backbone is incorrectly increased since the pyramidalization may be switched from one side to another by backbone strain, giving large backbone amide omega torsion. This variation is not limited to Pople basis sets, and is seen to a lesser extent and in small ranges of C-CA torsion with def2-TZVP, def2-QZVPP and aug-cc-pVTZ. This is not exclusively an MP2 issue, for the quite recent DFT method M11-L also demonstrates this pyramidalization at antiparallel beta strand psi torsion with most basis sets tested. This pyramidalization error is not unambiguously present in geometry-optimized protein coordinates since this pyramidalization may arise by other means such as backbone strain and hydrogen bonding, so validation of methods with respect to this error is best done on small amides such as ethanamide or N-methylethanamide. In ethanamide, this second problem can be corrected by geometry optimization on the Counterpoise Corrected potential energy surface by definition of fragment boundaries at the C-N bond, but not at the C-C bond.

We recommend that non-wavefunction electronic structure calculation of protein be accompanied by NBO analysis of backbone amide carbonyl local symmetry and bond bending. Since hyperconjugations involving C-O(s)* and C-O(p)* are available at various C-CA torsions, this recommendation extends beyond the secondary structures tested here, antiparallel beta sheets, to other protein secondary structures that have hydrogen bonded chains of backbone amides. While the consequences of inaccurate calculation of amide resonances are not amplified in random coil as in secondary structure, they may be expected to be significant in determining random coil structure and hydrogen bonding networks.

NBO-based assessment of maintenance of local symmetry and bond bending is proposed for evaluation of existing or development of new electronic structure system methods. The tests used here involving vinylamine and ethanamide could serve as standard benchmarks. Measurement of pyramidalization and NBO assessment of sp hybridization at the amide nitrogen in ethanamide with carbonyl hyperconjugation at C-C torsion could similarly serve as a standard benchmark.

## 8   Acknowledgements

Prof. Peter M. W. Gill is acknowledged for reading the earliest draft of this manuscript and pointing out the resemblance of the second mentioned problem, pyramidalization at amide nitrogen at wavefunction method/Pople basis set, to the benzene non-planarity problem [17].

eResearch South Australia is acknowledged for hosting and administering machines provided under Australian Government Linkage, Infrastructure, Equipment and Facilities grants for Supercomputing in South Australia, directing funds to the acquisition of Nvidia Tesla GPU nodes and allocating 64 CPU cores and 256 GB RAM of the NeCTAR Research Cloud (a collaborative Australian research platform supported by the National Collaborative Research Infrastructure Strategy) to the present work.

# 10 Appendix 1

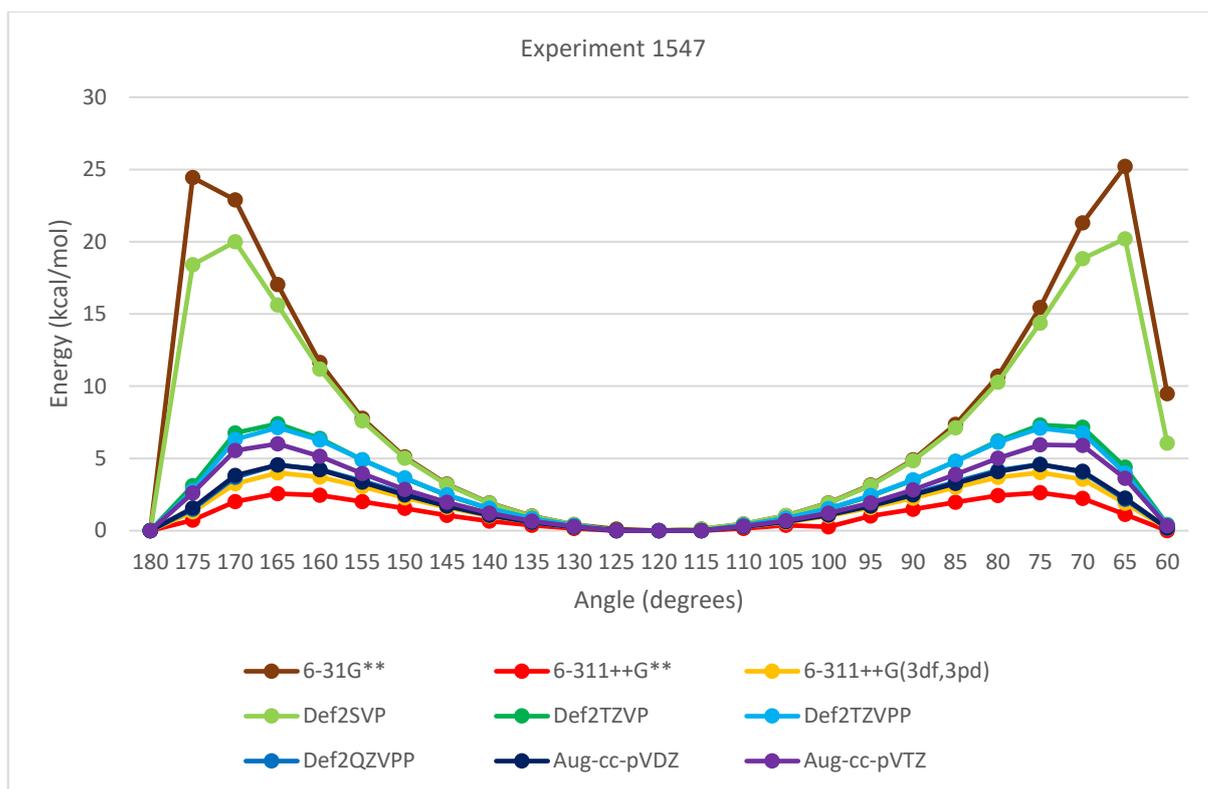

Figure 19. C-O(p)|C-O(s) Steric Exchange Energy in Ethanamide at O-C-CA-H Dihedral Angle with B3LYP

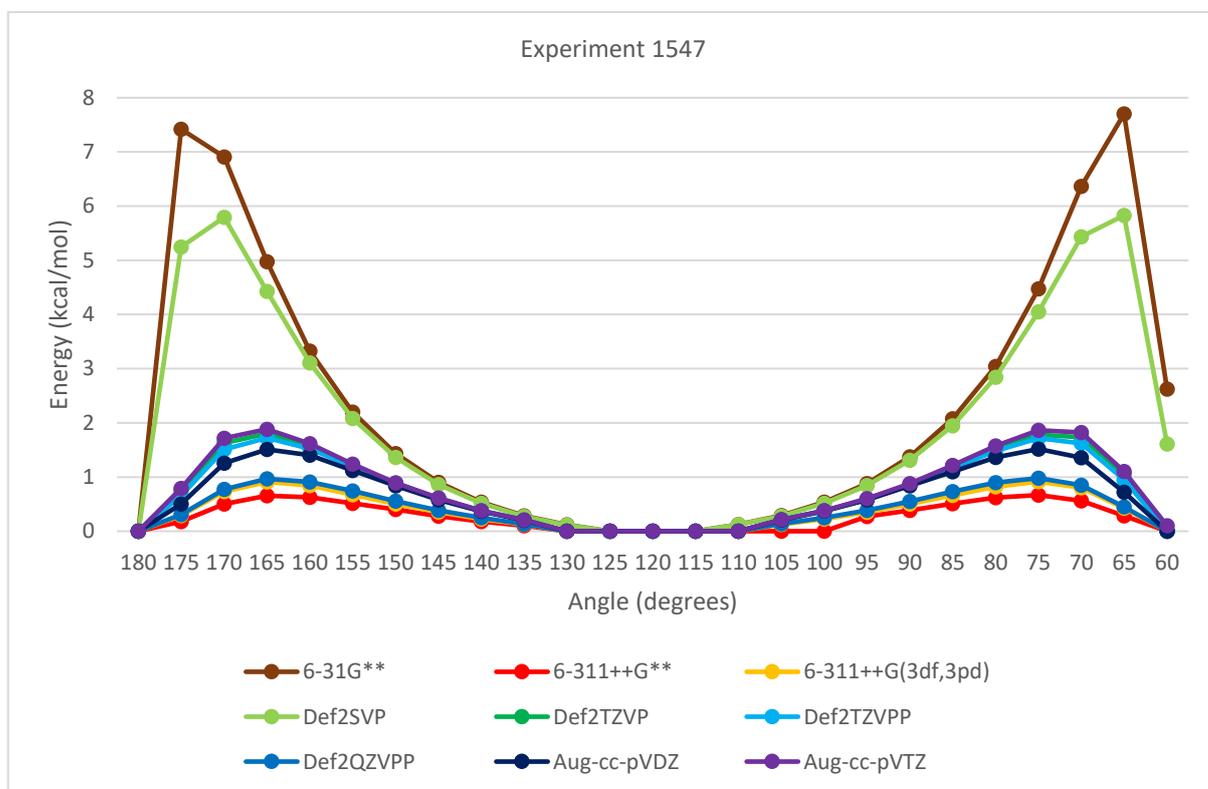

Figure 20. O(lp-s)|C-O(p) Steric Exchange Energy in Ethanamide at O-C-CA-H Dihedral Angle with B3LYP

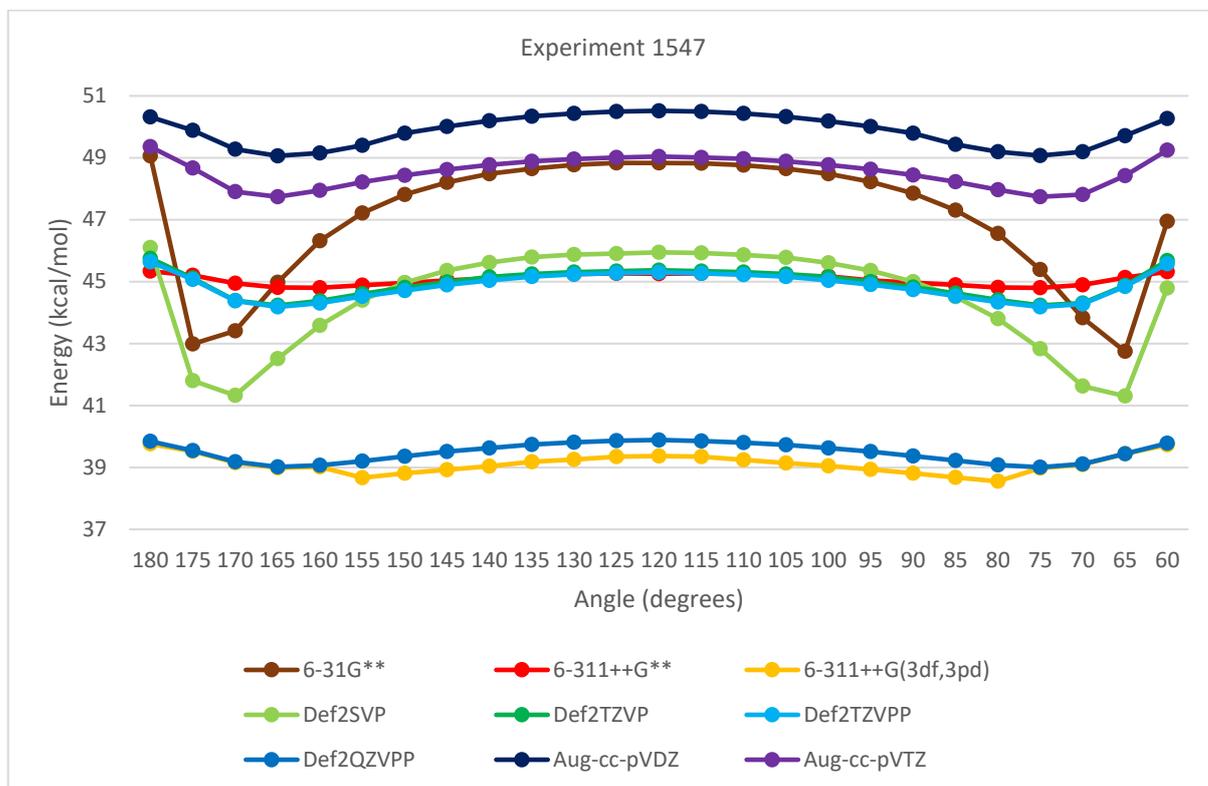

Figure 21. O(lp-s)|C-O(s) Steric Exchange Energy in Ethanamide at O-C-CA-H Dihedral Angle with B3LYP

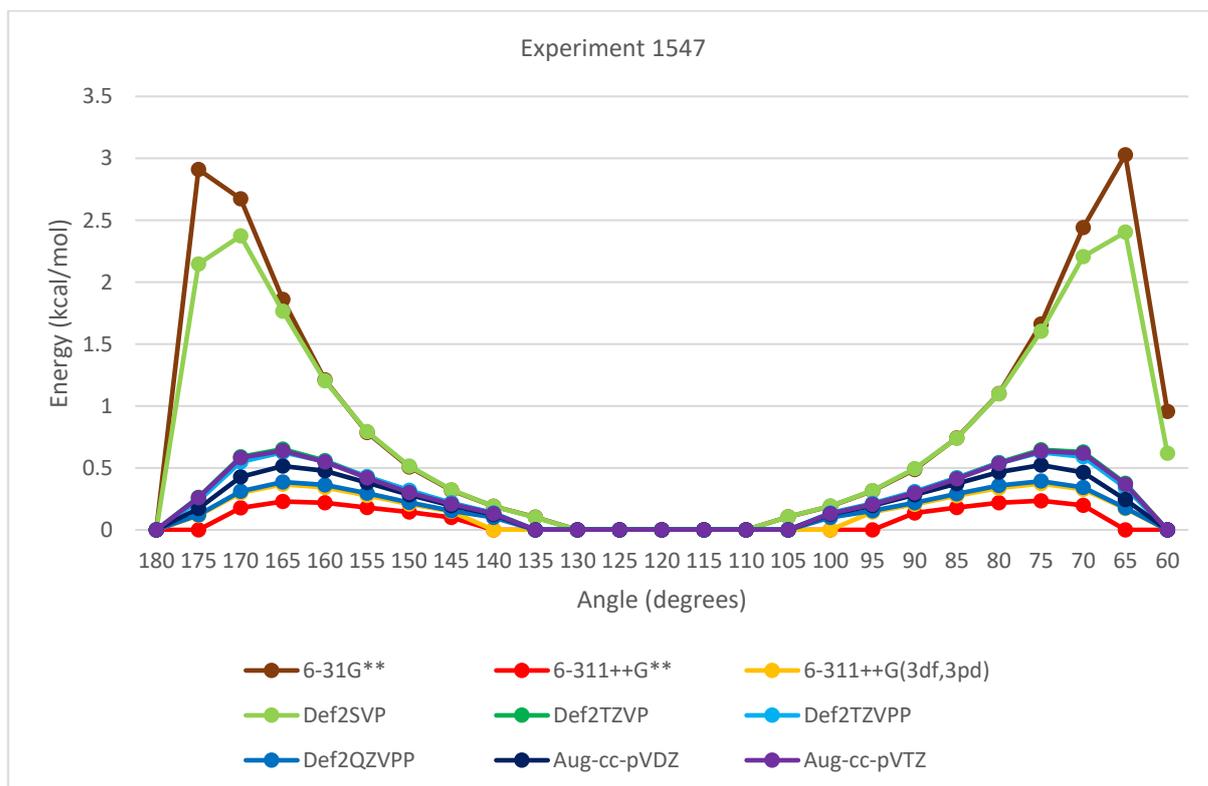

Figure 22. N(lp)->C-O(s)* SOPT Energy in Ethanamide at O-C-CA-H Dihedral Angle with B3LYP. Values less than 0.01 as zero.

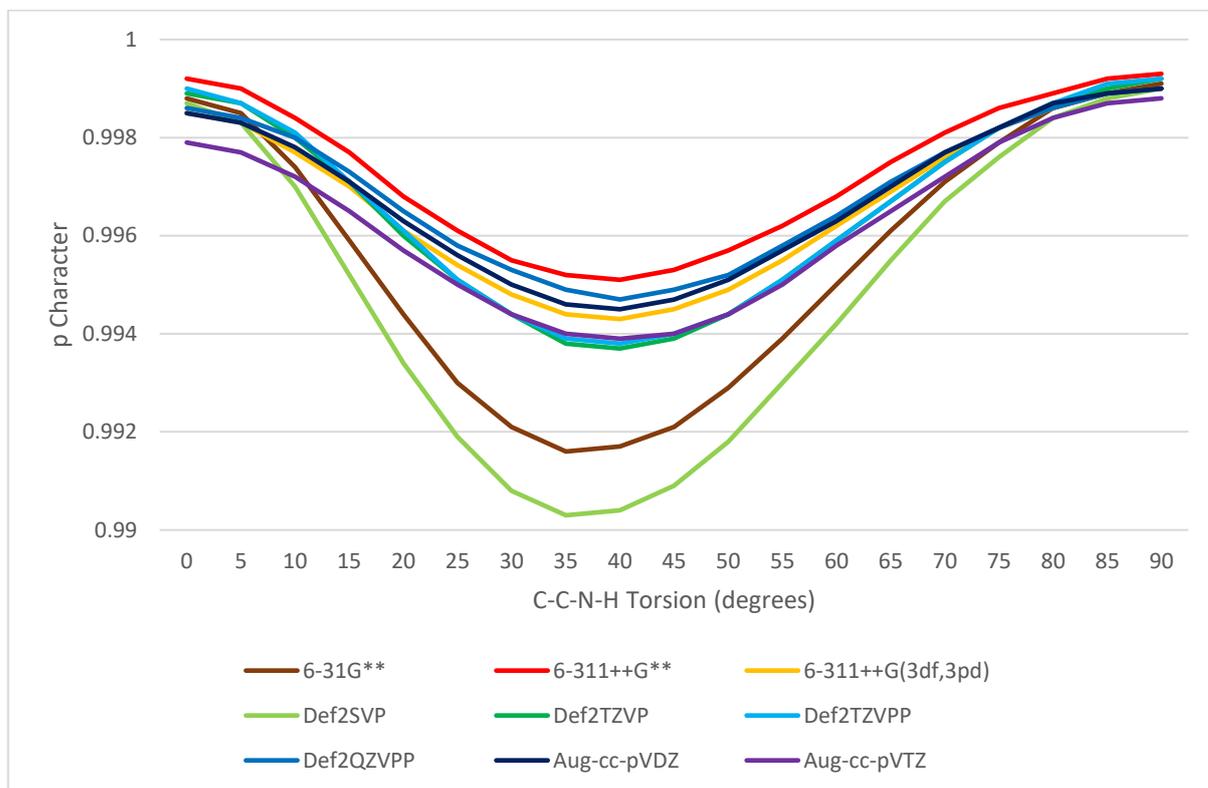

Figure 23. C-C(pi) Central Carbon NHO p Character in Vinylamine at Varying C-C-N-H Torsion and Basis Set at MP2

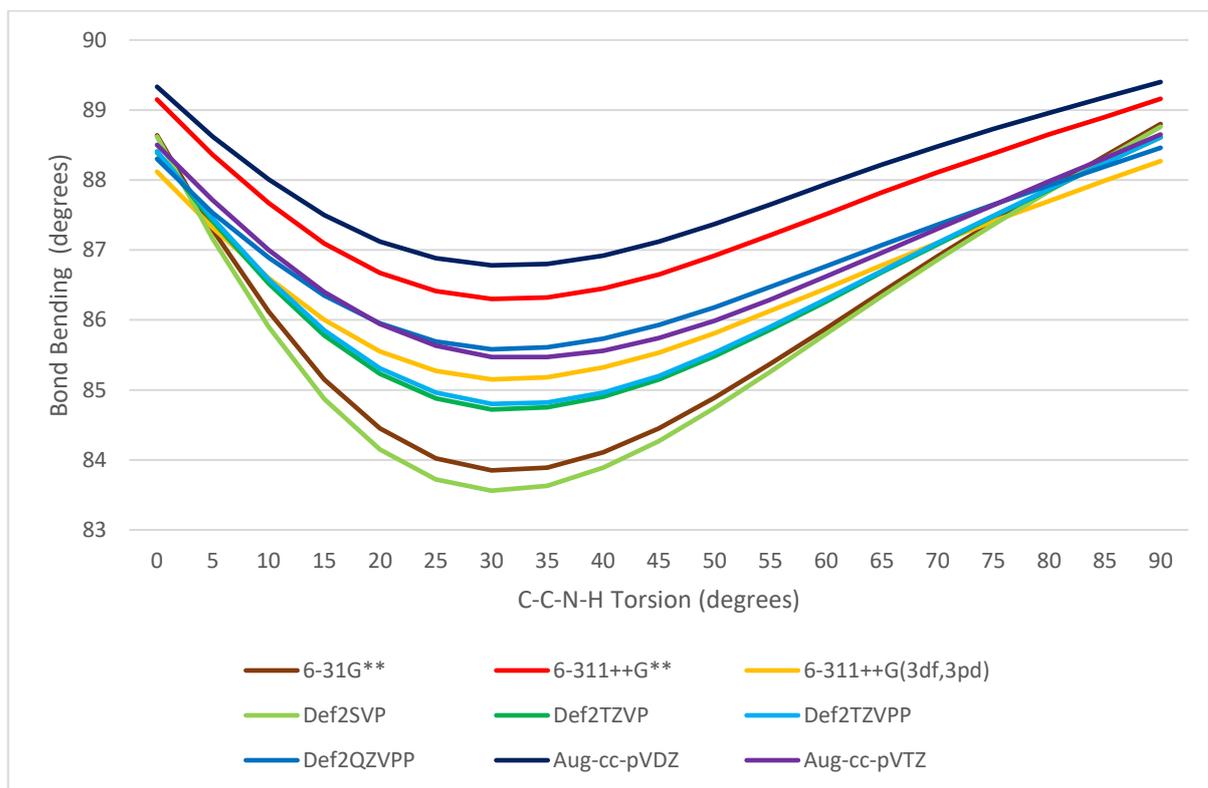

Figure 24. C-C(pi) Central Carbon NHO Bond Bending in Vinylamine at Varying C-C-N-H Torsion and Basis Set at MP2

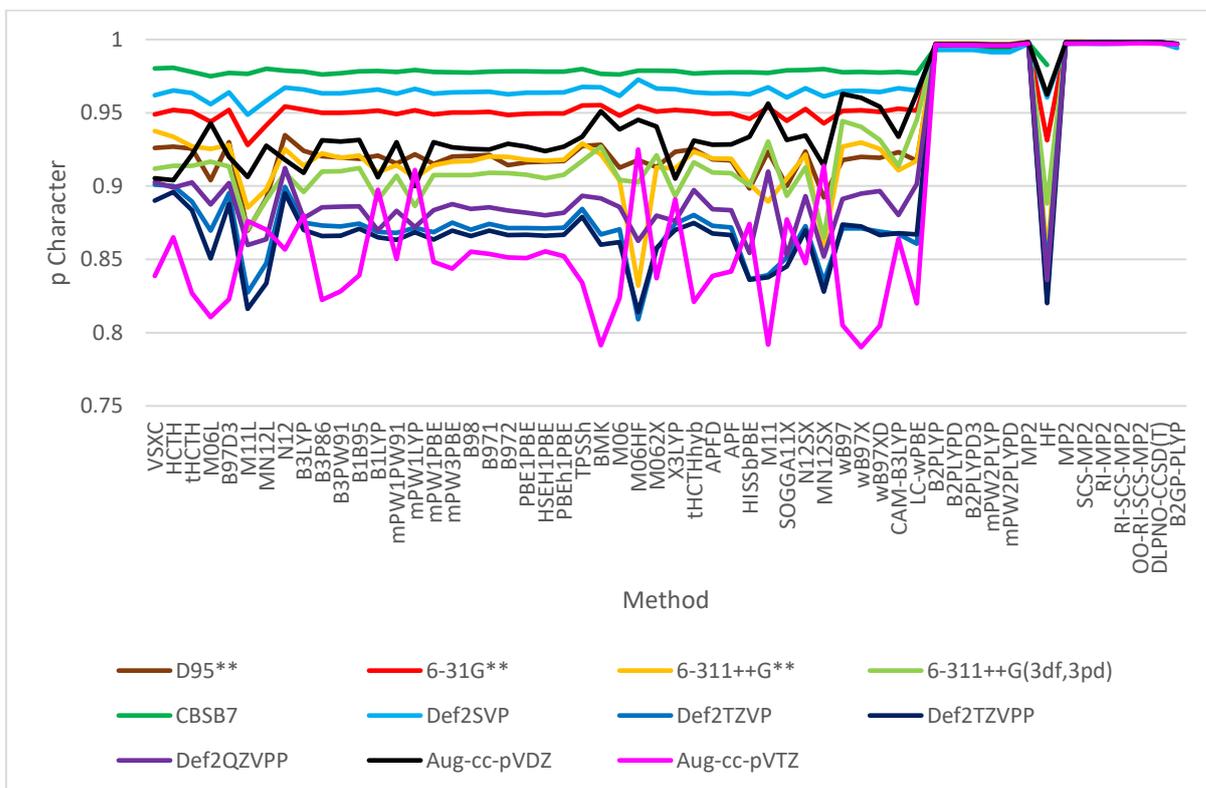

Figure 25. C-C(pi) Central Carbon NHO p Character in Vinylamine at 10 Degree C-C-N-H Torsion with Varying Method and Basis Set

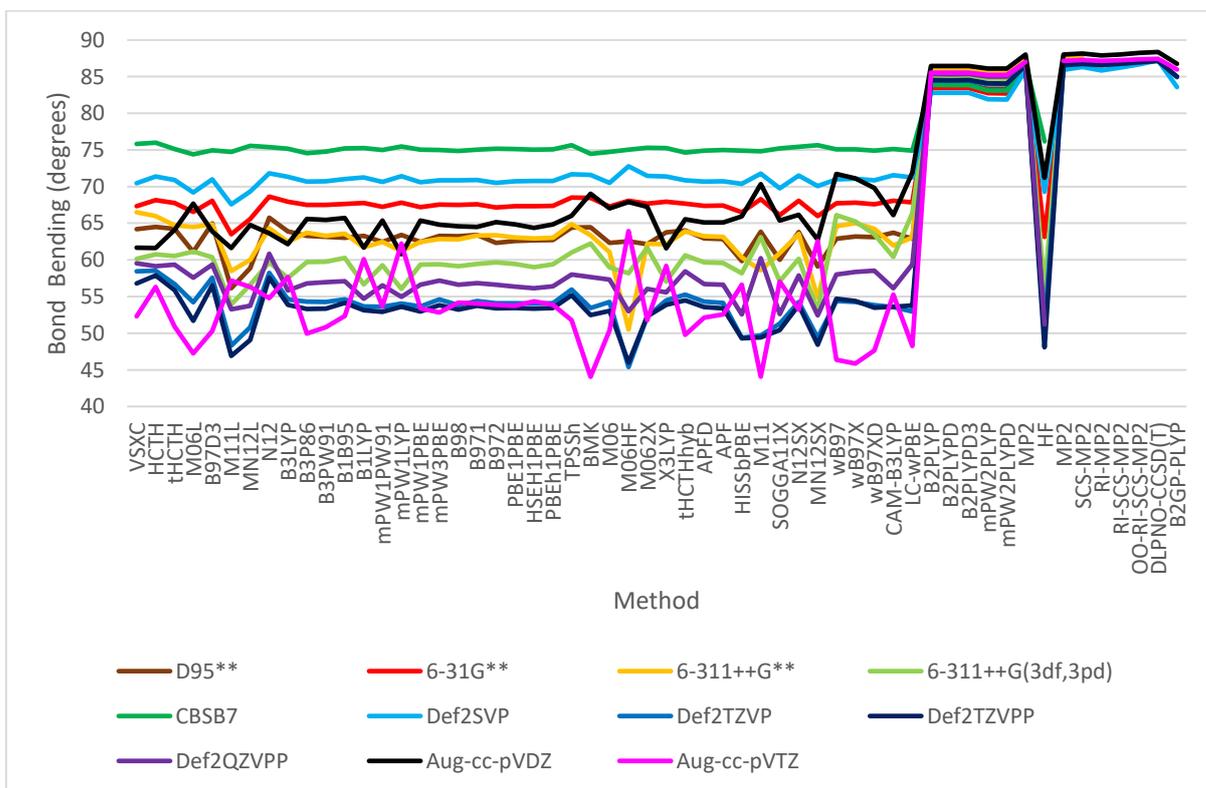

Figure 26. C-C(pi) Central Carbon NHO Bond Bending in Vinylamine at 10 degree C-C-N-H Torsion with Varying Method and Basis Set

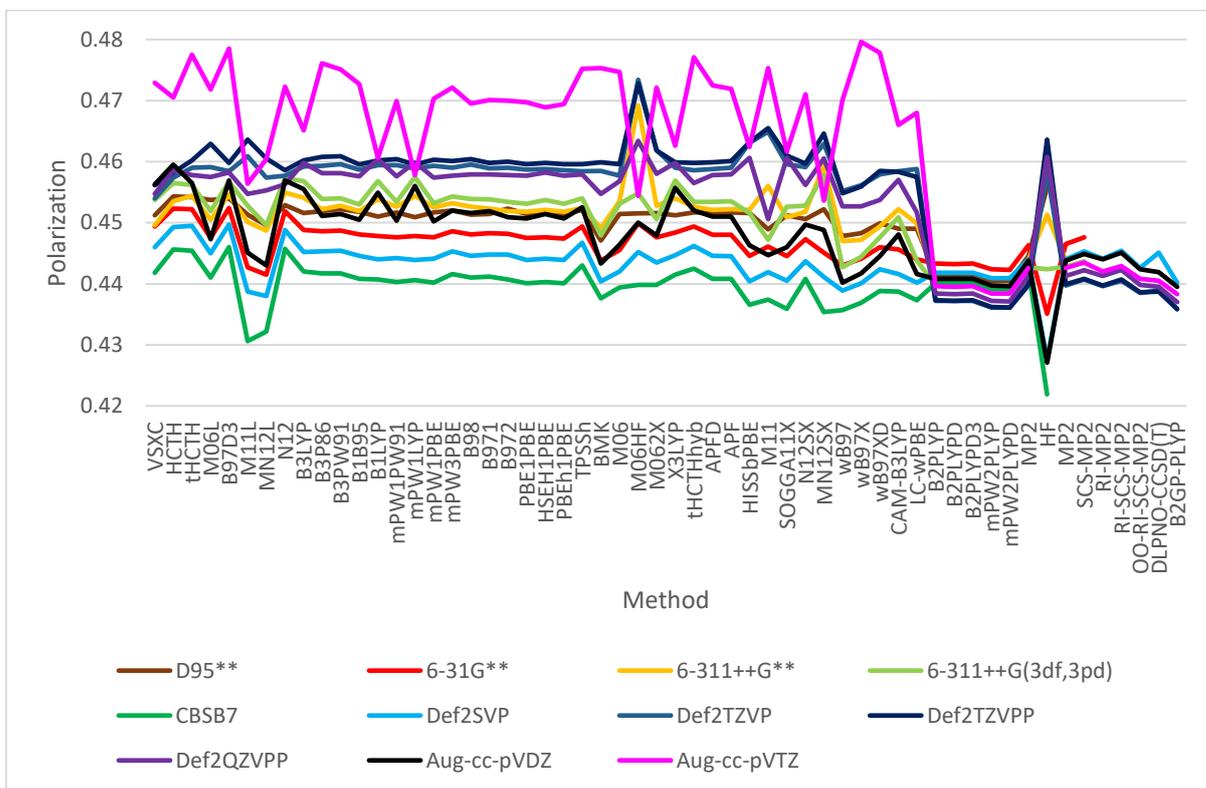

Figure 27. C-C(pi) Central Carbon NHO Polarization in Vinylamine at 10 degree C-C-N-H Torsion with Varying Method and Basis Set

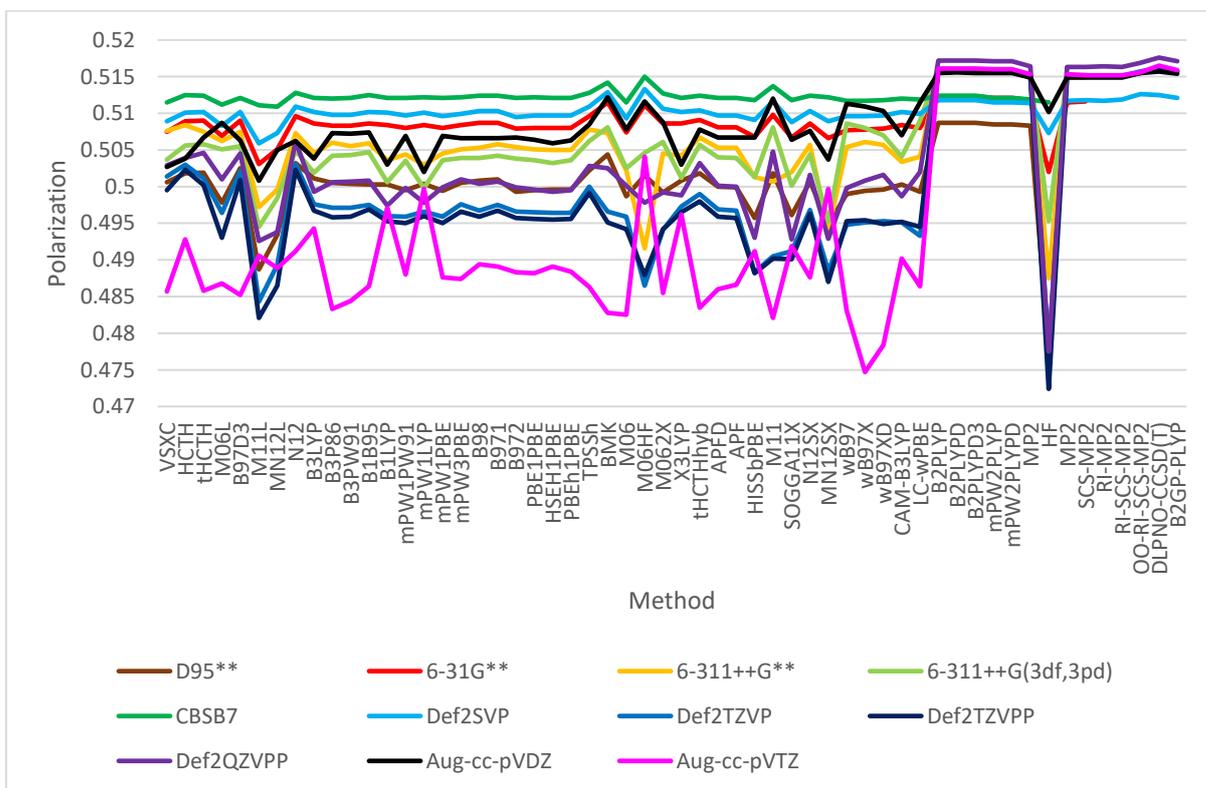

Figure 28. C-C(sigma) Central Carbon NHO Polarization in Vinylamine at 10 Degree C-C-N-H Torsion with Varying Method and Basis Set

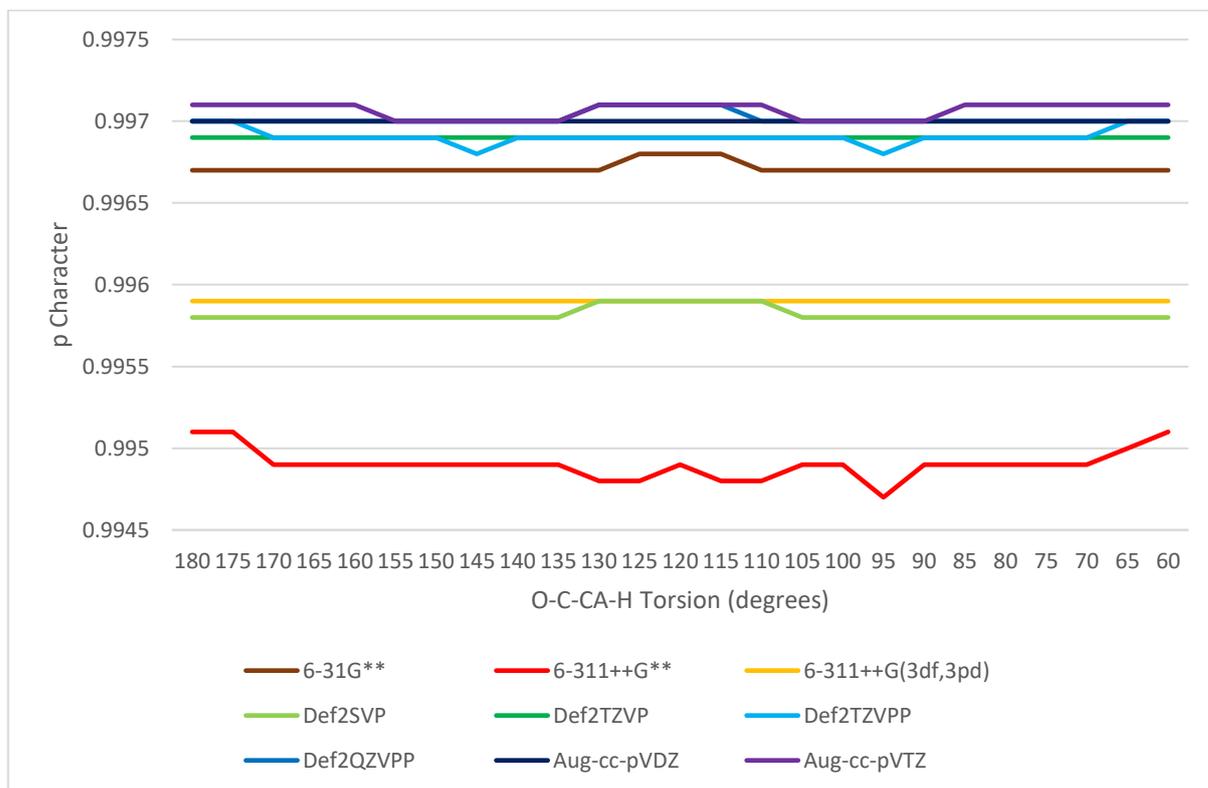

Figure 29. C-O(pi) Carbon NHO p Character in Ethanamide at Varying O-C-CA-H Torsion and Basis Set at MP2

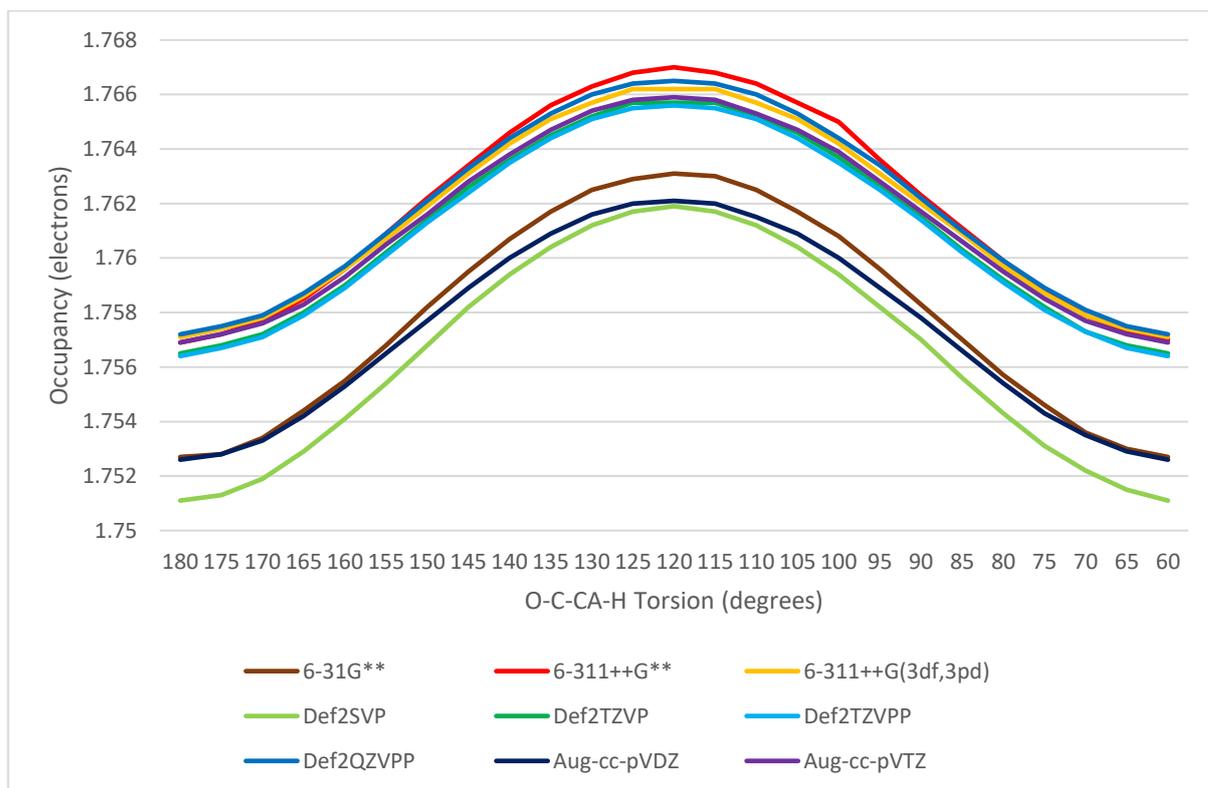

Figure 30. Nitrogen Lone Pair NBO Occupancy in Ethanamide at Varying O-C-CA-H Torsion and Basis Set at B3LYP

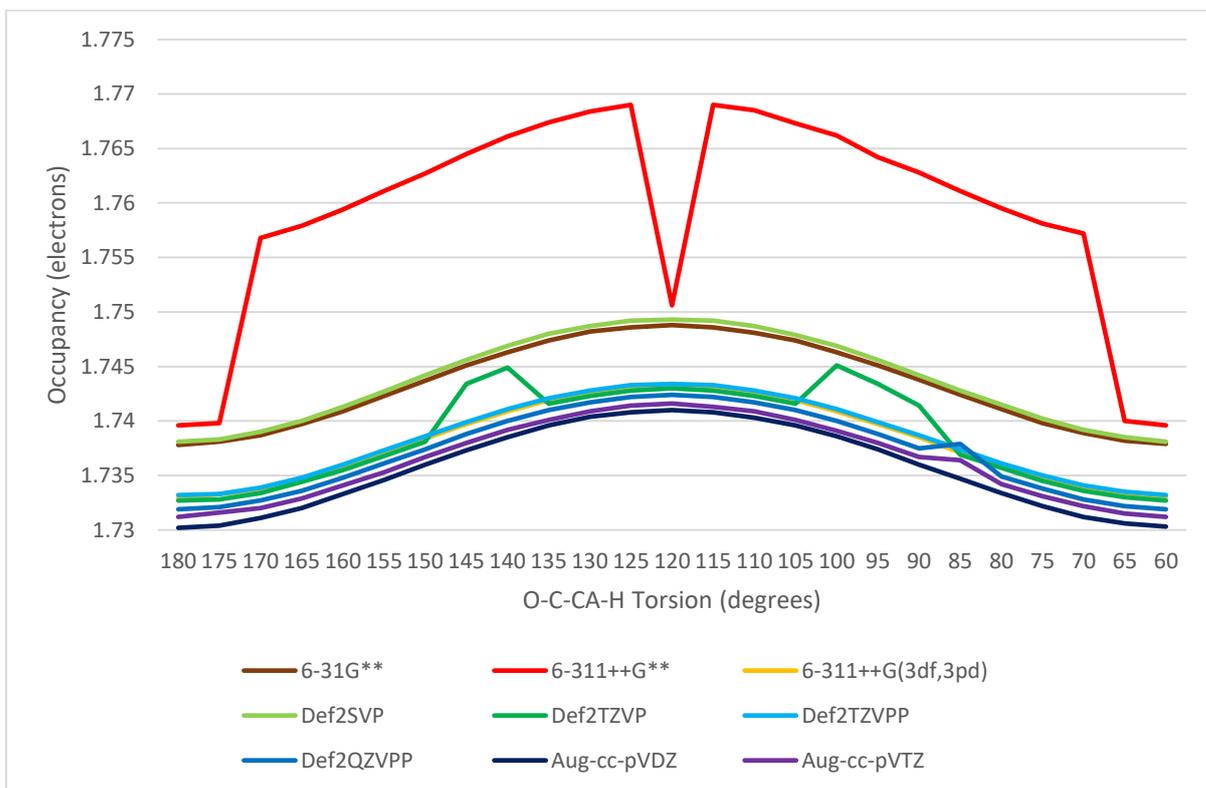

Figure 31. Nitrogen Lone Pair NBO Occupancy in Ethanamide at Varying O-C-CA-H Torsion and Basis Set at MP2

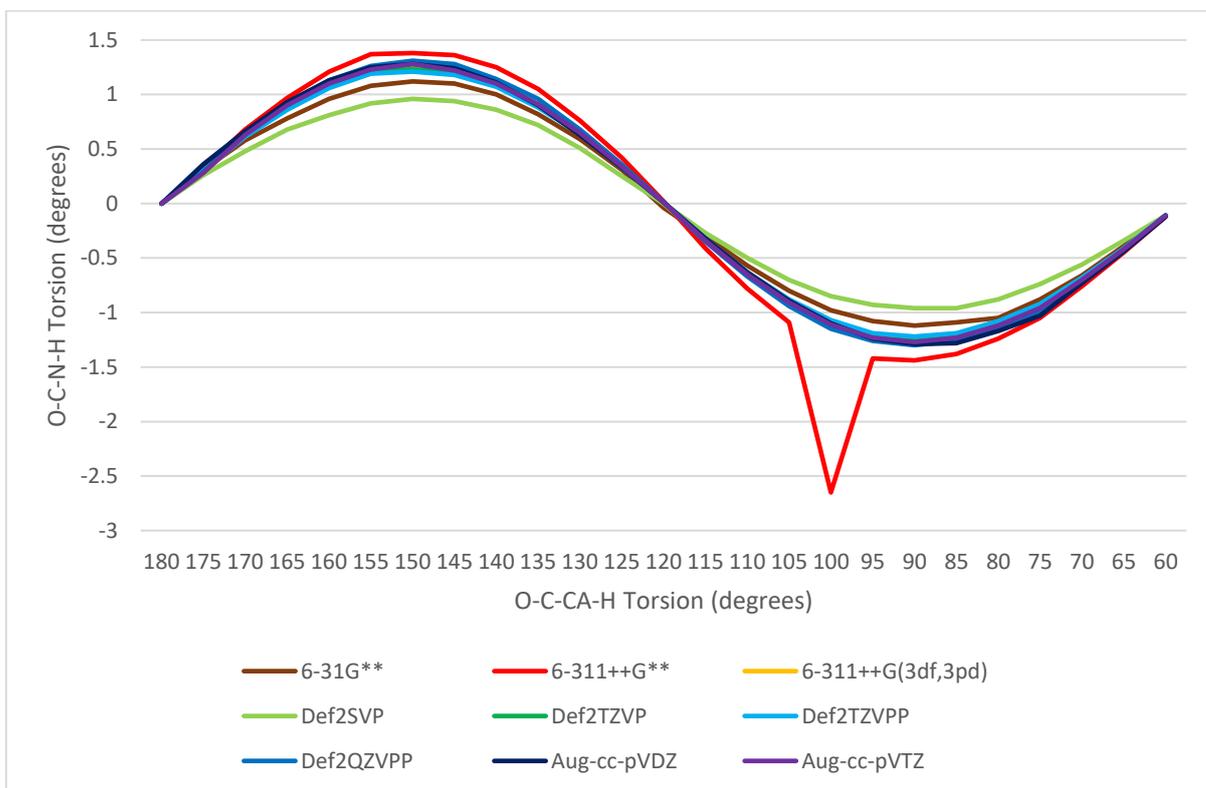

Figure 32. O-C-N-H Torsion in Ethanamide with Varying O-C-CA-H Torsion and Basis Set at B3LYP

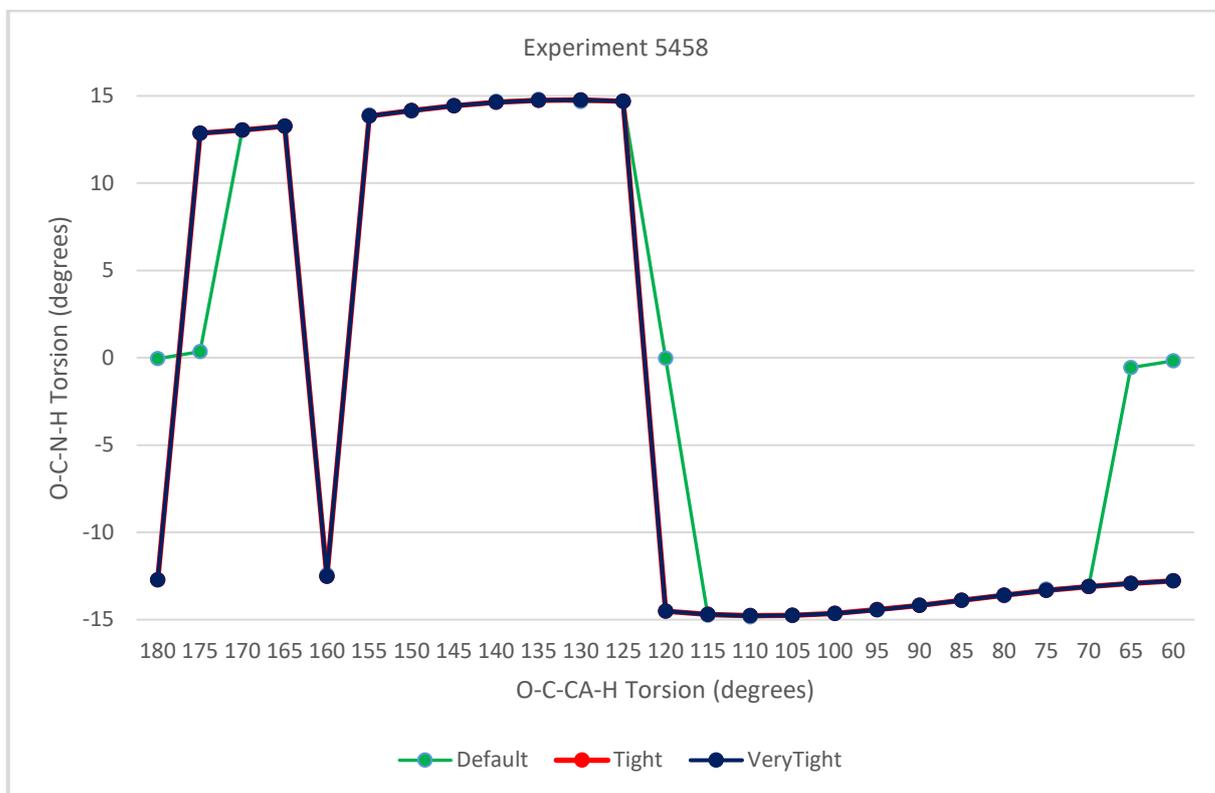

Figure 33. Ethanamide O-C-N-H Torsion with MP2/6-311++G** UltraFineGrid at Varying O-C-CA-H Torsion and Gaussian Optimization Convergence Label

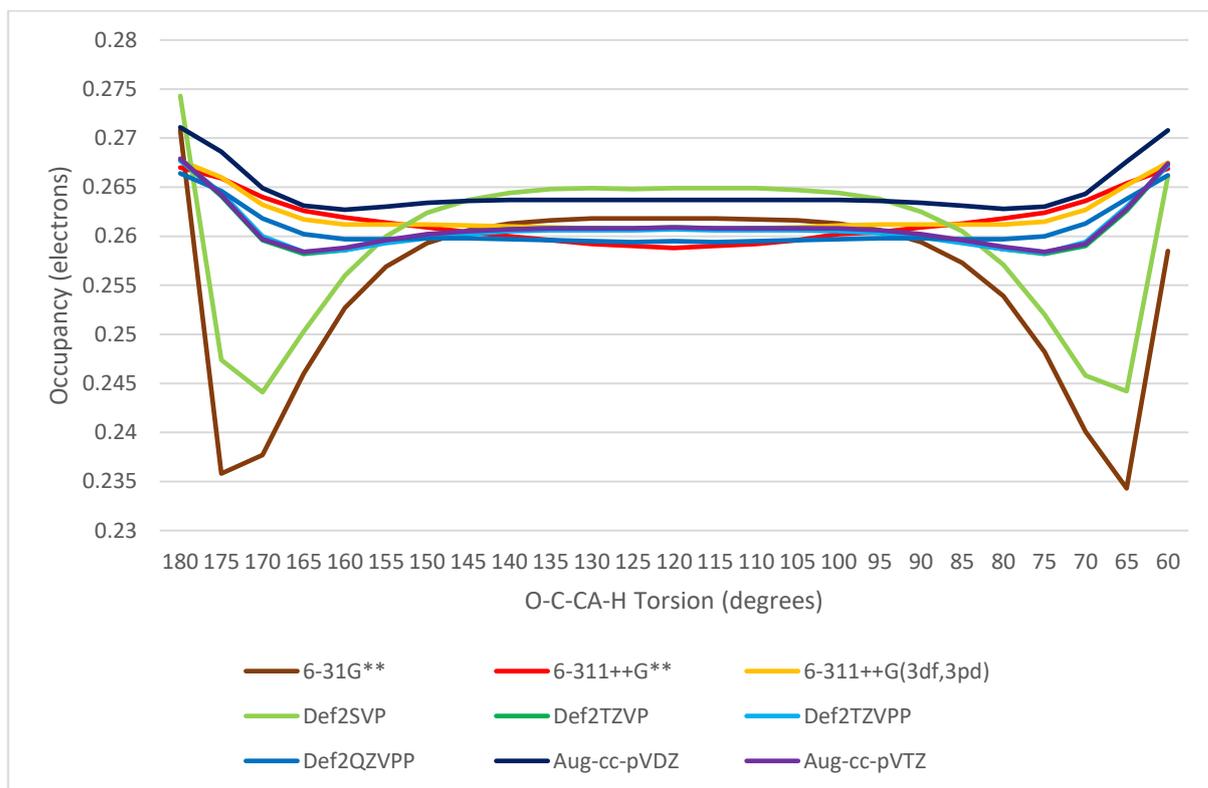

Figure 34. C-O(pi)* NBO Occupancy at Varying O-C-CA-H Torsion and Basis Set at B3LYP

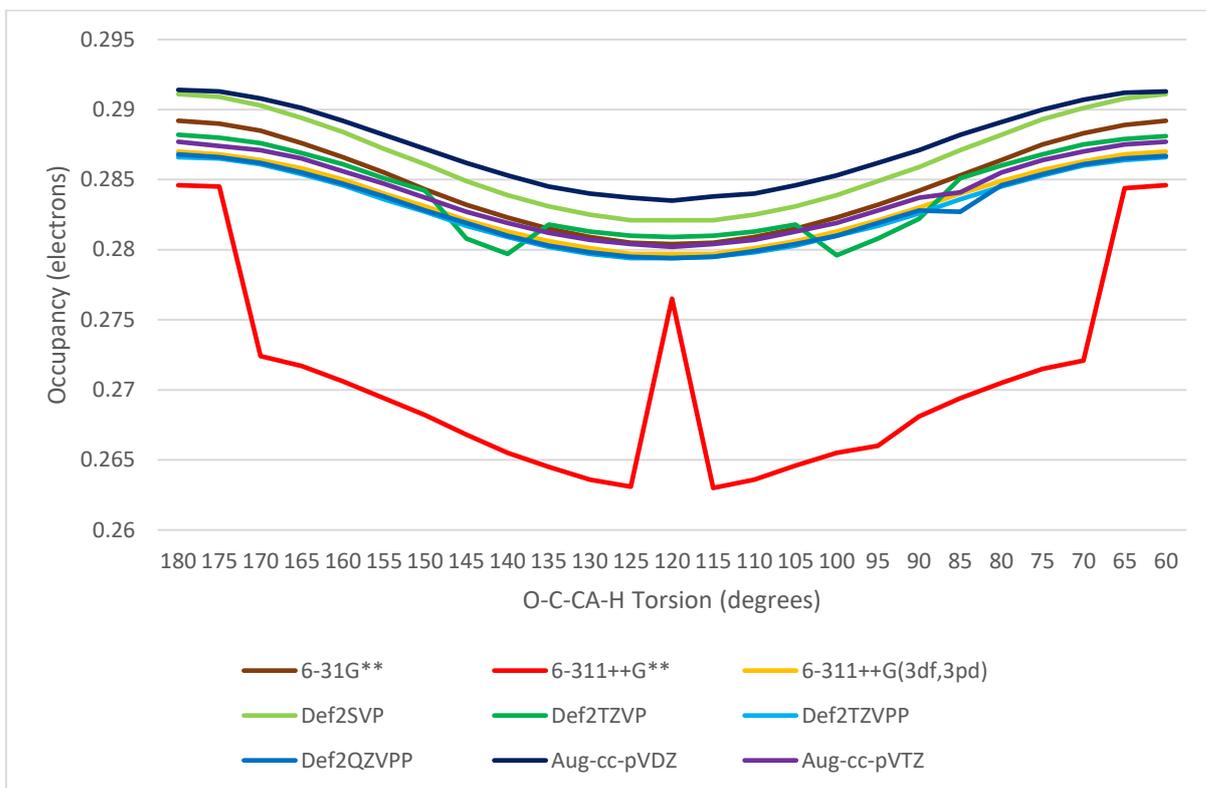

Figure 35. C-O(pi)* NBO Occupancy at Varying O-C-CA-H Torsion and Basis Set at MP2

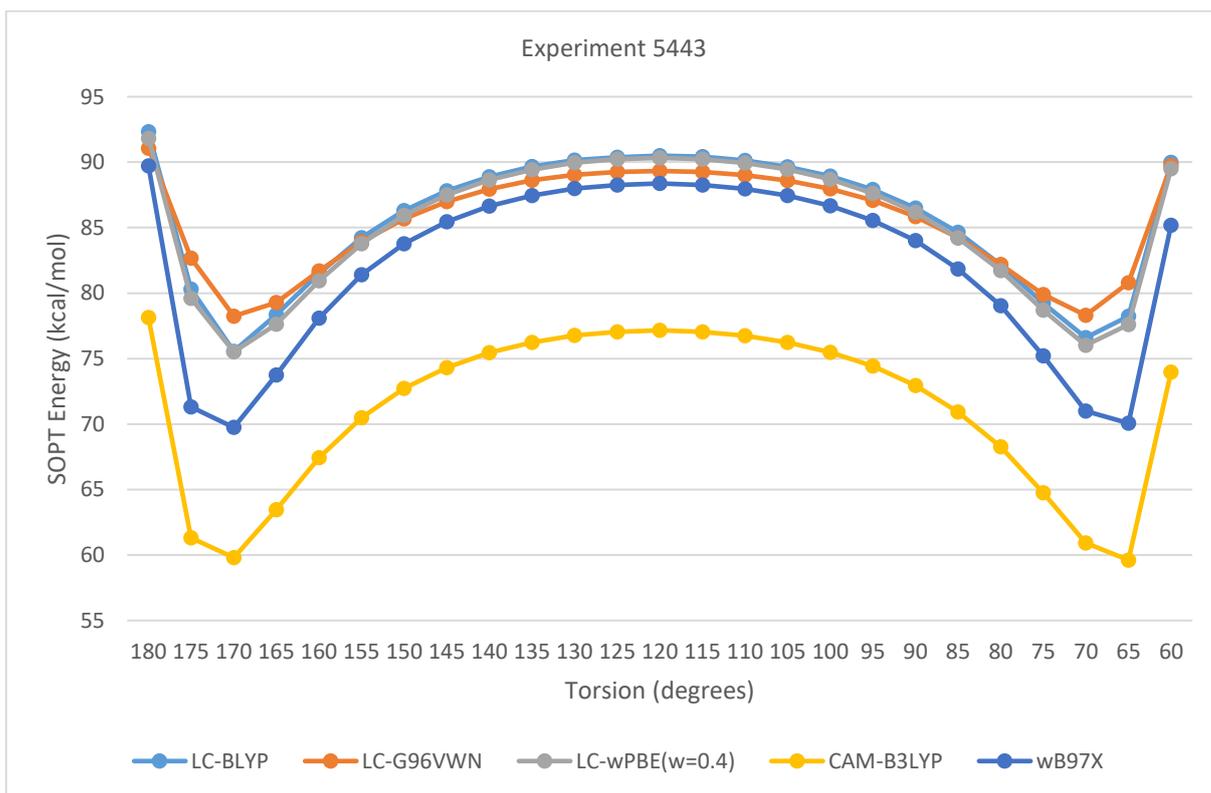

Figure 36. Ethanamide N(lp)->C-O(p)* SOPT Energy with 6-31G** UltraFineGrid and Varying O-C-CA-H Torsion and Long Range Corrected Method

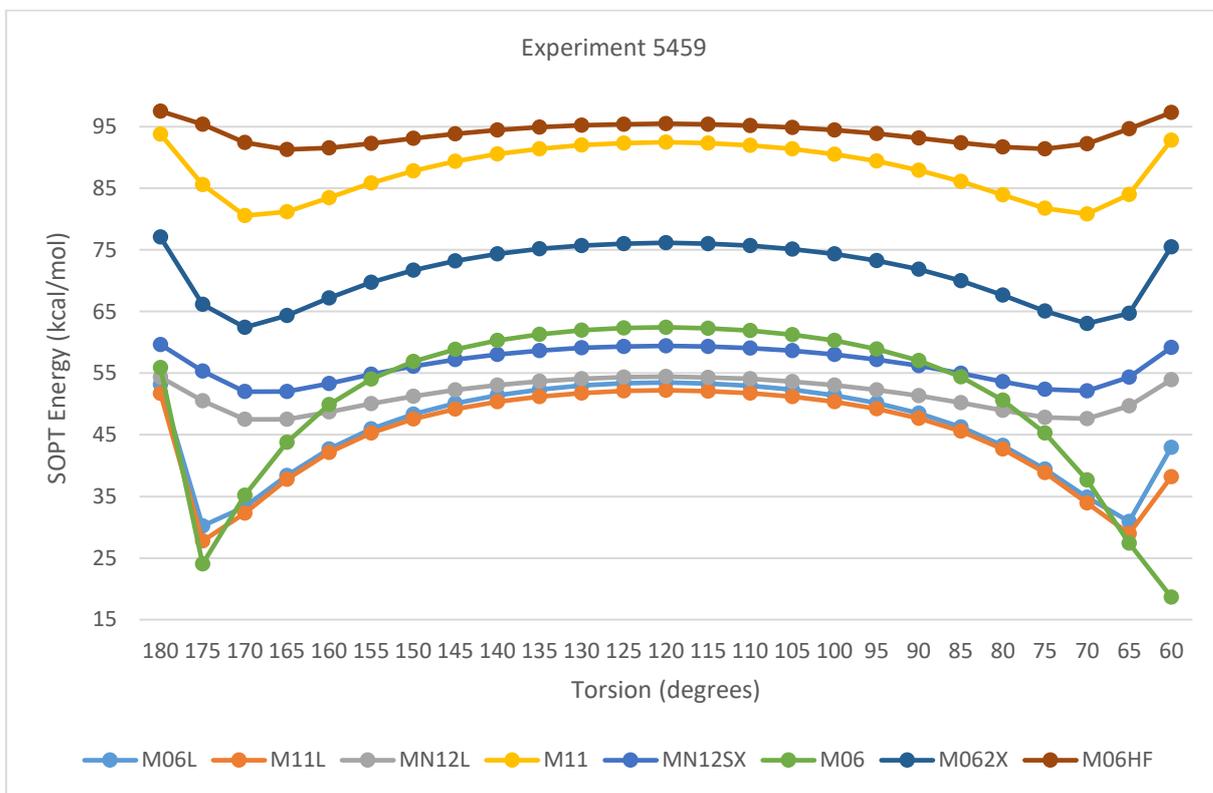

Figure 37. Ethanamide N(lp)->C-O(p)* SOPT Energy with 6-31G** UltraFineGrid and SCS-MP2/def2-QZVPP Optimized Geometry and Varying O-C-CA-H Torsion and Method

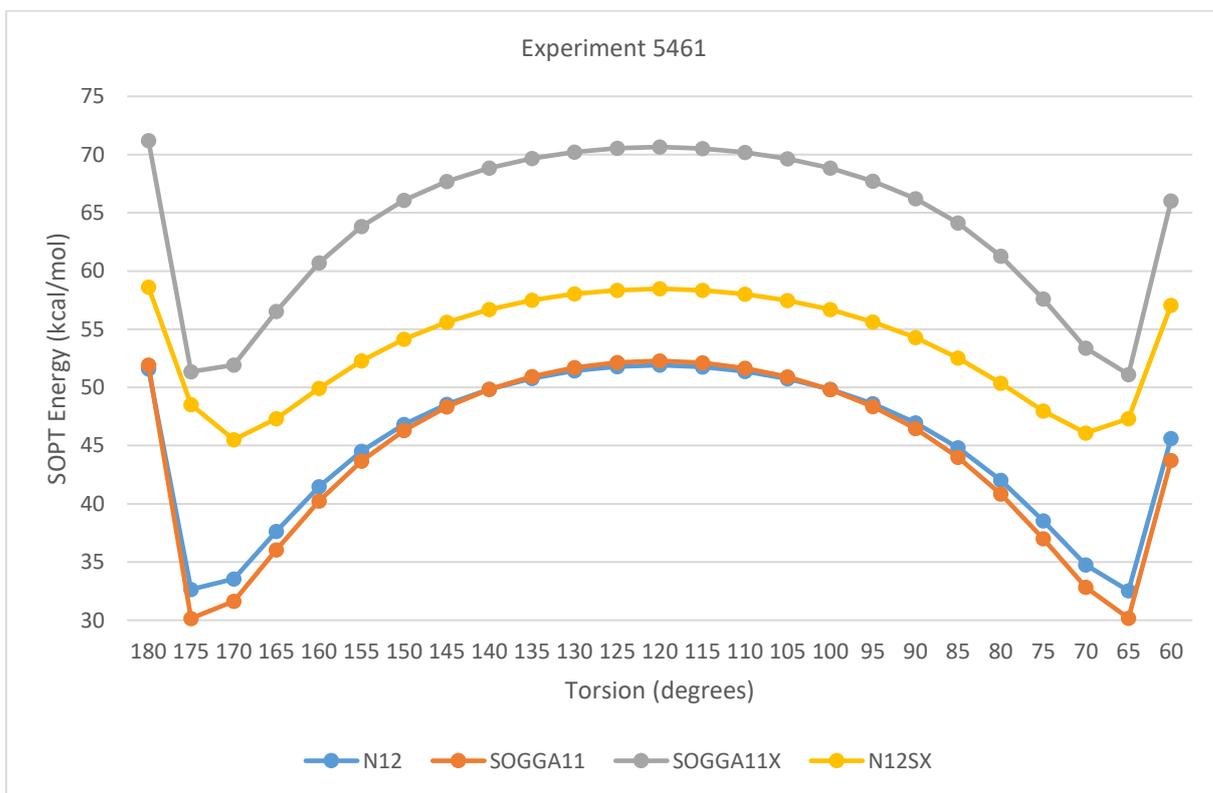

Figure 38. Ethanamide N(lp)->C-O(p)* SOPT Energy with 6-31G** UltraFineGrid and SCS-MP2/def2-QZVPP Optimized Geometry and Varying O-C-CA-H Torsion and Method

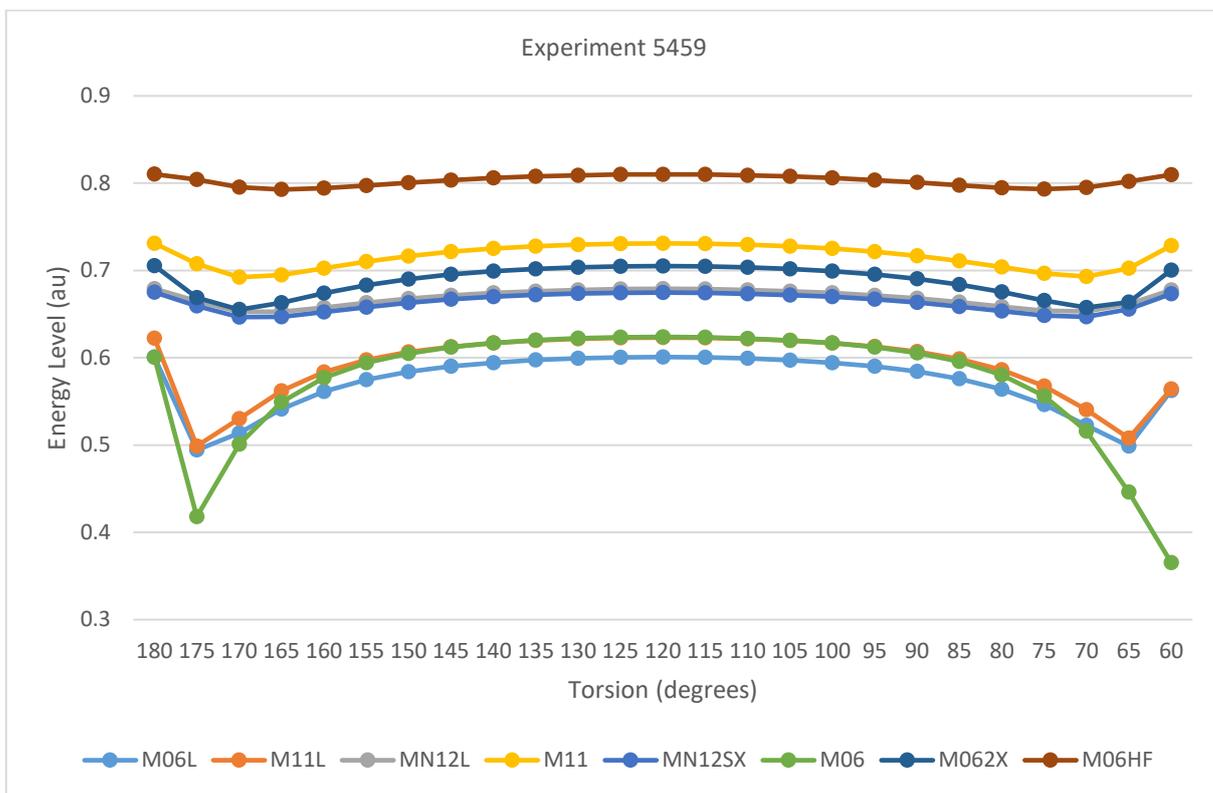

Figure 39. Ethanamide C-O(s)* Energy Level with 6-31G** UltraFineGrid and Varying O-C-CA-H Torsion and Method

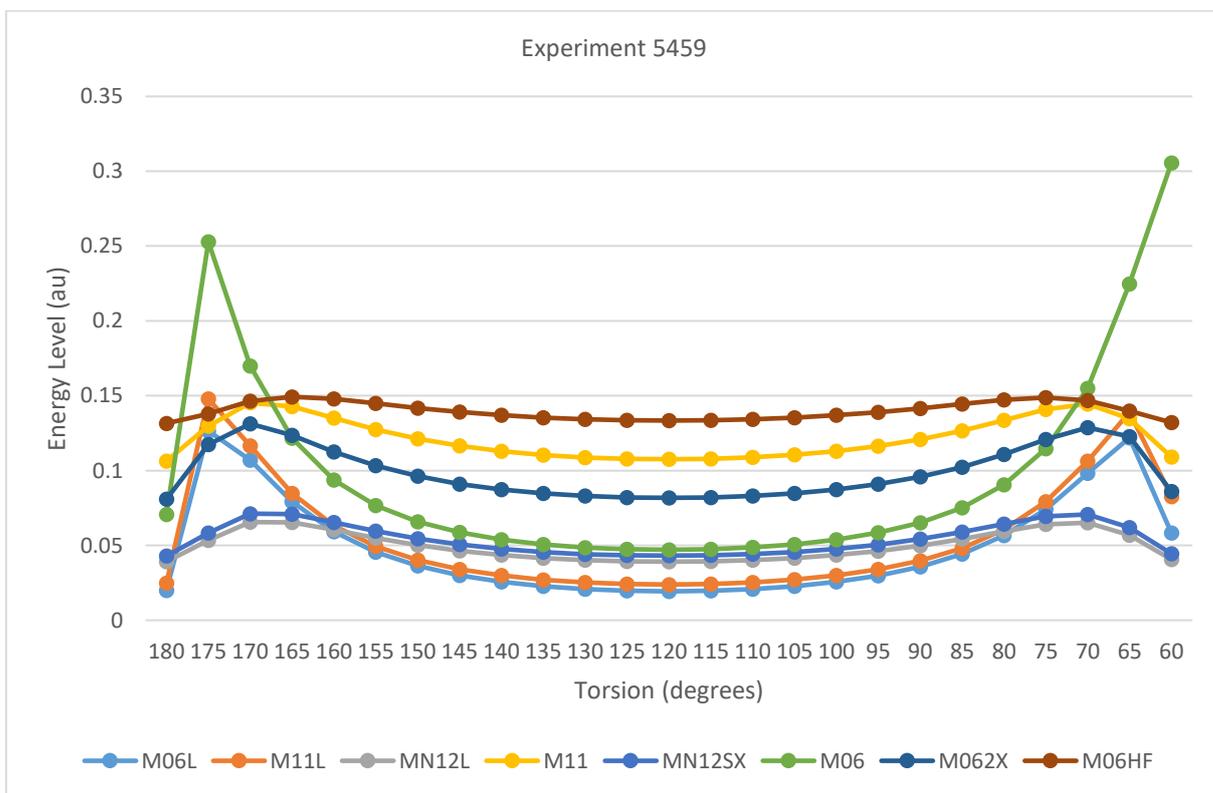

Figure 40. Ethanamide C-O(p)* Energy Level with 6-31G** UltraFineGrid and Varying O-C-CA-H Torsion and Method

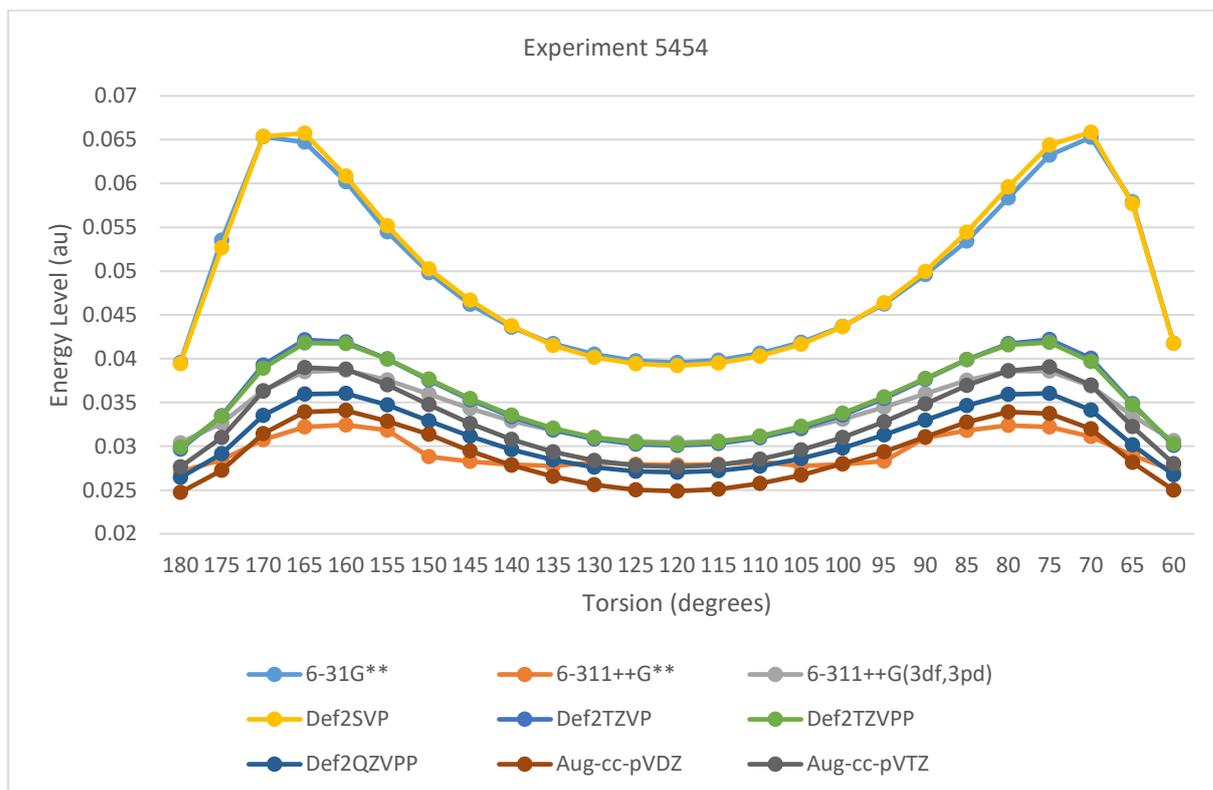

Figure 41. Ethanamide C-O(p)* Energy Level with MN12-L UltraFineGrid and Varying O-C-CA-H Torsion and Basis Set

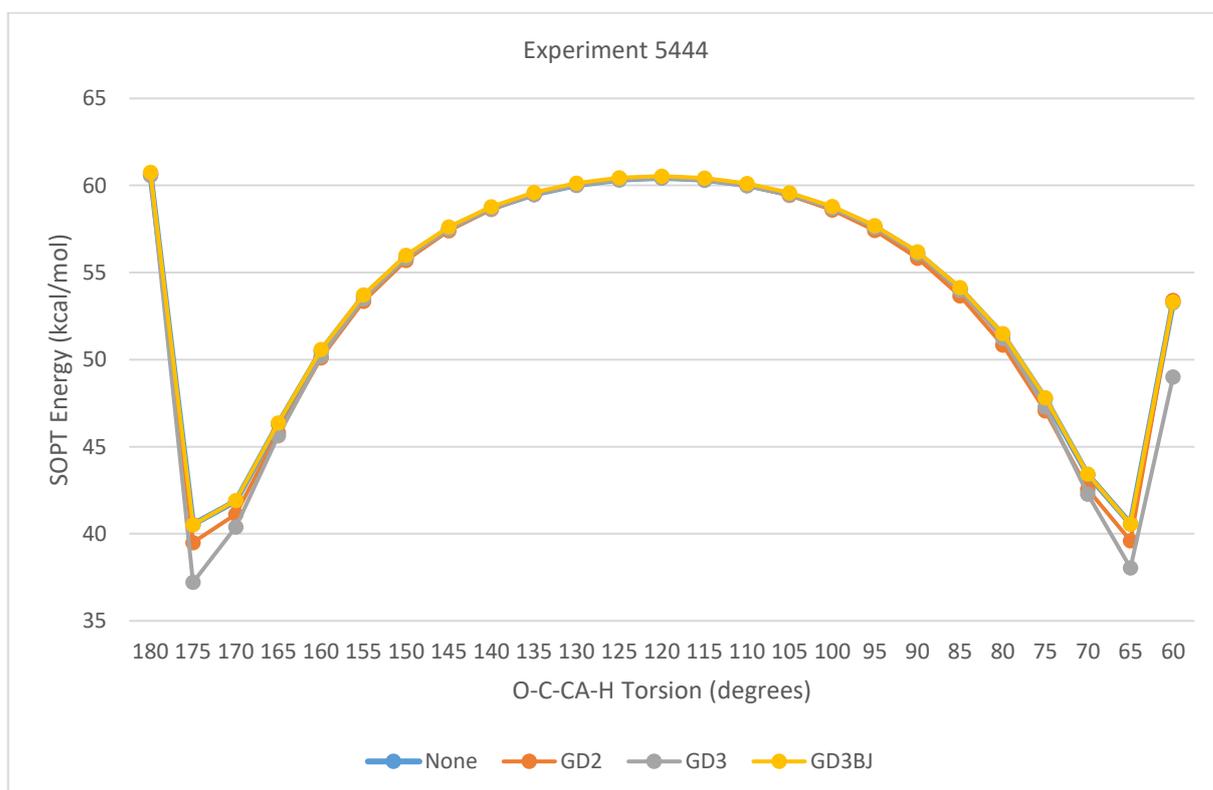

Figure 42. Ethanamide N(lp)->C-O(p)* SOPT Energy with B3LYP/6-31G** and UltraFineGrid at Varying O-C-CA-H Torsion and Empirical Dispersion Scheme

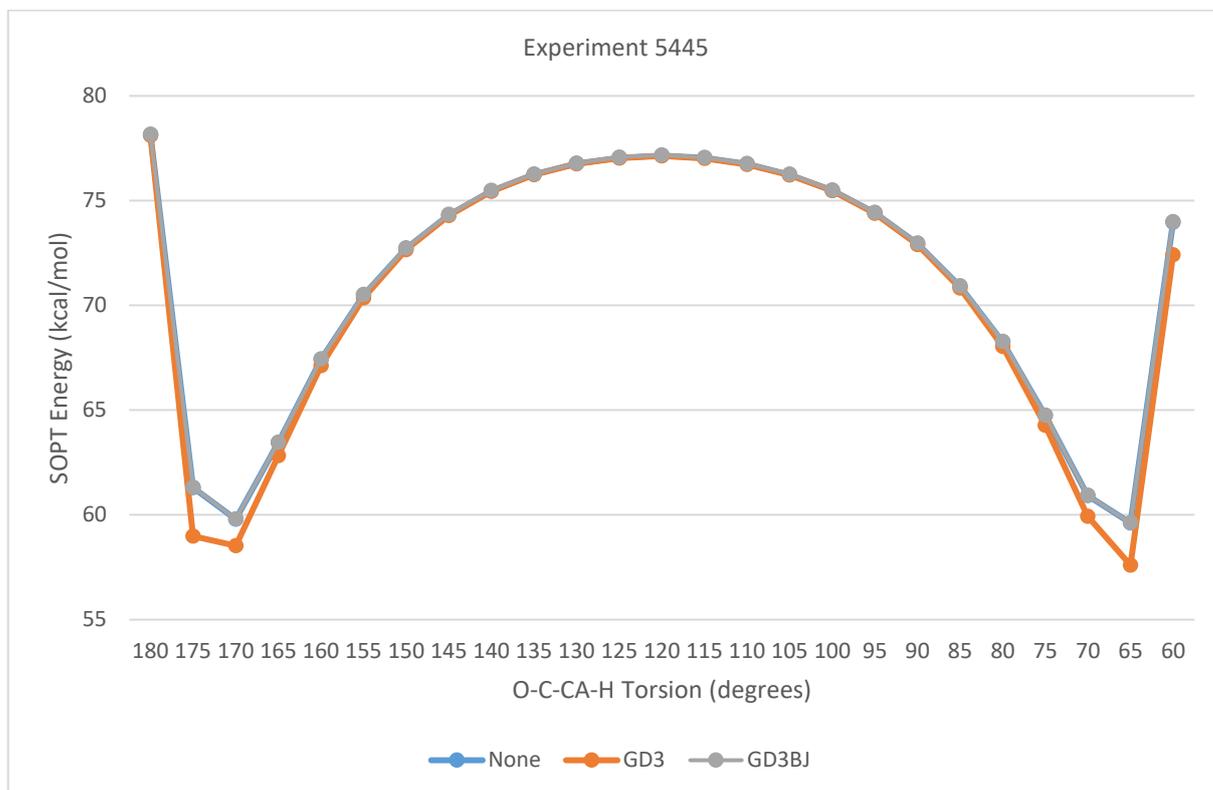

Figure 43. Ethanamide N(lp)->C-O(p)* SOPT Energy with CAM-B3LYP/6-31G** and UltraFineGrid at Varying O-C-CA-H Torsion and Empirical Dispersion Scheme

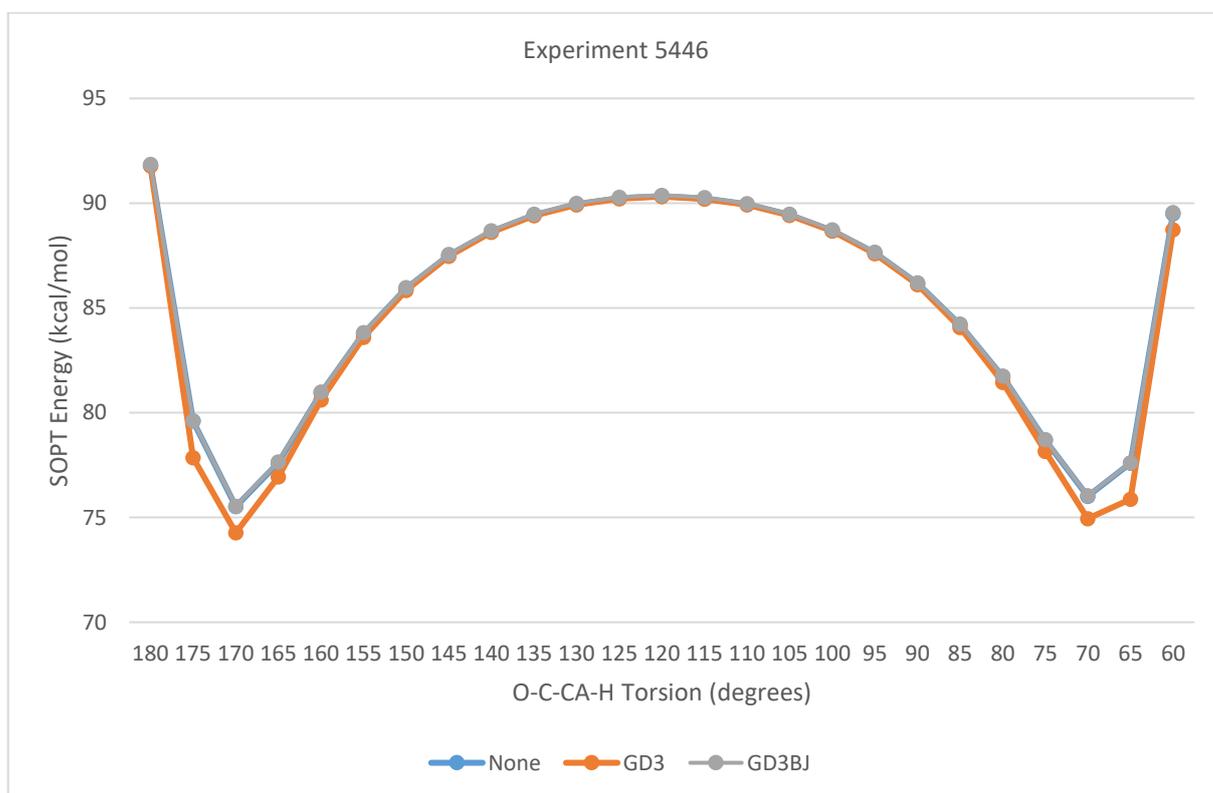

Figure 44. Ethanamide N(lp)->C-O(p)* SOPT Energy with LC-wPBE(w=0.4)/6-31G** and UltraFineGrid at Varying O-C-CA-H Torsion and Empirical Dispersion Scheme

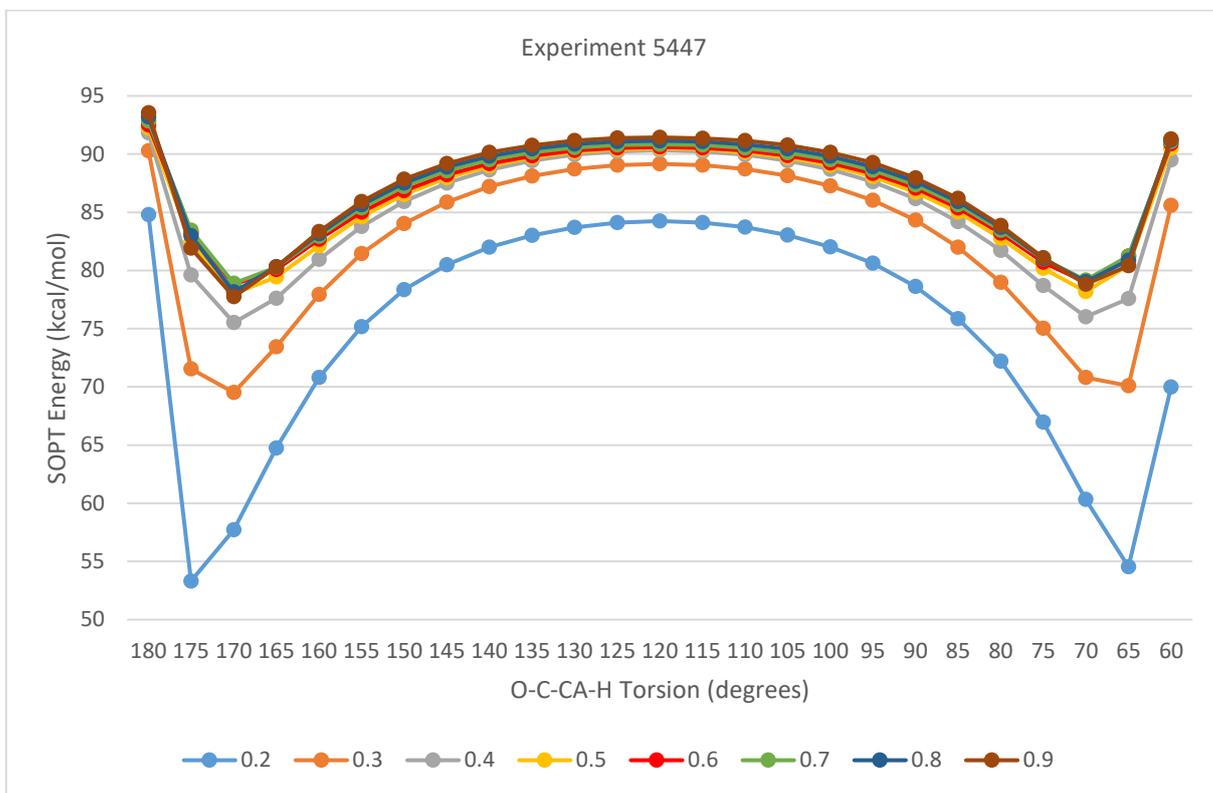

Figure 45. Ethanamide N(lp)->C-O(p)* SOPT Energy with LC-wPBE/6-31G** and UltraFineGrid at Varying O-C-CA-H Torsion and Omega

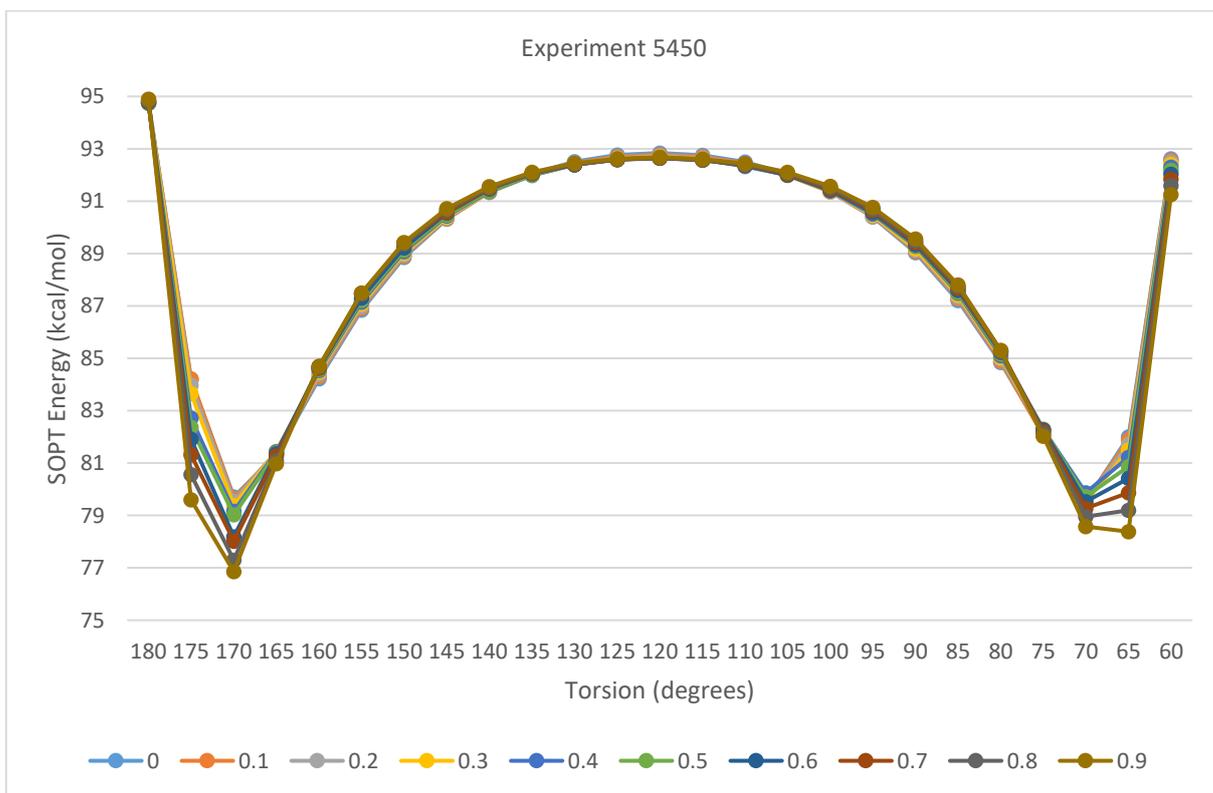

Figure 46. Ethanamide N(lp)->C-O(p)* SOPT Energy with LC-PBEhPBE at Omega=0.6 and Varying O-C-CA-H Torsion and HF Exchange Coefficient

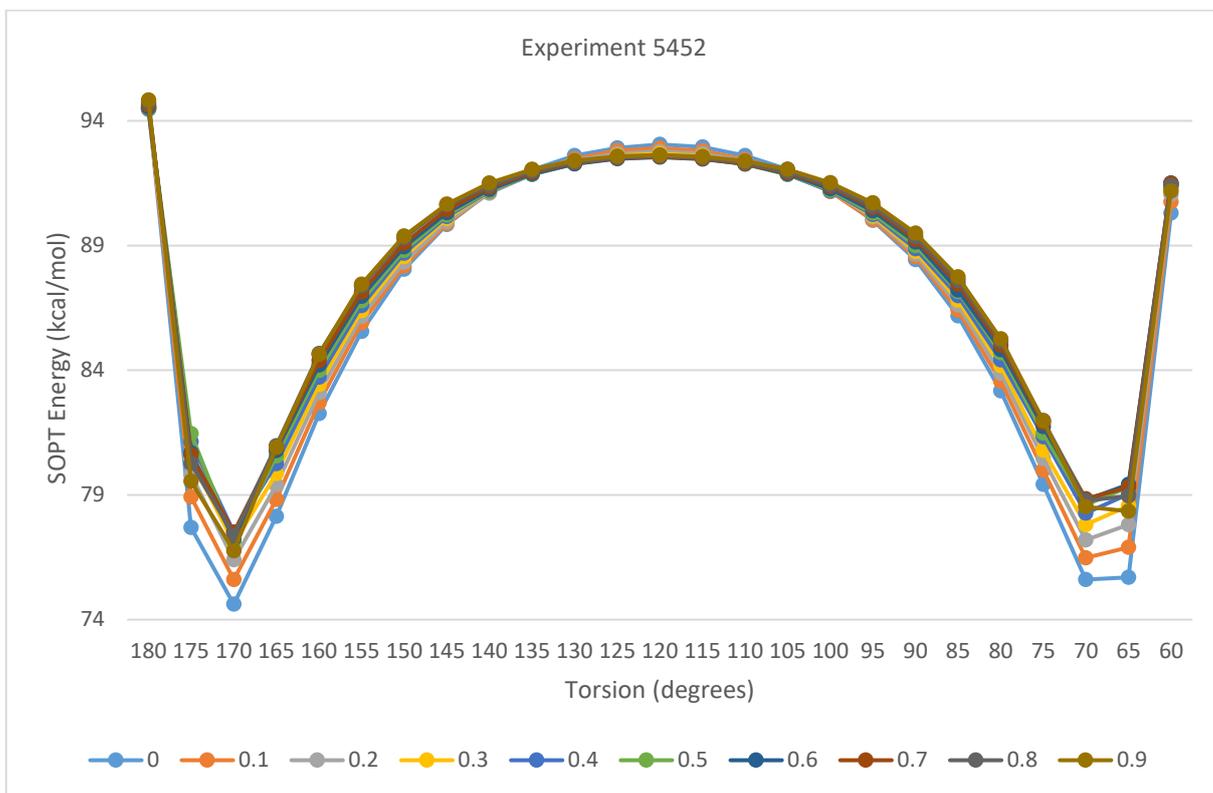

Figure 47. Ethanamide N(lp)->C-O(p)* SOPT Energy with LC-PBEhPBE at Omega=0.4 and Varying O-C-CA-H Torsion and HF Exchange Coefficient

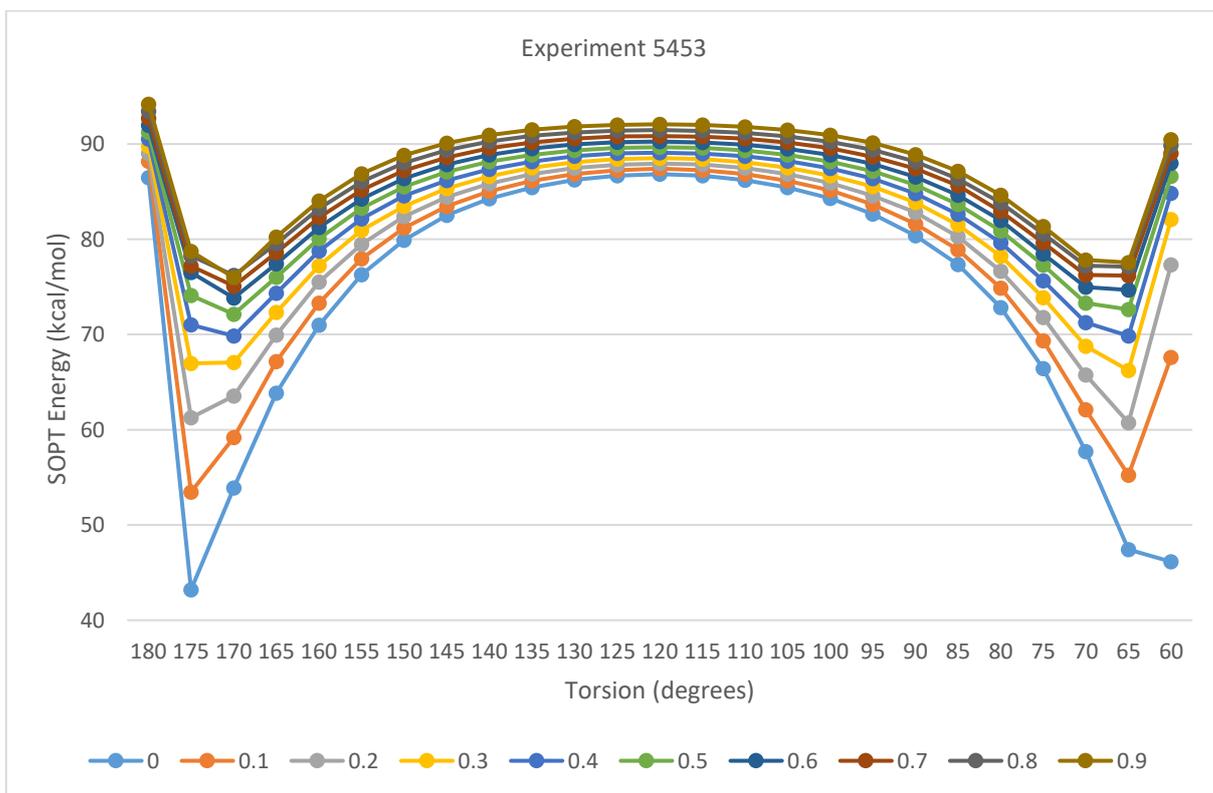

Figure 48. Ethanamide N(lp)->C-O(p)* SOPT Energy with LC-PBEhPBE Omega=0.2 and Varying O-C-CA-H Torsion and HF Exchange Coefficient

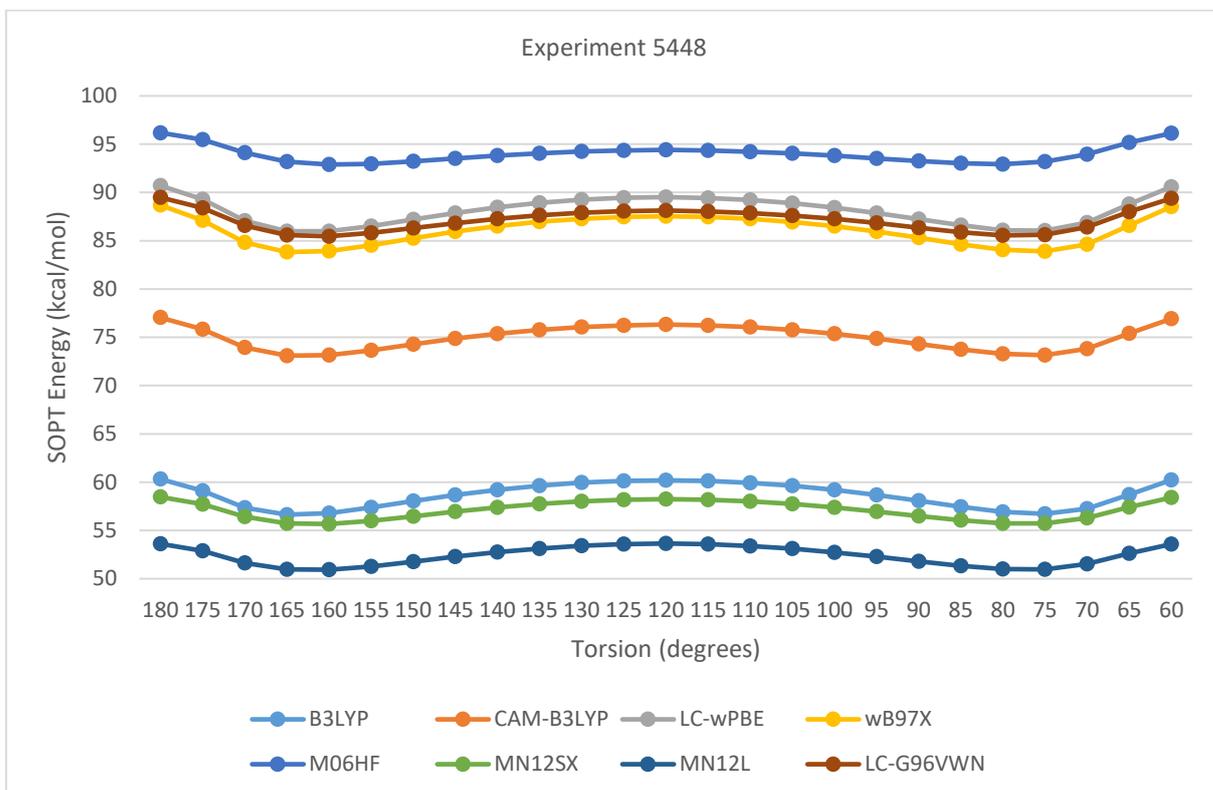

Figure 49. Ethanamide N(lp)->C-O(p)* SOPT Energy with def2-QZVPP UltraFineGrid over SCS-MP2/def2-QZVPP Optimized Geometry at Varying O-C-CA-H Torsion and Method

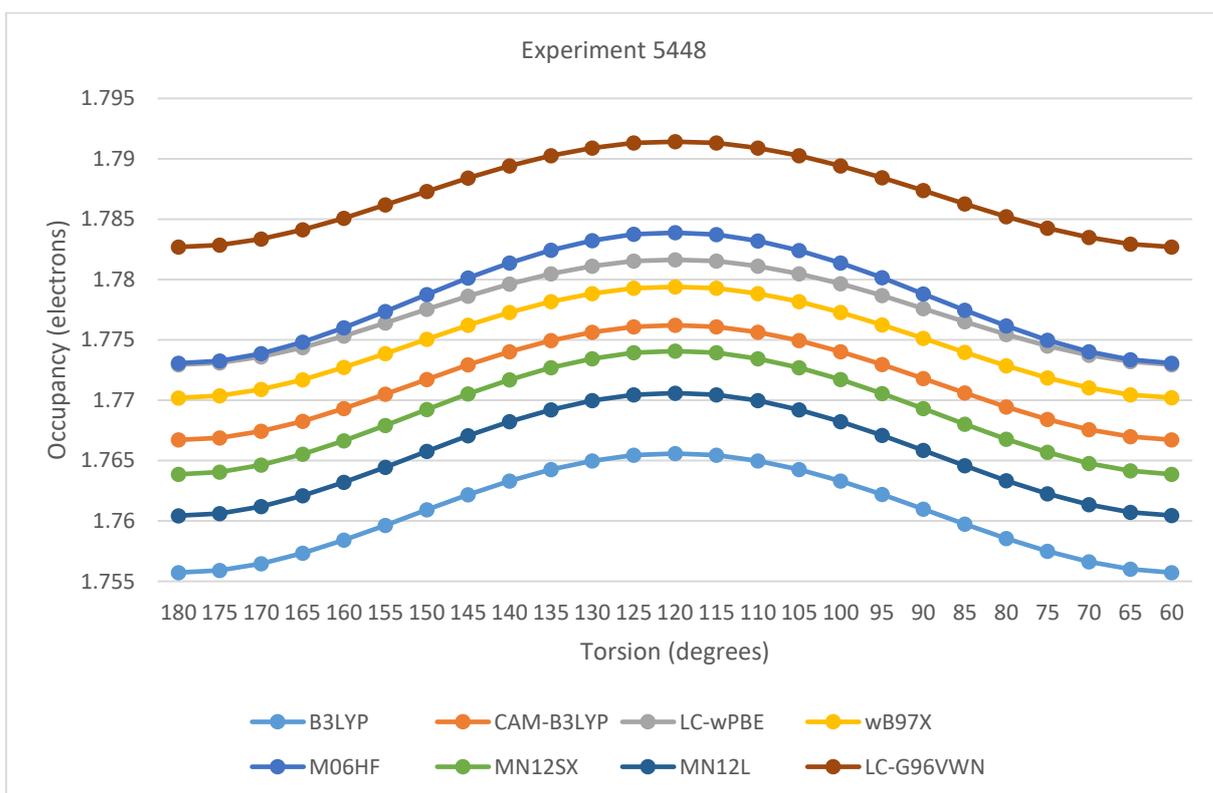

Figure 50. Ethanamide N(lp) Occupancy with def2-QZVPP UltraFineGrid and SCS-MP2/def2-QZVPP Optimized Geometry with Varying O-C-CA-H Torsion and Method

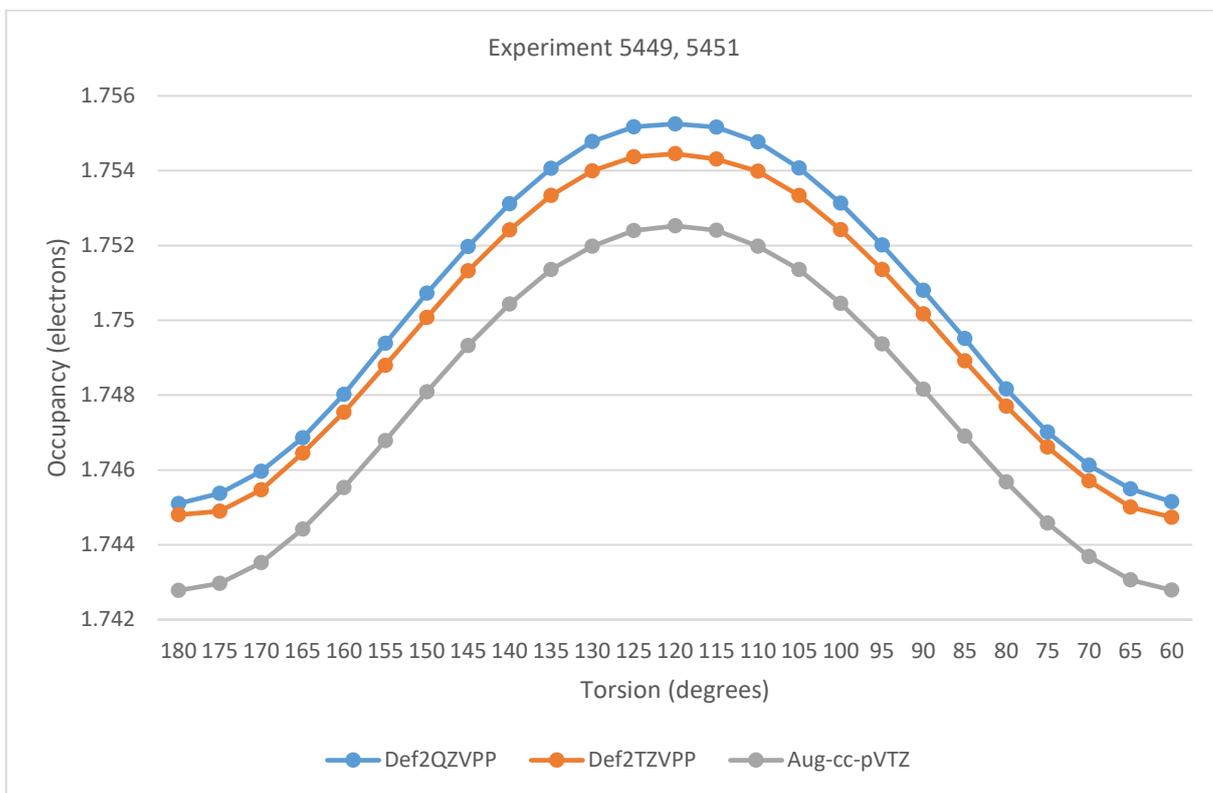

Figure 51. Ethanamide N(lp) Occupancy with DLPNO-CCSD(T) over SCS-MP2/def2-QZVPP Optimized Geometry with Varying O-C-CA-H Torsion and Basis Set

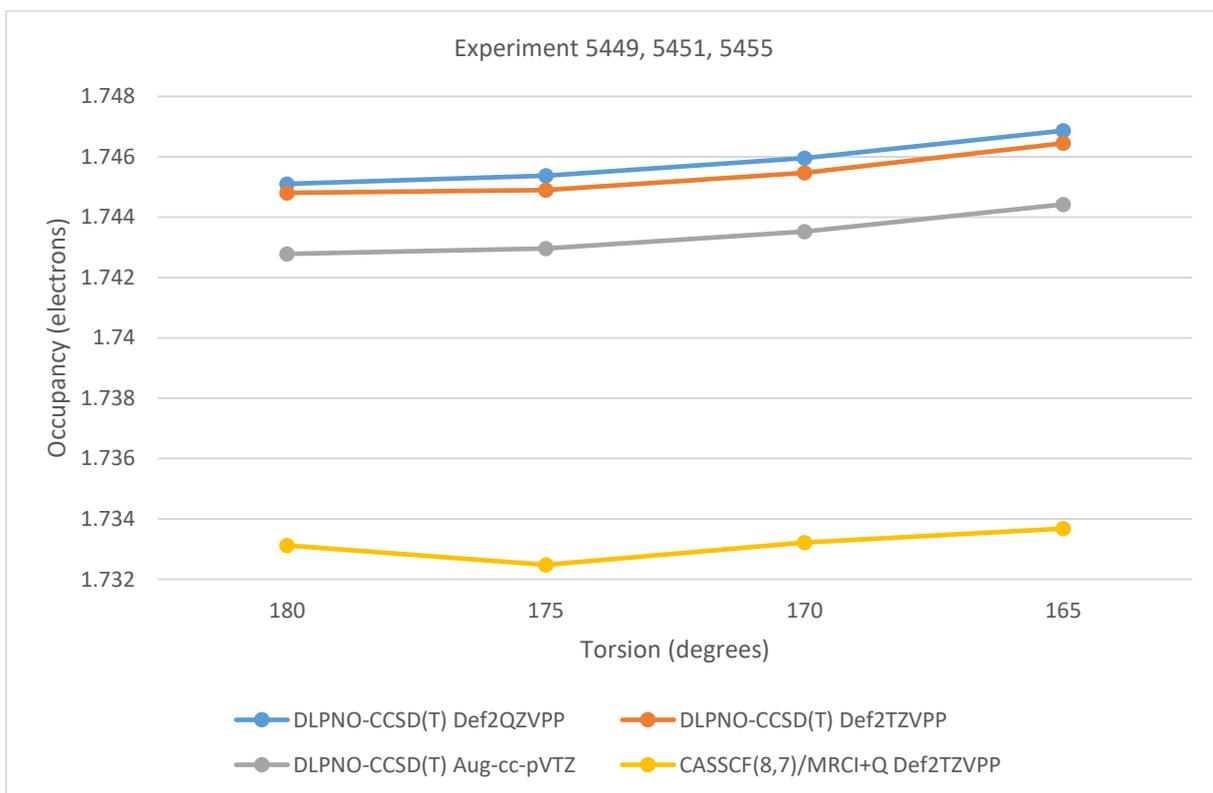

Figure 52. Ethanamide N(lp) Occupancy at SCS-MP2/def2-QZVPP Optimized Geometry and Varying O-C-CA-H Torsion and Method/Basis Set

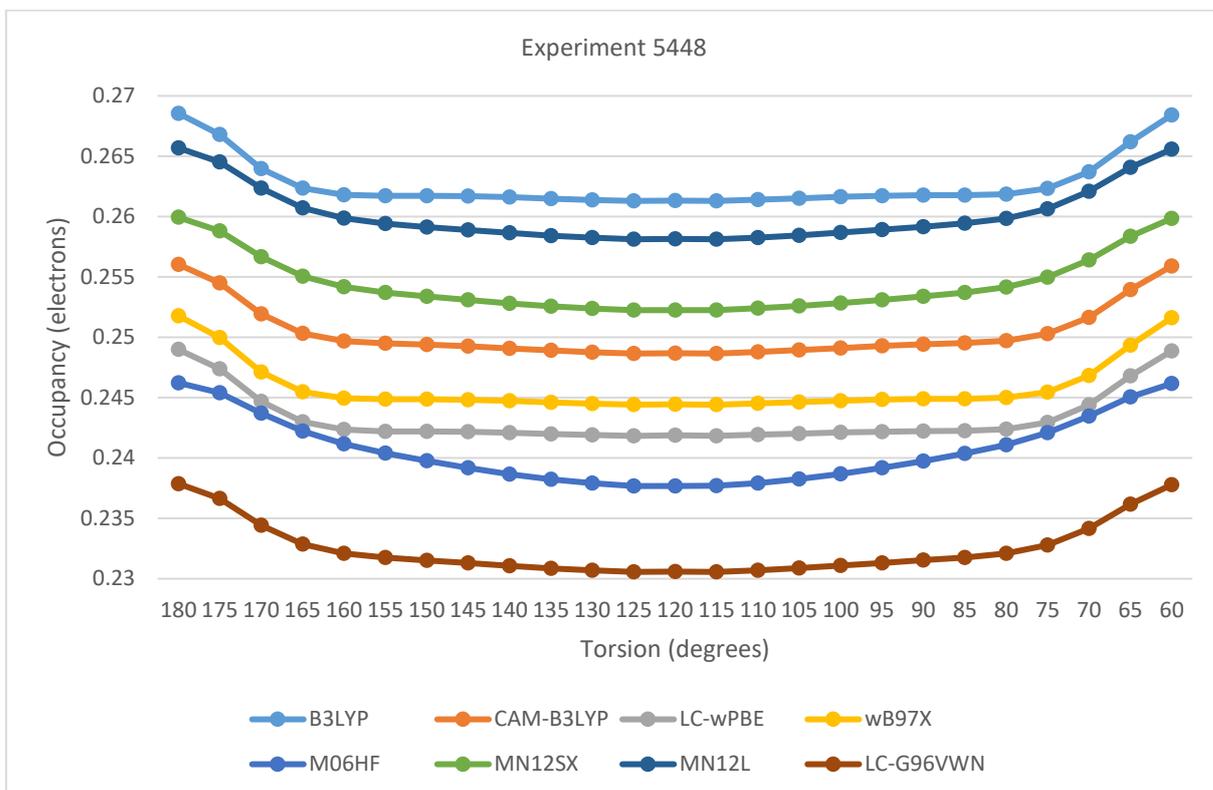

Figure 53. Ethanamide C-O(p)* Occupancy with def2-QZVPP UltraFineGrid and SCS-MP2/def2-QZVPP Optimized Geometry and Varying O-C-CA-H Torsion and Method

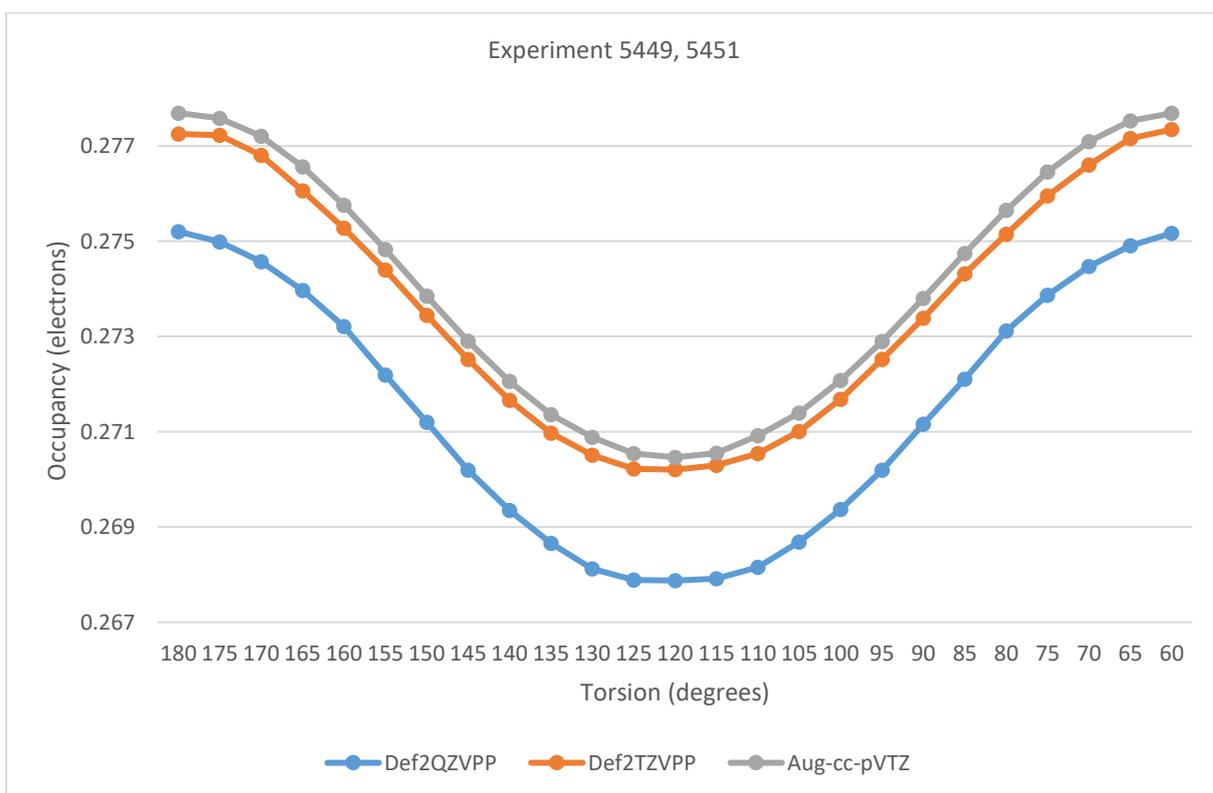

Figure 54. Ethanamide C-O(p)* Occupancy with DLPNO-CCSD(T) at SCS-MP2/def2-QZVPP Optimized Geometry and Varying O-C-CA-H Torsion Basis Set

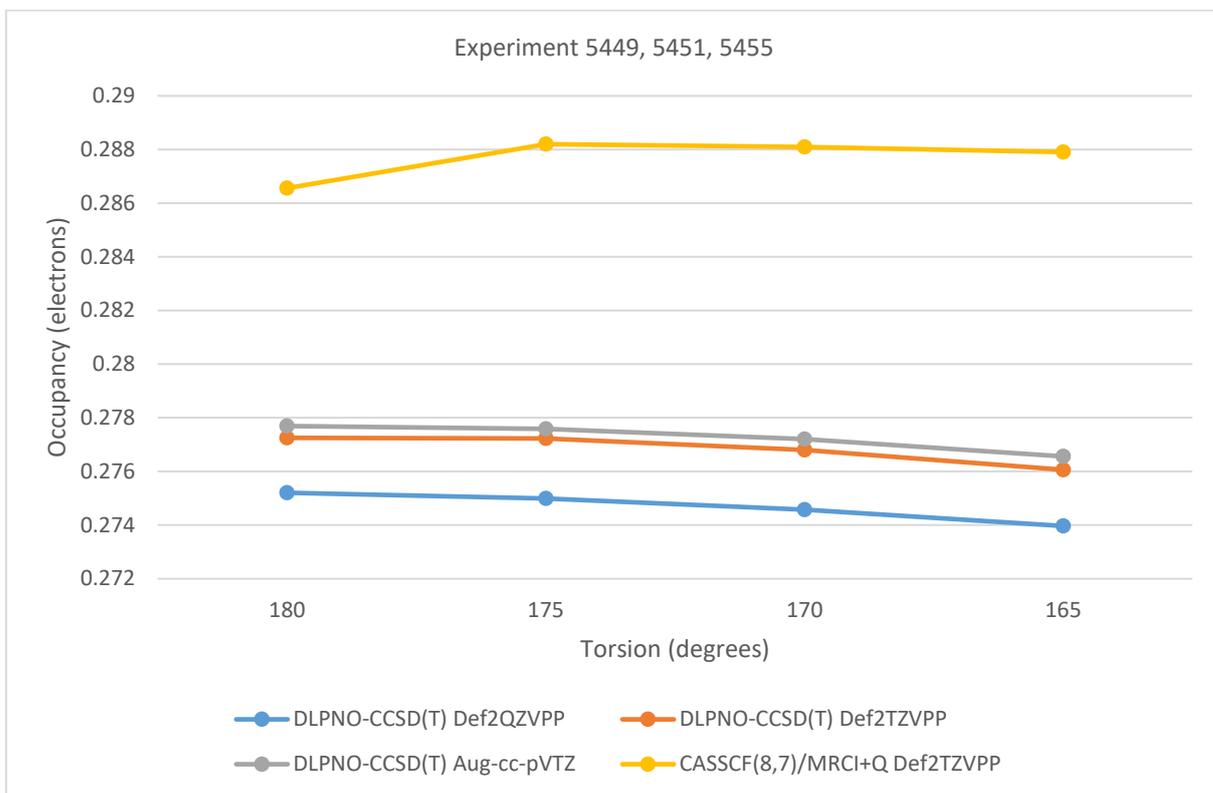

Figure 55. Ethanamide C-O(p)* Occupancy with SCS-MP2/def2-QZVPP Optimized Geometry and Varying O-C-CA-H Torsion and Method/Basis

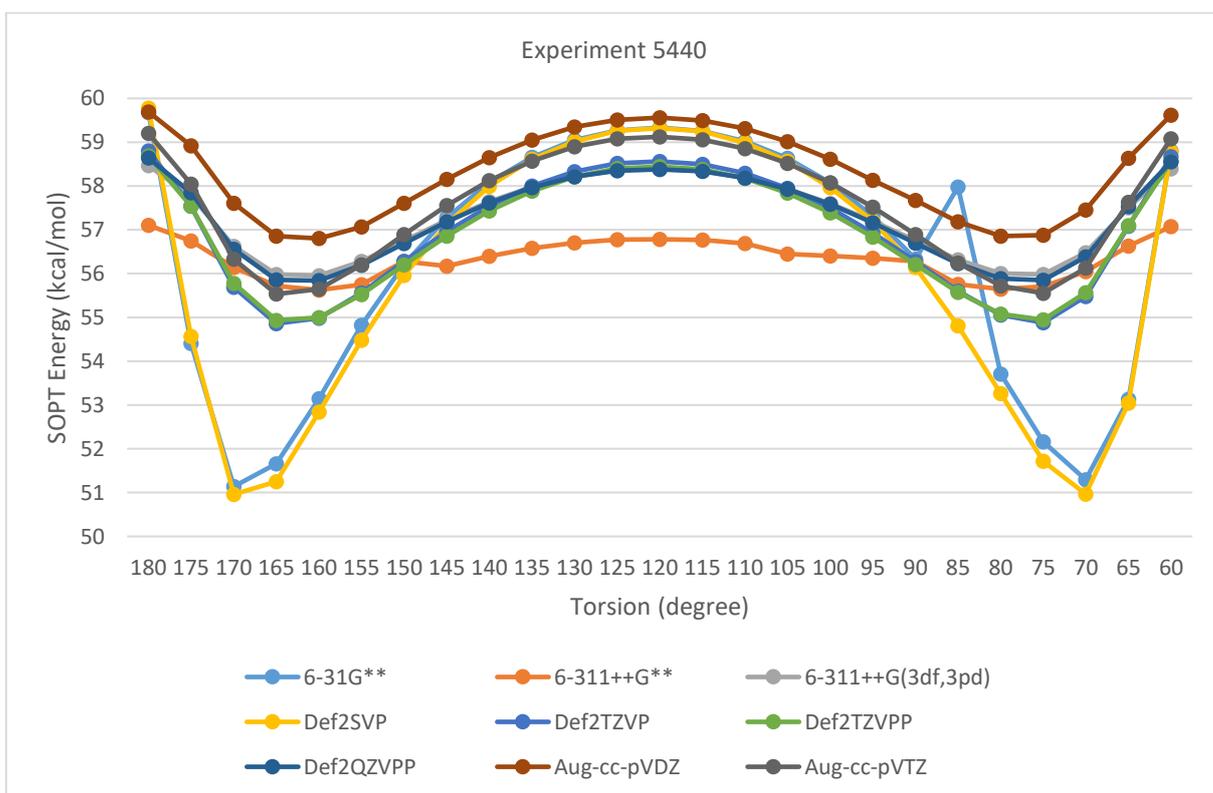

Figure 56. Ethanamide N(lp)->C-O(p)* SOPT Energy with MN12-SX UltraFineGrid and Varying O-C-CA-H Torsion and Basis Set

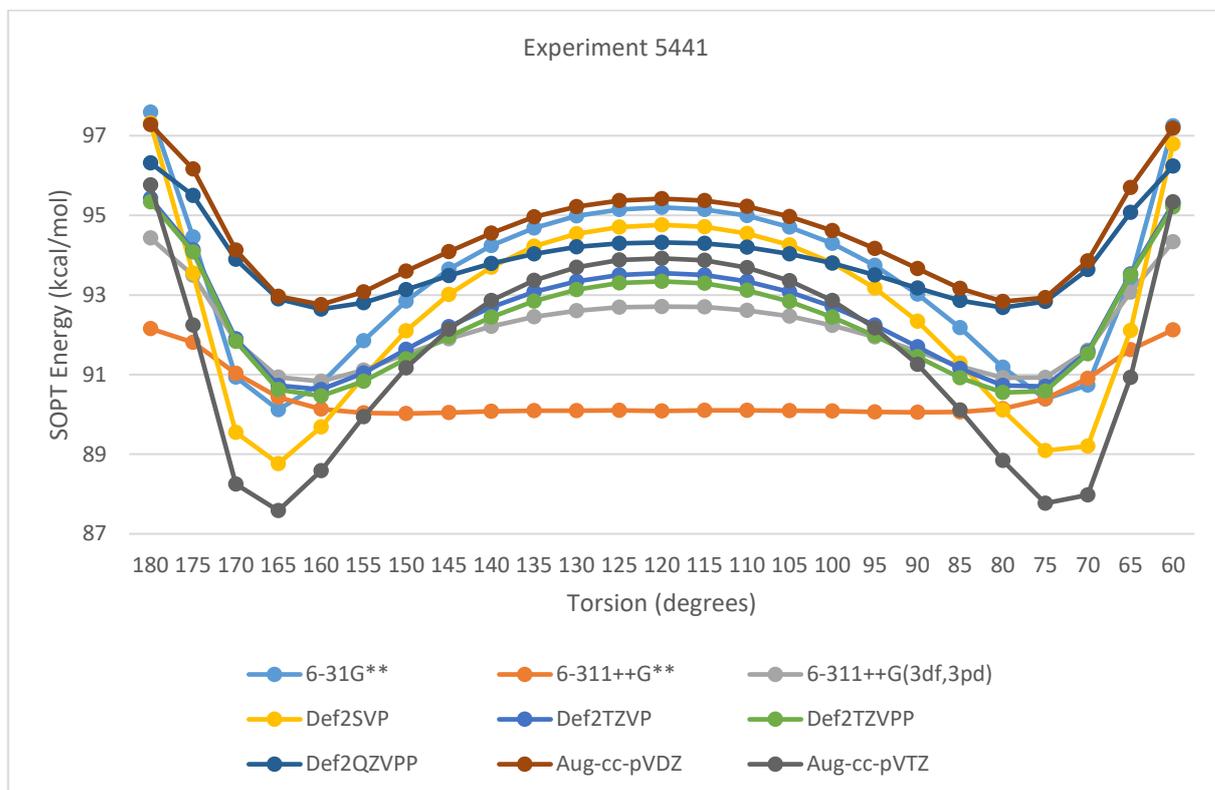

Figure 57. Ethanamide N(lp)->C-O(p)* SOPT Energy with M06HF UltraFineGrid and Varying O-C-CA-H Torsion and Basis Set

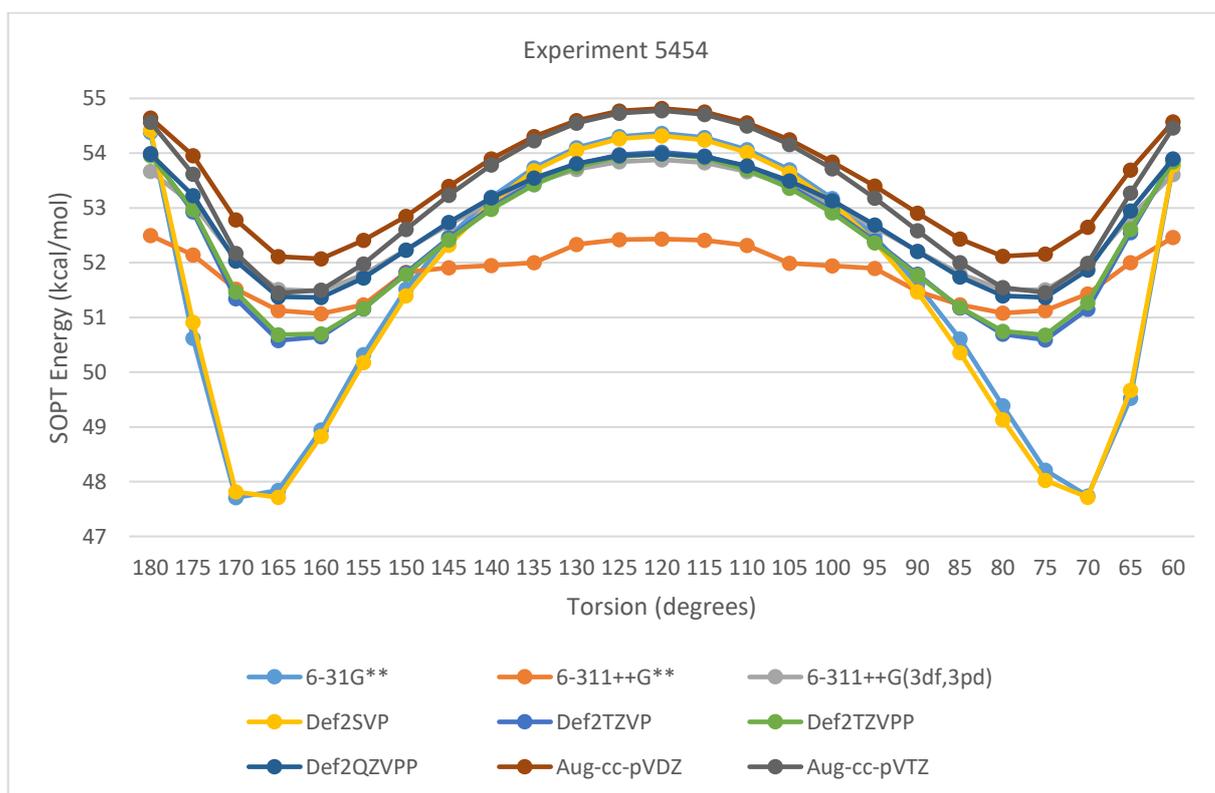

Figure 58. Ethanamide N(lp)->C-O(p)* SOPT Energy with MN12-L UltraFineGrid and Varying O-C-CA-H Torsion and Basis Set

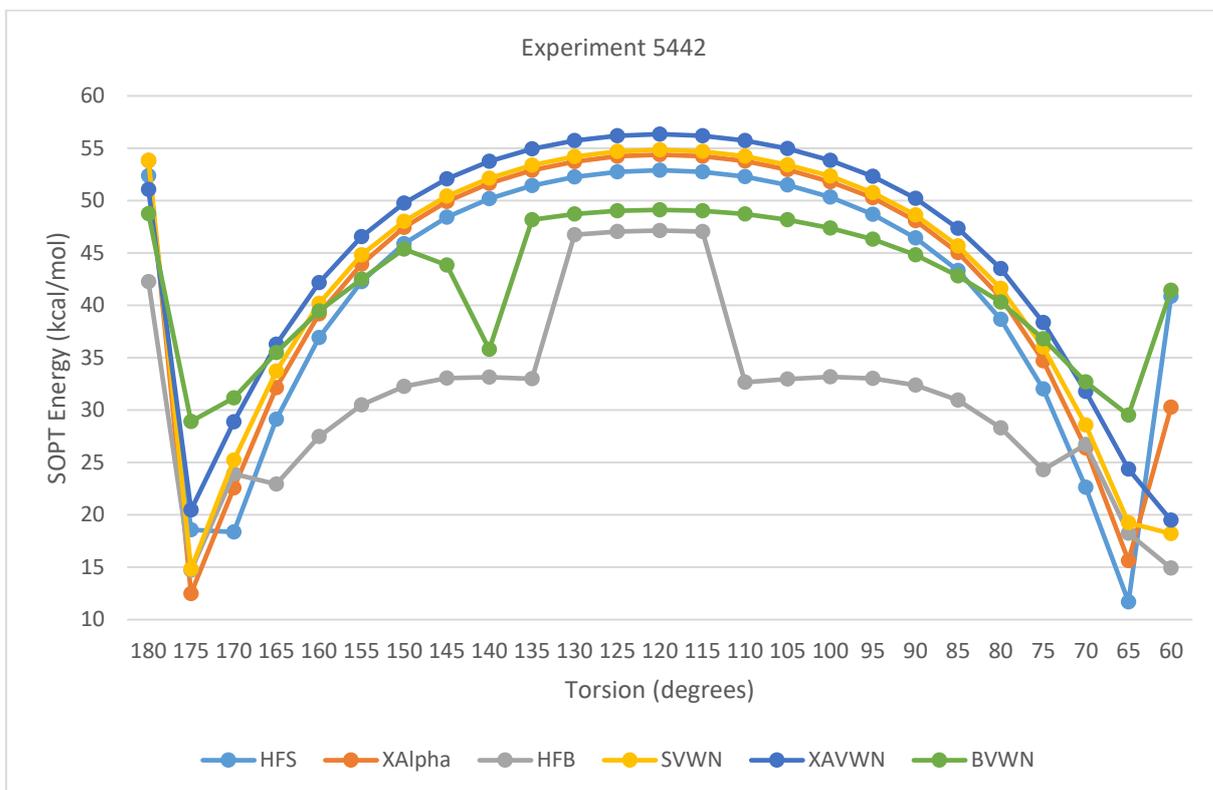

Figure 59. Ethanamide N(lp)->C-O(p)* SOPT Energy at 6-31G** UltraFineGrid at Varying O-C-CA-H Torsion and Method

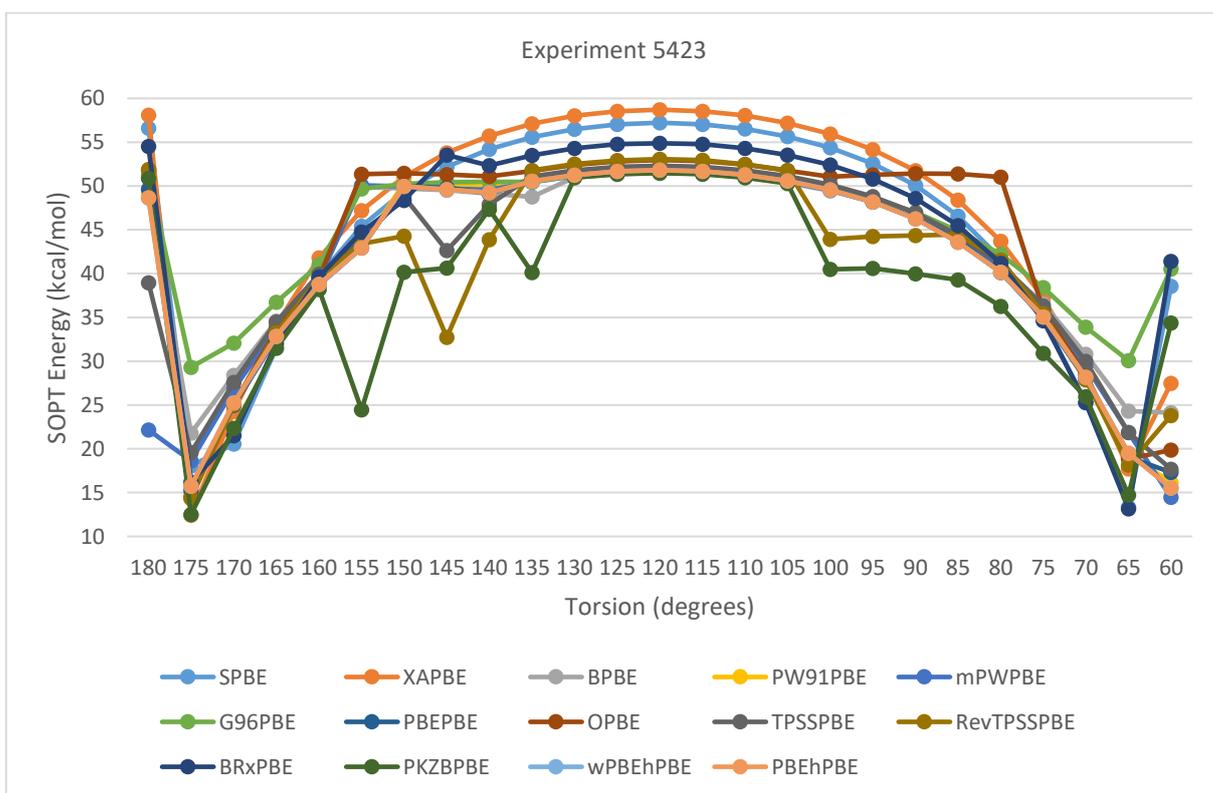

Figure 60. Ethanamide N(lp)->C-O(p)* SOPT Energy with 6-31G** and UltraFineGrid and Varying O-C-CA-H Torsion and Method having PBE Correlation

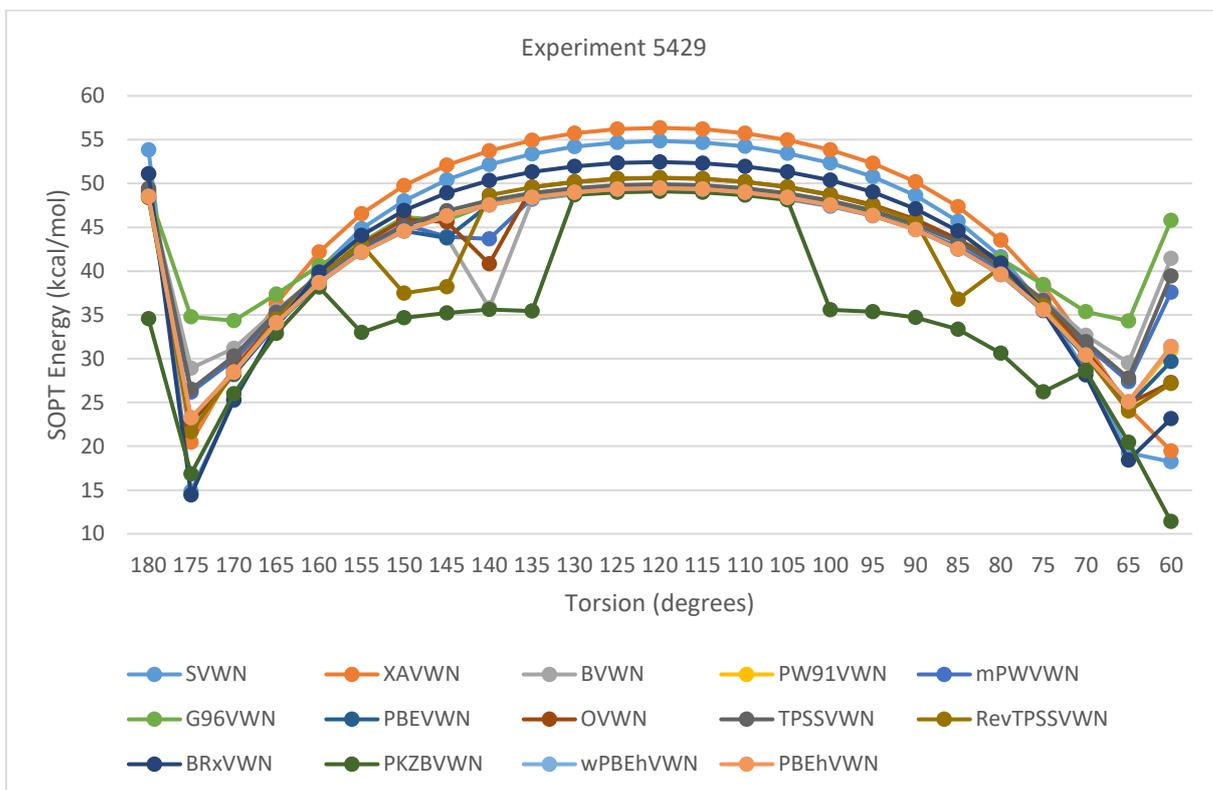

Figure 61. Ethanamide N(lp)->C-O(p)* SOPT Energy with 6-31G** UltraFineGrid and Varying O-C-CA-H Torsion and Method having VWN Correlation

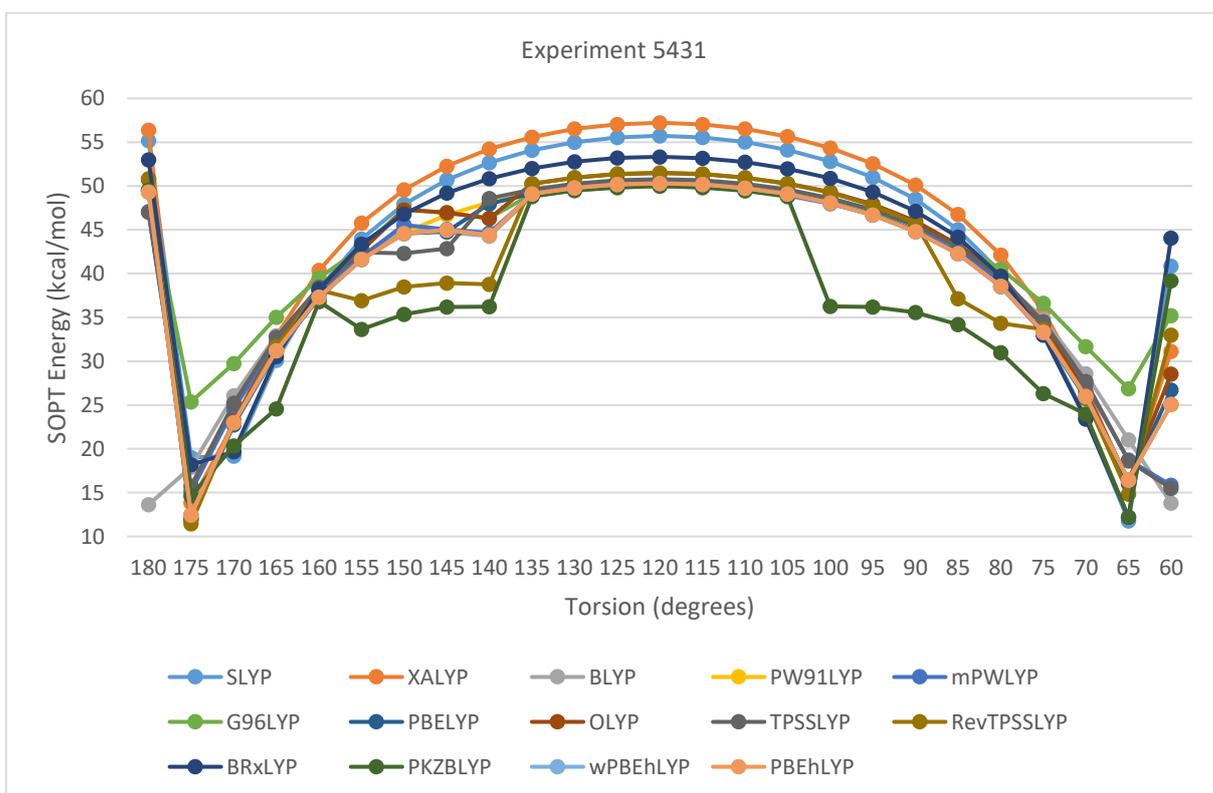

Figure 62. Ethanamide N(lp)->C-O(p)* SOPT Energy with 6-31G** UltraFineGrid and Varying O-C-CA-H Torsion and Method having LYP Correlation

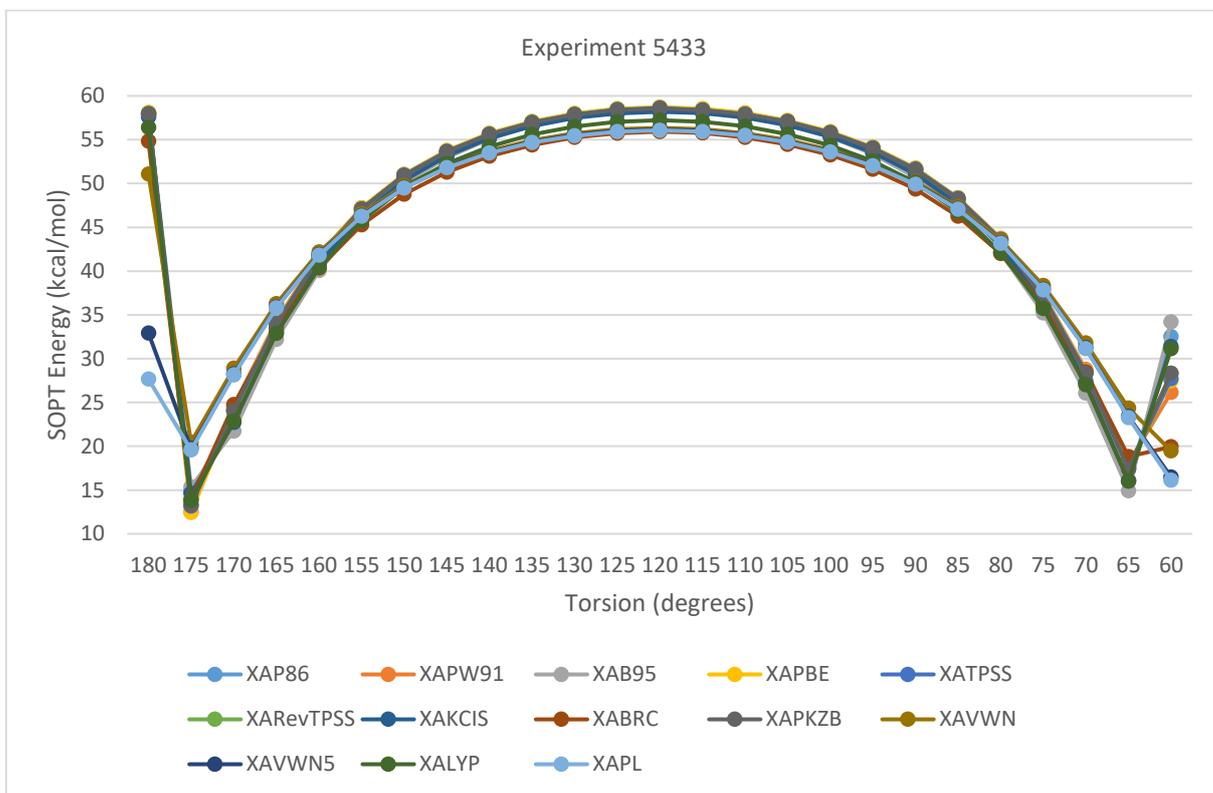

Figure 63. Ethanamide N(lp)->C-O(p)* SOPT Energy with 6-31G** UltraFineGrid and Varying O-C-CA-H Torsion and Method having XA Exchange

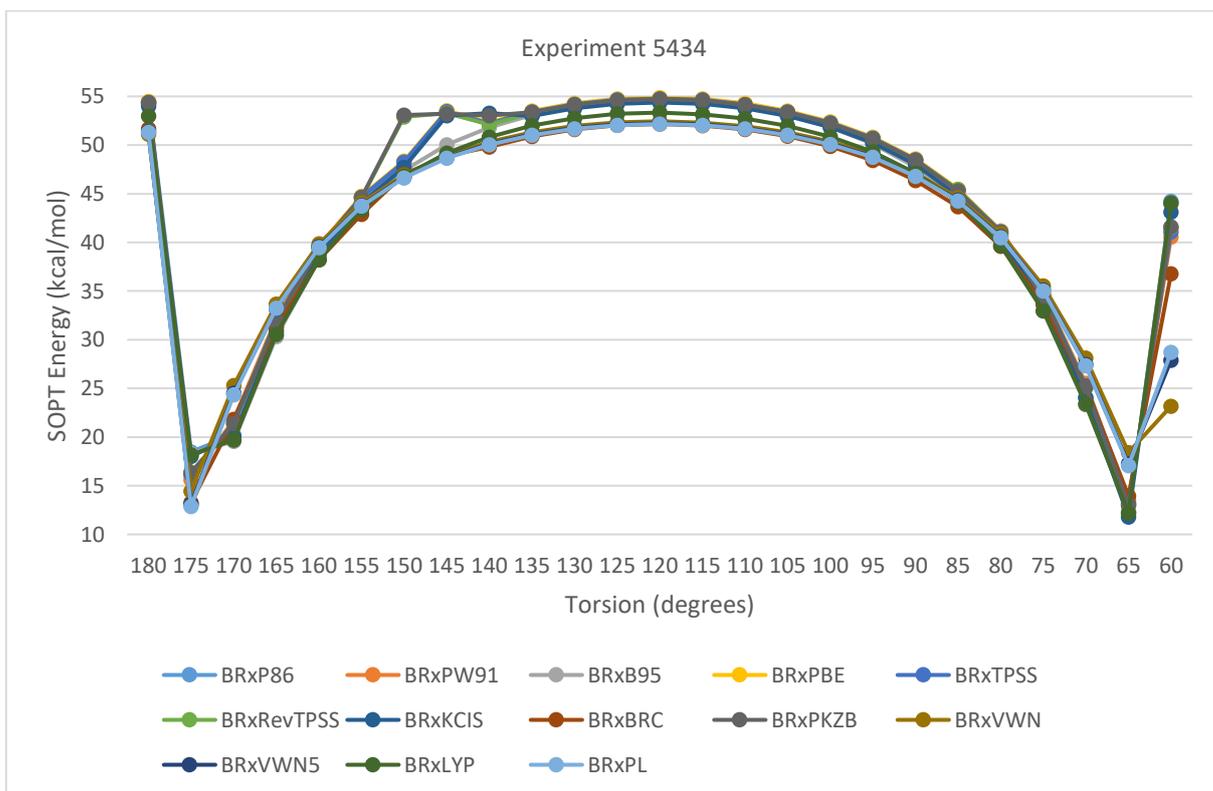

Figure 64. Ethanamide N(lp)->C-O(p)* SOPT Energy with 6-31G** UltraFineGrid and Varying O-C-CA-H Torsion and Method having BRx Exchange

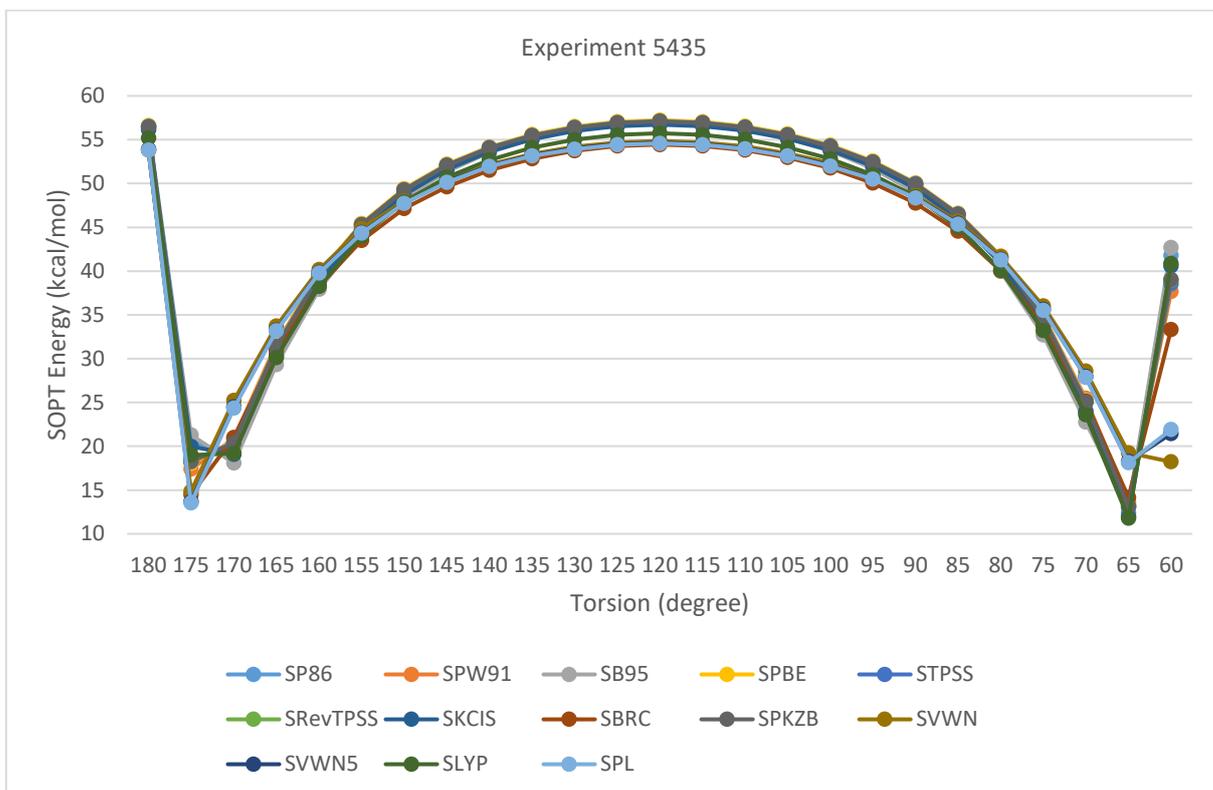

Figure 65. Ethanamide N(lp)->C-O(p)* SOPT Energy with 6-31G** UltraFineGrid and Varying O-C-CA-H Torsion and Method having Slater Exchange

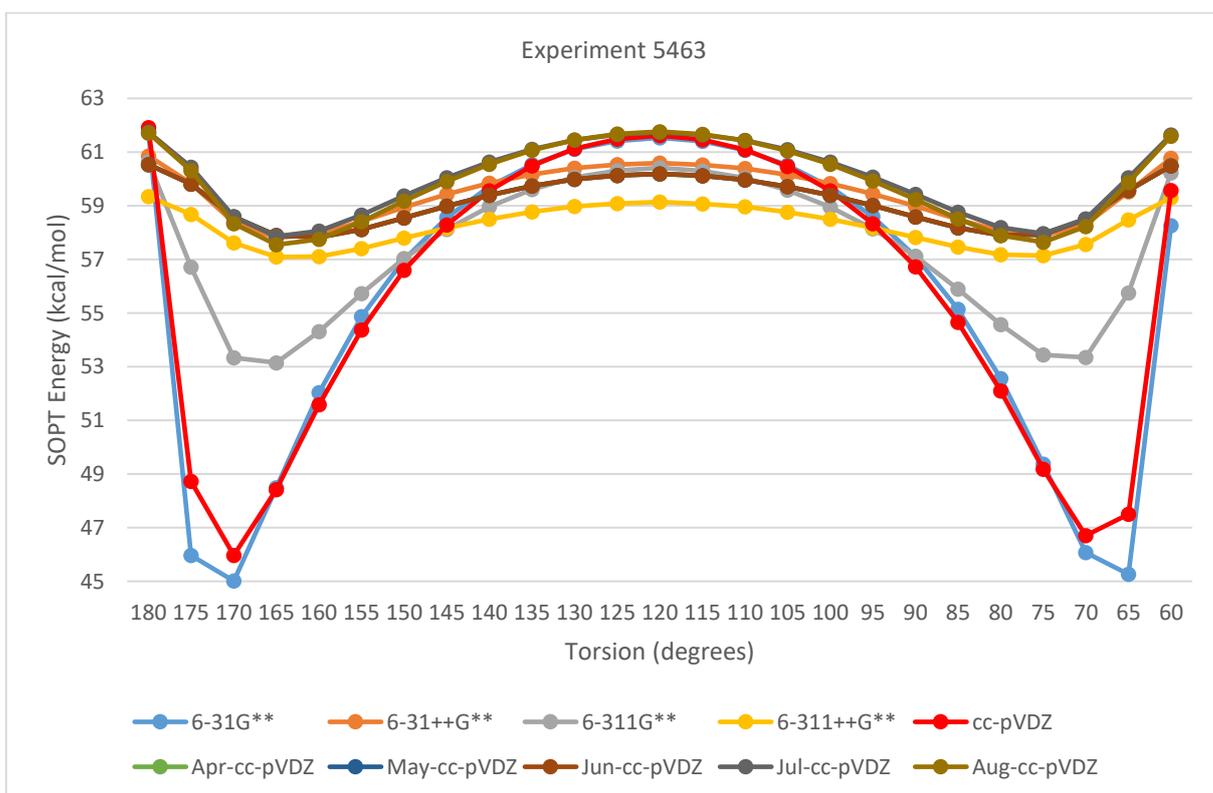

Figure 66. Ethanamide N(lp)->C-O(p)* SOPT Energy with B3LYP over SCS-MP2/def2-QZVPP Optimized Geometry, UltraFineGrid and Varying O-C-CA-H Torsion and Basis Set

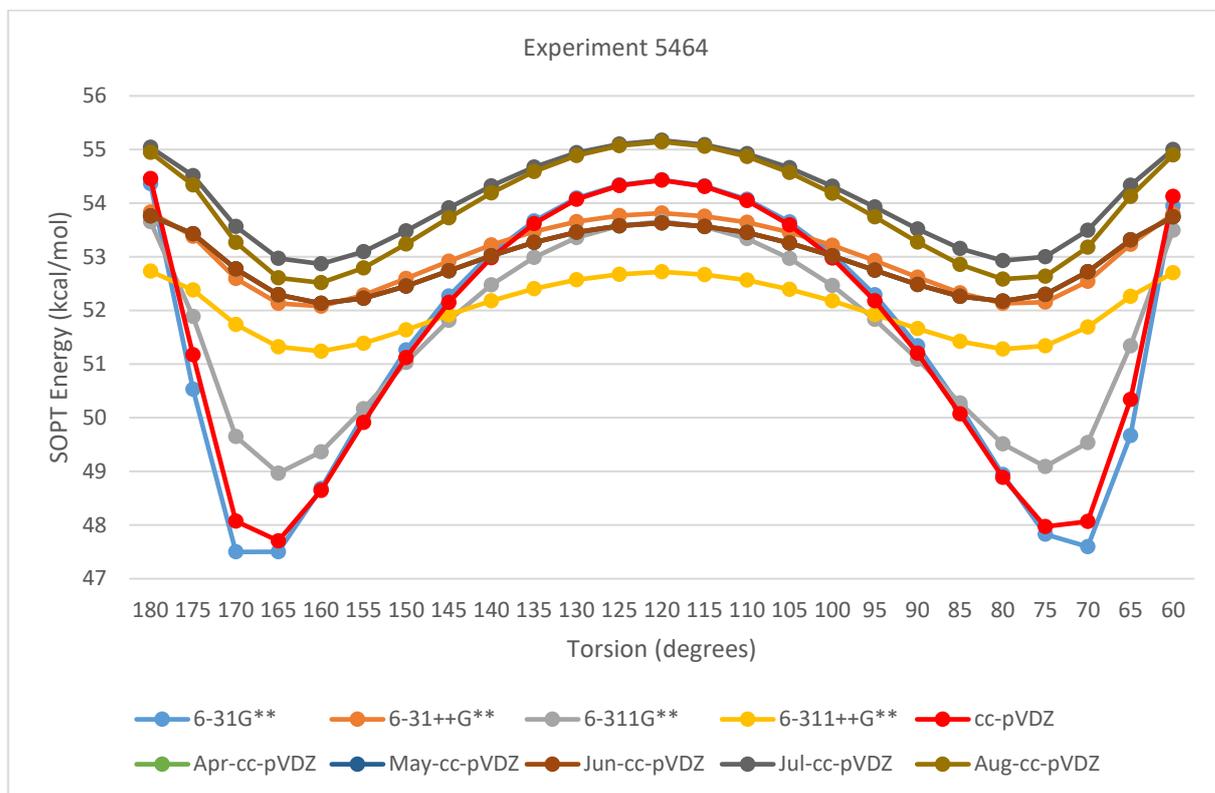

Figure 67. Ethanamide N(lp)->C-O(p)* SOPT Energy with MN12-L and SCS-MP2/def2-QZVPP Optimized Geometry, UltraFineGrid and Varying O-C-CA-H Torsion and Basis Set

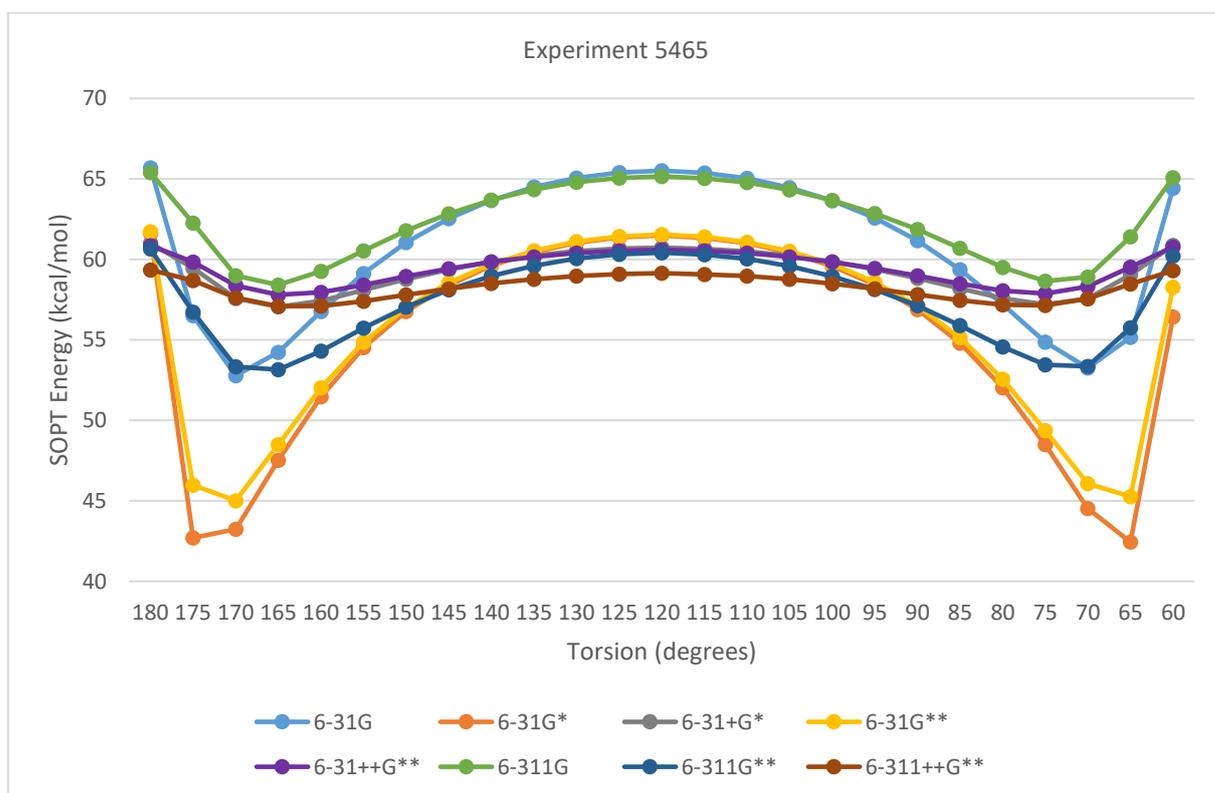

Figure 68. Ethanamide N(lp)->C-O(p)* SOPT Energy with B3LYP over SCS-MP2/def2-QZVPP Optimized Geometry, UltraFineGrid and Varying O-C-CA-H Torsion and Basis Set

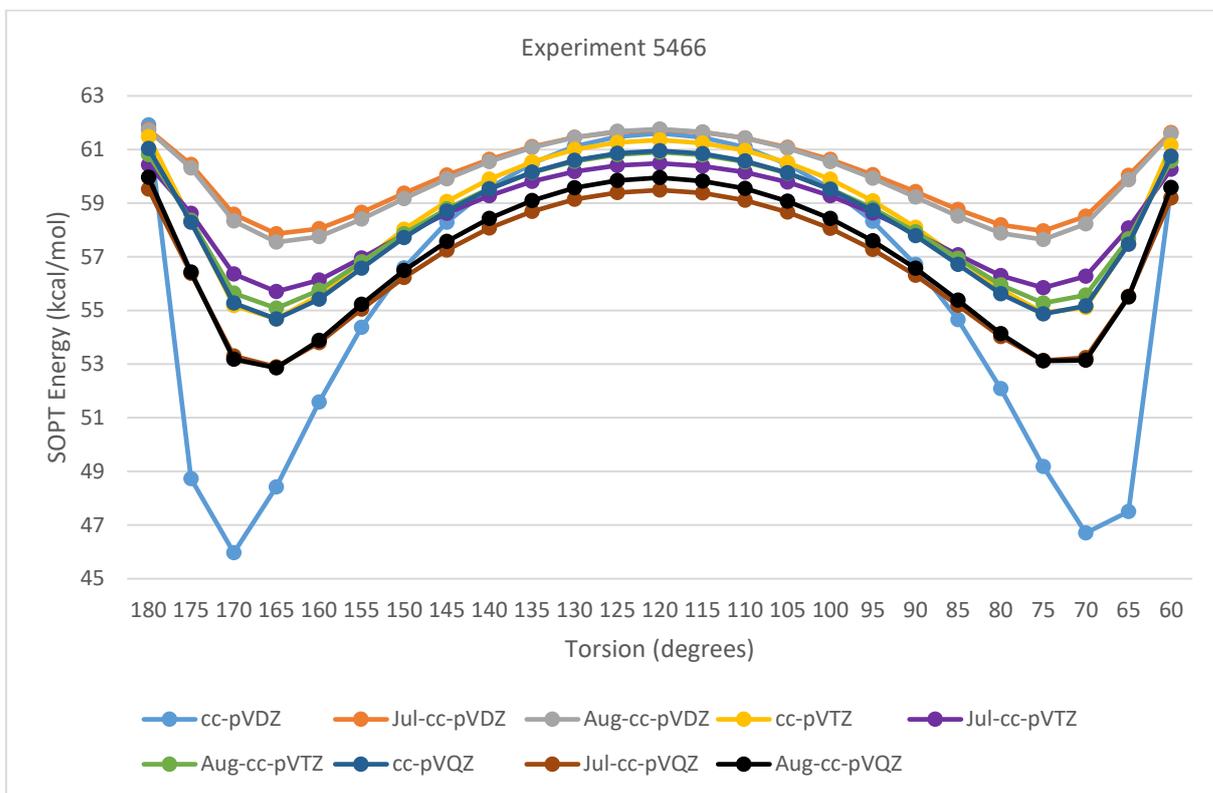

Figure 69. Ethanamide N(lp)->C-O(p)* SOPT Energy with B3LYP over SCS-MP2/def2-QZVPP Optimized Geometry, UltraFineGrid and Varying O-C-CA-H Torsion and Basis Set

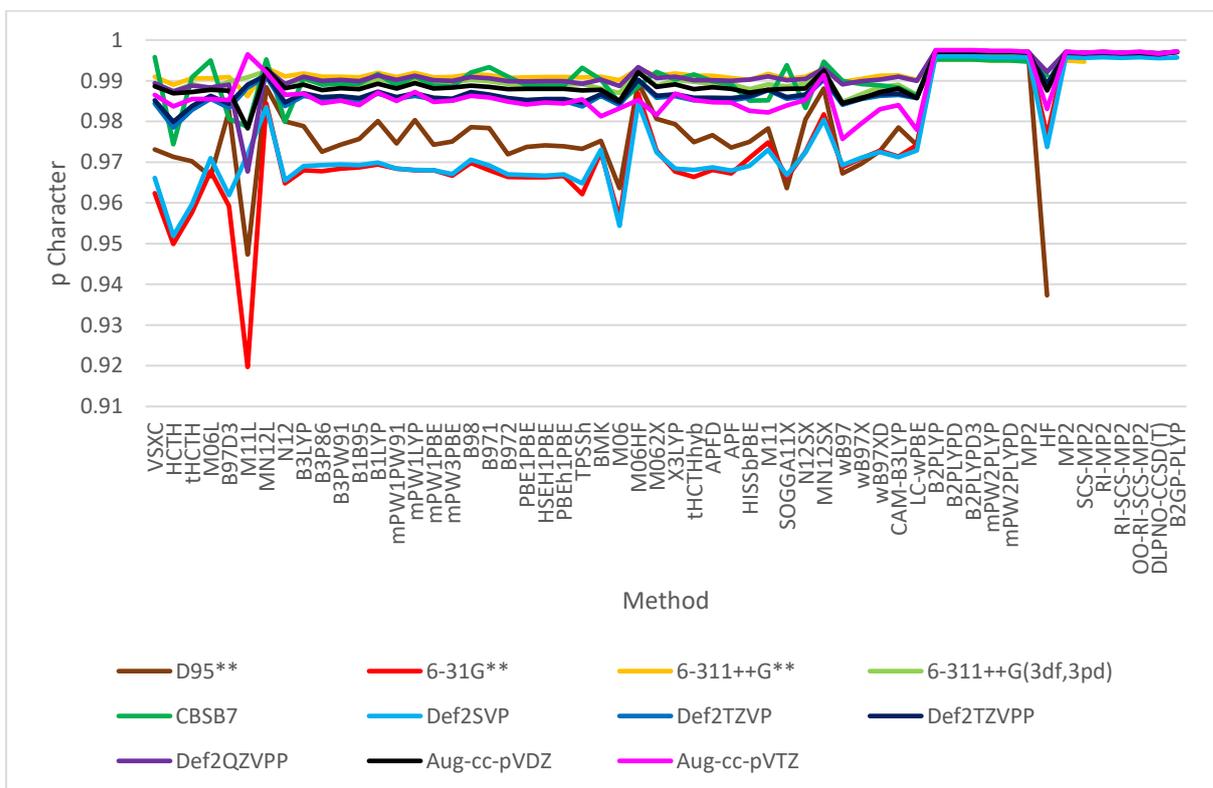

Figure 70. C-O(pi) Carbon NHO p Character in Ethanamide at Antiparallel Beta Strand Psi Torsion with Varying Method and Basis Set

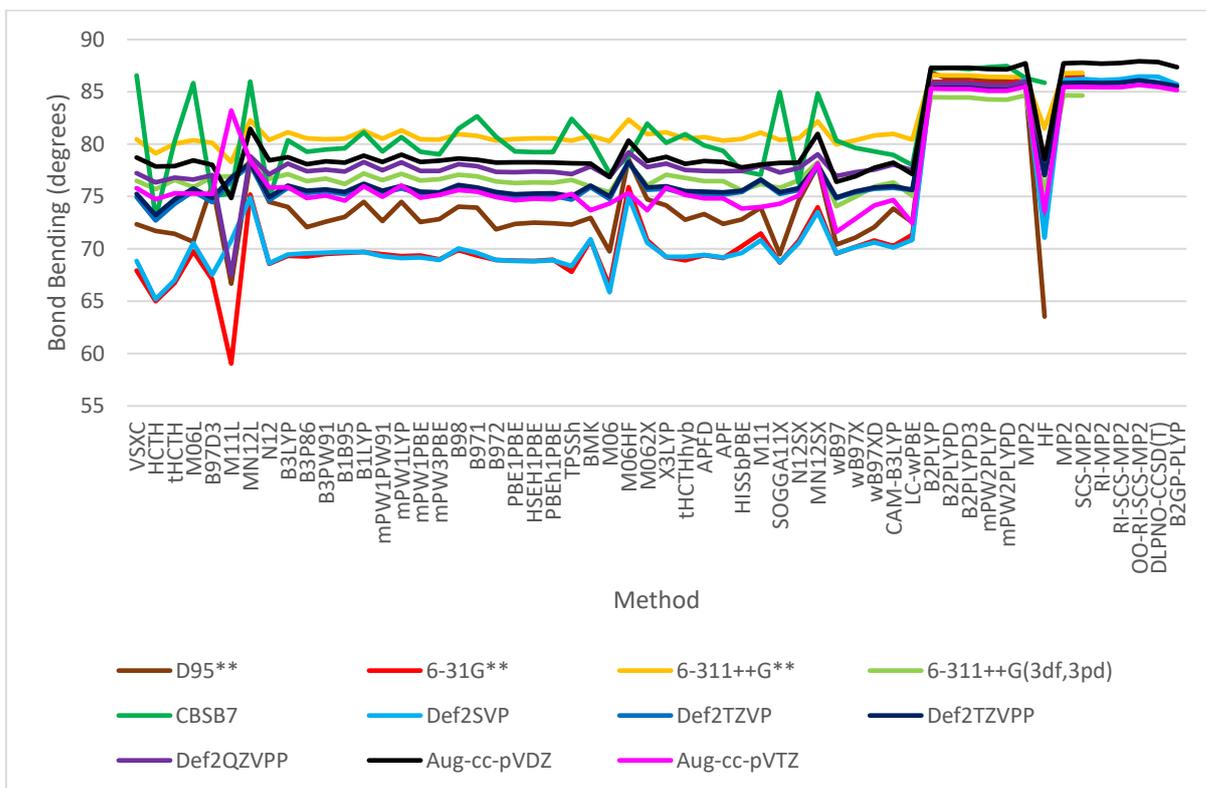

Figure 71. C-O(pi) Carbon NHO Bond Bending at Antiparallel Beta Strand Psi Torsion with Varying Method and Basis Set

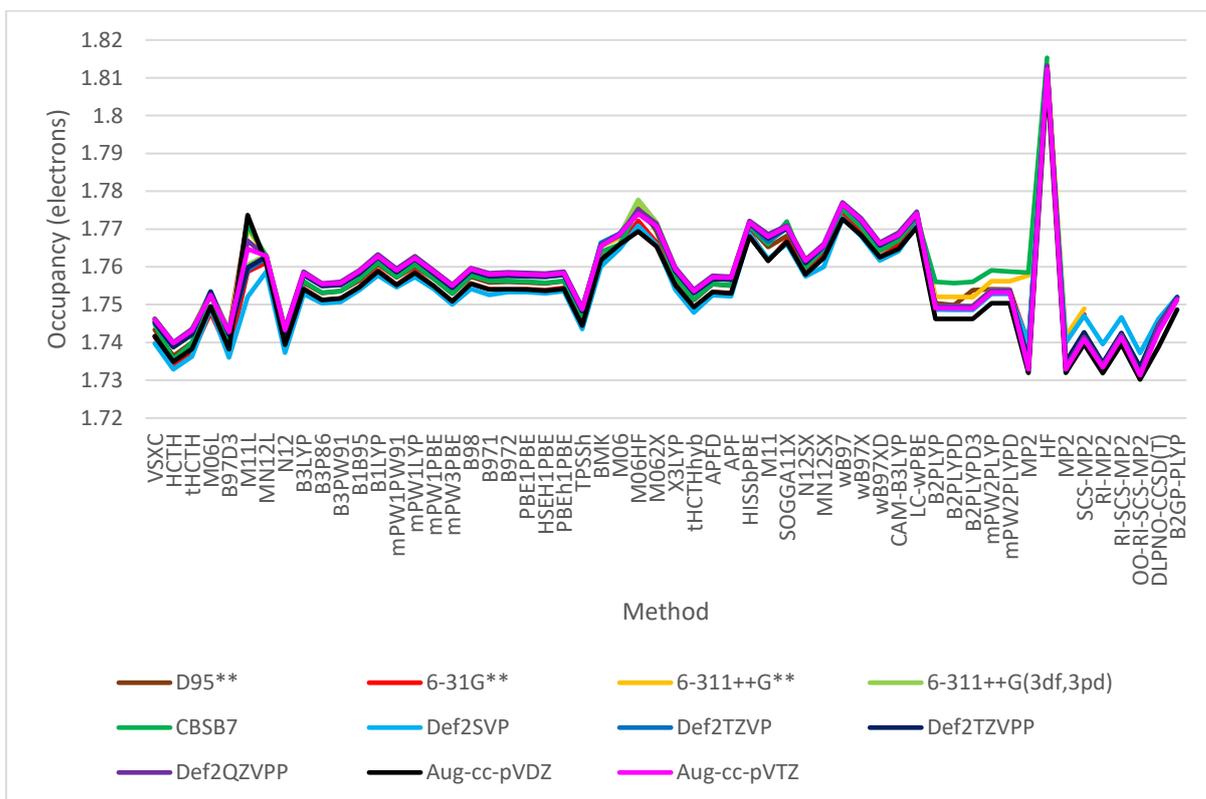

Figure 72. Nitrogen Lone Pair NBO Occupancy in Ethanamide at Antiparallel Beta Strand Psi Torsion with Varying Method and Basis Set

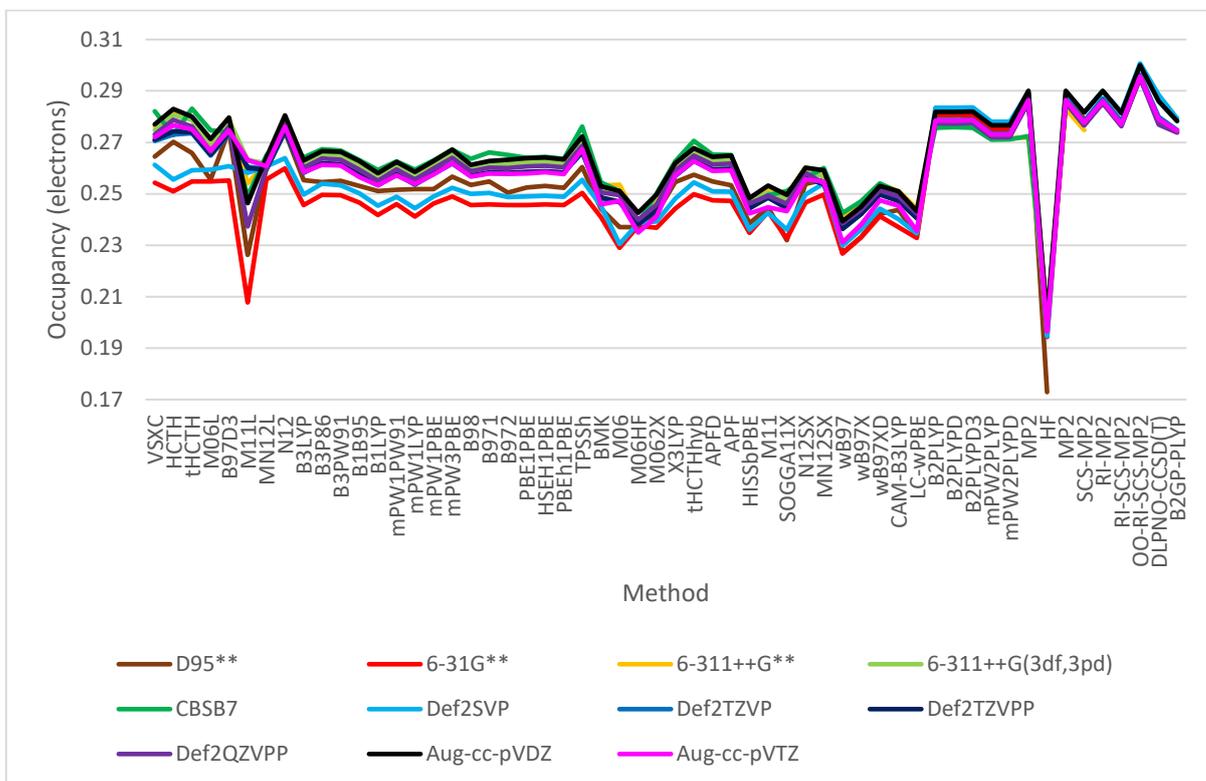

Figure 73. C-O(pi)* NBO Occupancy in Ethanamide at Antiparallel Beta Strand Psi Torsion with Varying Method and Basis Set

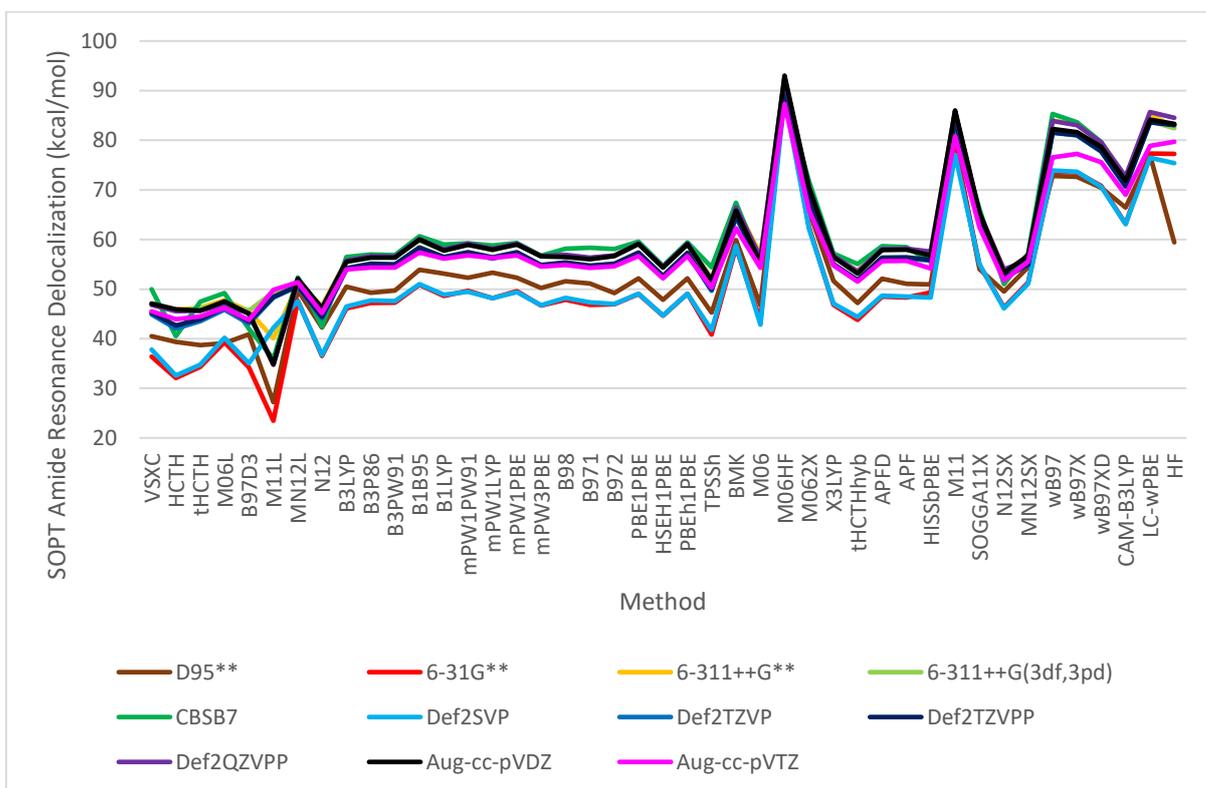

Figure 74. Second-Order Perturbation Theory Amide Resonance Delocalization (N(LP)->C-O(pi)*) in Ethanamide at Antiparallel Beta Strand Psi Torsion and Varying Method and Basis Set

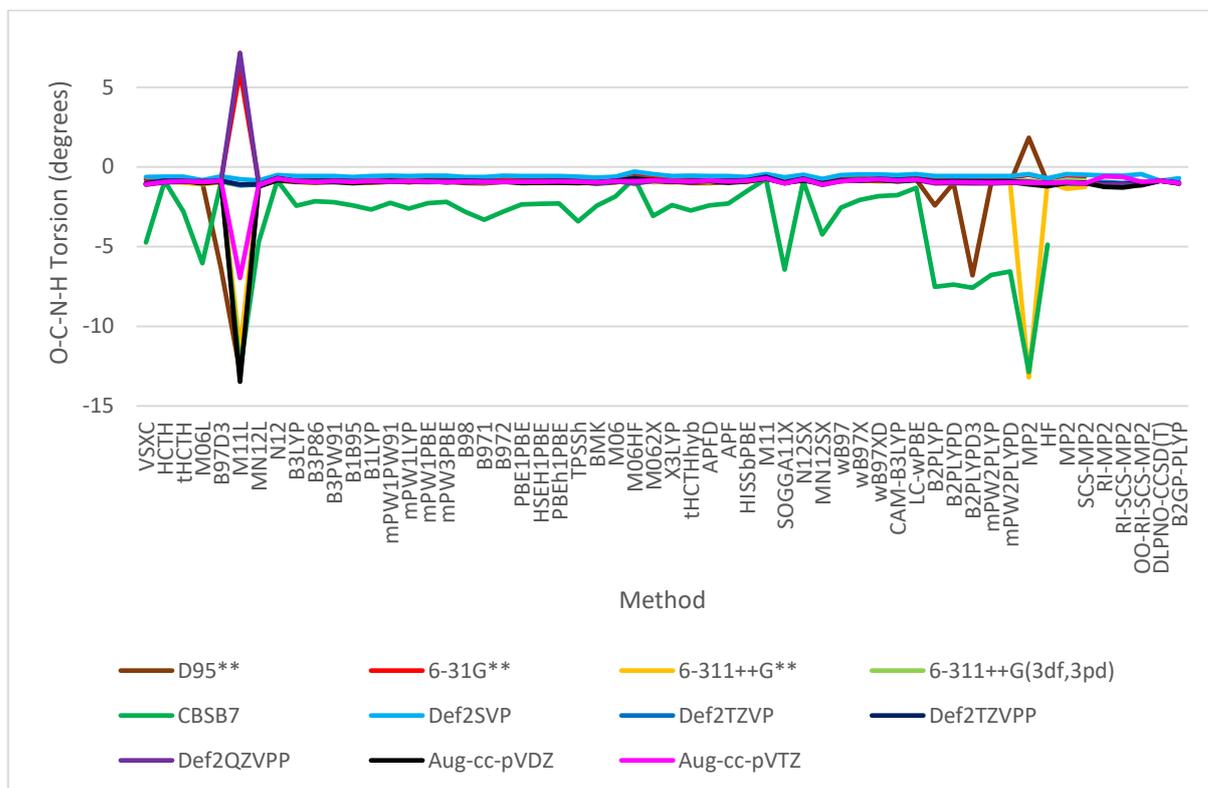

Figure 75. O-C-N-H Torsion in Ethanamide at Antiparallel Beta Strand Psi Torsion with Varying Method and Basis Set

## 11  Appendix 2

Table 1. CASSCF(8,7)/MRCI+Q TZVPP Ethanamide Symmetry and Bond Bending at Beta Sheet Psi Torsions

| Beta Psi Torsion | Energy at Ground | Largest Config Weight | Second Config Weight | p Character of C-O(p) C NHO | Bond Bending of C-O(p) C NHO |
|---|---|---|---|---|---|
| Parallel | -208.862366 | 0.8435 | 0.013 | 0.9968 | 86.2 |
| Antiparallel | -208.871637 | 0.8387 | 0.0186 | 0.9968 | 86.2 |

Table 2. CASSCF(8,7)/MRCI+Q TZVPP Ethanamide N(lp) Occupancy at Varying O-C-CA-H Torsion

| O-C-CA-H deg | Energy at Ground | Largest Config Weight | Second Config Weight | N(lp) Occ |
|---|---|---|---|---|
| 180 | -208.868559 | 0.8372 | 0.0183 | 1.73312 |
| 175 | -208.870524 | 0.839 | 0.0181 | 1.73248 |
| 170 | -208.871203 | 0.8397 | 0.0181 | 1.73322 |
| 165 | -208.871525 | 0.8401 | 0.018 | 1.73368 |

Table 3. Dihedral Angles (degrees) in Ethanamide with O-C-CA-H Constrained to -165 at MP2/6-311++G** with CP Fragment Boundaries at Bond Shown

| Fragment | O-C-N-H1 | O-C-N-H2 |
|---|---|---|
| none | -163.359 | -13.248 |
| C-C | -165.651 | -11.899 |
| C-N | 178.312 | -0.373 |

**Column Names for Table 4 and Table 5**

s: s character of C-O(pi)* Carbon NHO

p: p character of C-O(pi)* Carbon NHO

d: d character of C-O(pi)* Carbon NHO

Dev: bond bending of C-O(pi)* Carbon NHO (degrees)

Dih: O-C-CA-HA dihedral angle

PolP: polarization of C-O(pi) NBO

PolS: polarization of C-O(sigma) NBO

CoapOcc: occupancy of C-O(pi)* NBO

CoasOcc: occupancy of C-O(sigma)* NBO

SoptP: N(LP)->C-O(pi)* SOPT (kcal/mol)

SoptS: N(LP)->C-O(sigma)* SOPT (kcal/mol)

CaHaCos: CA-HA->C-O(sigma)* SOPT (kcal/mol)

CaCbCop: CA-CB->C-O(pi)* SOPT (kcal/mol)

CaNCop: CA-N->C-O(pi)* SOPT (kcal/mol)

CopCaN: C-O(pi)>Ca-N* SOPT (kcal/mol)

Cid: identity of Carbon atom central to backbone amide group

Strand: beta strand in sheet

Chain: cross-strand hydrogen bonding chain

Table 4. Polyvaline Antiparallel Beta Sheet Backbone Amide Resonance Delocalization, Bond Bending and Loss of sigma/pi Symmetry at LC-wPBE/6-31G**

| s | p | d | Dev | Dih | PolP | PolS | CoasOcc | CoapOcc | SoptP | SoptS | CaHaCos | CaCbCop | CaNCop | CopCaN | Cid | Strand | Chain |
|---|---|---|---|---|---|---|---|---|---|---|---|---|---|---|---|---|---|
| 0.0452 | 0.9514 | 0.0034 | 65.33 | 179.3 | 0.2698 | 0.3348 | 0.058 | 0.2841 | 88.32 | 7.33 | 5.16 | 2.98 | | | 80 | 2 | 1 |
| 0.0436 | 0.953 | 0.0034 | 65.66 | -180 | 0.2693 | 0.335 | 0.0566 | 0.2854 | 89.32 | 7.05 | 5.23 | 3.12 | | | 158 | 3 | 4 |
| 0.0226 | 0.9742 | 0.0032 | 70.86 | 158.1 | 0.2895 | 0.3465 | 0.0347 | 0.2583 | 85.52 | 4.28 | 3.71 | | | | 2 | 1 | 4 |
| 0.0214 | 0.9754 | 0.0032 | 71.29 | 158.7 | 0.2891 | 0.3469 | 0.0333 | 0.2603 | 87.24 | 3.85 | 3.8 | 1.03 | | | 236 | 4 | 1 |
| 0.0105 | 0.9859 | 0.0036 | 76.24 | -165.6 | 0.2651 | 0.3424 | 0.0241 | 0.3 | 110.91 | 1.36 | 5.38 | 4.42 | 2.01 | | 252 | 4 | 1 |
| 0.0096 | 0.9869 | 0.0036 | 76.78 | -166 | 0.2644 | 0.3427 | 0.0229 | 0.3013 | 111.94 | 1.18 | 5.38 | 4.42 | 2.01 | | 18 | 1 | 3 |
| 0.0034 | 0.993 | 0.0036 | 82.1 | -171.8 | 0.2699 | 0.3453 | 0.0177 | 0.2961 | 112.24 | | 4.83 | 4.1 | 2.22 | 1.05 | 50 | 1 | 1 |
| 0.0023 | 0.9942 | 0.0036 | 83.4 | -170.6 | 0.2695 | 0.3454 | 0.0166 | 0.2968 | 113.23 | | 4.83 | 4.27 | 2.16 | 1.01 | 284 | 4 | 4 |
| 0.0009 | 0.9957 | 0.0034 | 83.84 | -167.5 | 0.2821 | 0.3504 | 0.0149 | 0.2901 | 111.74 | | 4.53 | 4.41 | 1.85 | | 268 | 4 | 3 |
| 0.0006 | 0.996 | 0.0034 | 84.45 | -168.4 | 0.2821 | 0.3505 | 0.0145 | 0.2907 | 112.24 | | 4.57 | 4.34 | 1.96 | | 34 | 1 | 2 |
| 0.0005 | 0.9958 | 0.0037 | 85.07 | -164.9 | 0.2586 | 0.3443 | 0.0166 | 0.3258 | 128.3 | | 4.76 | 5.03 | 1.62 | | 96 | 2 | 2 |
| 0.0002 | 0.9961 | 0.0037 | 86.16 | -164.1 | 0.2582 | 0.3445 | 0.016 | 0.3265 | 128.93 | | 4.81 | 5.06 | 1.6 | | 174 | 3 | 3 |
| 0.0002 | 0.9961 | 0.0037 | 86.35 | -165.3 | 0.2608 | 0.3447 | 0.0159 | 0.3235 | 127.29 | | 4.8 | 5.17 | 1.59 | | 190 | 3 | 2 |
| 0.0001 | 0.9962 | 0.0037 | 86.93 | -165.7 | 0.2607 | 0.3447 | 0.0156 | 0.3235 | 127.48 | | 4.84 | 5.21 | 1.58 | | 112 | 2 | 3 |
| 0.0003 | 0.9961 | 0.0036 | 87.46 | -173 | 0.2639 | 0.3455 | 0.0154 | 0.3234 | 128.47 | | 5.07 | 4.2 | 2.1 | 1.12 | 128 | 2 | 4 |
| 0.0002 | 0.9962 | 0.0036 | 87.84 | -173.6 | 0.2639 | 0.3456 | 0.0153 | 0.3233 | 128.45 | | 5.05 | 4.13 | 2.17 | 1.14 | 206 | 3 | 1 |

Table 5. Polyalanine Antiparallel Beta Sheet Backbone Amide Resonance Delocalization, Bond Bending and Loss of sigma/pi Symmetry at LC-wPBE/6-31G**

| s | p | d | Dev | Dih | PolP | PolS | CoasOcc | CoapOcc | SoptP | SoptS | CaHaCos | CaCbCop | CopCaCb | Cid | Strand | Chain |
|---|---|---|---|---|---|---|---|---|---|---|---|---|---|---|---|---|
| 0.0215 | 0.9753 | 0.0032 | 71.5 | -169.16 | 0.2922 | 0.3476 | 0.0344 | 0.2602 | 86.01 | 4.07 | 4.71 | | | 64 | 1 | 6 |
| 0.0158 | 0.981 | 0.0032 | 73.75 | -171.07 | 0.2905 | 0.3488 | 0.0288 | 0.2662 | 91.11 | 2.91 | 4.94 | | | 168 | 4 | 1 |
| 0.0139 | 0.9826 | 0.0035 | 73.78 | -151.02 | 0.2784 | 0.3433 | 0.0292 | 0.2744 | 96.89 | 3.13 | 3.42 | 4.72 | 1.13 | 133 | 4 | 6 |
| 0.0124 | 0.9841 | 0.0035 | 74.47 | -149.44 | 0.2785 | 0.3437 | 0.0275 | 0.2753 | 98.12 | 2.79 | 3.36 | 4.74 | 1.16 | 26 | 1 | 1 |
| 0.0131 | 0.9833 | 0.0036 | 74.73 | -175.2 | 0.2615 | 0.3428 | 0.0275 | 0.3139 | 113.01 | 1.71 | 5.75 | | | 48 | 2 | 1 |
| 0.0099 | 0.9865 | 0.0036 | 76.42 | -173.57 | 0.2607 | 0.3434 | 0.0242 | 0.3163 | 115.78 | 1.19 | 5.75 | | | 184 | 3 | 6 |
| 0.0076 | 0.9888 | 0.0036 | 78.23 | -134.89 | 0.2674 | 0.3453 | 0.0204 | 0.302 | 112.74 | 1.36 | 4.24 | 6.38 | 1.66 | 123 | 4 | 4 |
| 0.0057 | 0.9906 | 0.0036 | 79.42 | -134.25 | 0.2665 | 0.3458 | 0.0187 | 0.3047 | 115.6 | | 4.08 | 6.2 | 1.64 | 32 | 1 | 3 |
| 0.0039 | 0.9926 | 0.0035 | 79.8 | -135.81 | 0.2817 | 0.3503 | 0.018 | 0.2859 | 106.46 | 1.54 | 2.67 | 4.34 | 1.44 | 128 | 4 | 5 |
| 0.0032 | 0.9933 | 0.0035 | 80.37 | -137.2 | 0.2813 | 0.3506 | 0.0172 | 0.2872 | 108.25 | 1.24 | 2.8 | 4.59 | 1.45 | 21 | 1 | 2 |
| 0.0025 | 0.994 | 0.0035 | 81.11 | -138.34 | 0.2815 | 0.3511 | 0.0164 | 0.2887 | 110.49 | | 2.98 | 4.71 | 1.49 | 55 | 1 | 4 |
| 0.0025 | 0.9941 | 0.0035 | 81.26 | -137.36 | 0.2815 | 0.351 | 0.0162 | 0.2887 | 110.48 | | 2.91 | 4.63 | 1.49 | 170 | 4 | 3 |
| 0.0035 | 0.9928 | 0.0037 | 81.37 | -141.14 | 0.2623 | 0.3454 | 0.0169 | 0.308 | 118.71 | | 4.55 | 6.59 | 1.57 | 163 | 4 | 2 |
| 0.0034 | 0.993 | 0.0037 | 81.59 | -143.92 | 0.262 | 0.3453 | 0.0169 | 0.3083 | 119.06 | | 4.68 | 6.63 | 1.54 | 59 | 1 | 5 |
| 0.0017 | 0.9946 | 0.0037 | 83.76 | -145.73 | 0.261 | 0.3458 | 0.0153 | 0.3218 | 127.37 | | 4.37 | 6.14 | 1.5 | 50 | 2 | 3 |
| 0.0013 | 0.995 | 0.0037 | 84.27 | -144.83 | 0.2611 | 0.3459 | 0.015 | 0.3217 | 127.45 | | 4.3 | 6.07 | 1.51 | 175 | 3 | 4 |
| 0.0012 | 0.9951 | 0.0037 | 84.47 | -143.88 | 0.262 | 0.3466 | 0.0149 | 0.3209 | 126.95 | | 4.25 | 6.21 | 1.53 | 152 | 3 | 3 |
| 0.0008 | 0.9955 | 0.0037 | 85.21 | -144.36 | 0.262 | 0.3465 | 0.0147 | 0.3213 | 127.31 | | 4.21 | 6.13 | 1.51 | 4 | 2 | 4 |
| 0.0007 | 0.9955 | 0.0037 | 85.4 | -145.35 | 0.257 | 0.3453 | 0.0151 | 0.3258 | 129.51 | | 4.37 | 6.17 | 1.46 | 179 | 3 | 5 |
| 0.0006 | 0.9956 | 0.0037 | 85.55 | -143.09 | 0.2576 | 0.3456 | 0.015 | 0.3248 | 128.93 | | 4.26 | 6.08 | 1.49 | 43 | 2 | 2 |
| 0.0004 | 0.9958 | 0.0038 | 86.19 | -147.33 | 0.2578 | 0.3455 | 0.0147 | 0.3286 | 130.93 | | 4.27 | 6.05 | 1.42 | 141 | 3 | 2 |
| 0 | 0.9964 | 0.0036 | 86.58 | -150.1 | 0.2696 | 0.347 | 0.0145 | 0.3103 | 122.22 | | 4.34 | 5.89 | 1.43 | 14 | 2 | 6 |
| 0.0003 | 0.9961 | 0.0036 | 86.67 | -150.96 | 0.2696 | 0.3472 | 0.0143 | 0.3108 | 122.79 | | 4.47 | 5.98 | 1.44 | 146 | 3 | 1 |
| 0.0001 | 0.9974 | 0.0025 | 88.02 | -146.83 | 0.7422 | 0.6546 | 0.0147 | 0.3285 | 130.83 | | 4.26 | 6.05 | 1.43 | 9 | 2 | 5 |